\documentclass[a4paper,12pt]{article}
\usepackage[a4paper,text={16.8cm,22.4cm}]{geometry}
\usepackage{amsmath,amssymb,latexsym,bm,braket}
\usepackage{cite}
\usepackage{graphicx}
\usepackage{psfrag}

\allowdisplaybreaks 
\newcommand{\Tr}{\mathrm{Tr}}
\newcommand{\Li}{\mathrm{Li}}
\newcommand{\ff}{f\hspace{-0.4em}f}

\newcommand{\scetopi}{\text{1PI}_{\text{SCET}}}
\newcommand{\scetpim}{\text{PIM}_{\text{SCET}}}

\usepackage{multirow}

\begin{document}

\begin{titlepage}

  \begin{flushright}
    MZ-TH/11-03 \\
    ZU-TH 02/11 \\
    \today \\
  \end{flushright}
  
  \vspace{5ex}
  
  \begin{center}
    \textbf{ \Large RG-improved single-particle
 inclusive cross sections  \\  and forward-backward asymmetry  in $t\bar t$
 production at hadron colliders} \vspace{7ex}
    
    \textsc{Valentin Ahrens$^a$, Andrea Ferroglia$^b$, Matthias Neubert$^a$,\\
      Ben D. Pecjak$^a$, and Li Lin Yang$^c$} \vspace{2ex}
  
    \textsl{${}^a$Institut f\"ur Physik (THEP), Johannes Gutenberg-Universit\"at\\
      D-55099 Mainz, Germany\\[0.3cm]
      ${}^b$New York City College of Technology, 300 Jay Street\\
      Brooklyn, NY 11201, USA\\[0.3cm]
      ${}^c$Institute for Theoretical Physics, University of Z\"urich\\
      CH-8057 Z\"urich, Switzerland}
  \end{center}

  \vspace{4ex}

  \begin{abstract}
    We use techniques from soft-collinear effective theory (SCET) to derive
    renormaliza\-tion-group improved predictions for single-particle inclusive (1PI)
    observables in top-quark pair production at hadron colliders. In particular, we study
    the top-quark trans\-verse-momentum and rapidity distributions, the forward-backward
    asymmetry at the Tevatron, and the total cross section at NLO+NNLL order in resummed
    perturbation theory and at approximate NNLO in fixed order. We also perform a
    detailed analysis of power corrections to the leading terms in the threshold expansion
    of the partonic hard-scattering kernels. We conclude that, although the threshold
    expansion in 1PI kinematics is susceptible to numerically significant power
    corrections, its predictions for the total cross section are in good agreement with
    those obtained by integrating the top-pair invariant-mass distribution in
    pair invariant-mass kinematics, as long as a certain set of subleading terms
    appearing naturally within the SCET formalism is included.
\end{abstract}

\end{titlepage}

\section{Introduction}
\label{sec:intro}

The top quark is the heaviest elementary particle known to date. Since its mass $m_t =
(173.3 \pm 1.1)$~GeV \cite{:1900yx} is on the order of the electroweak scale, the top quark
is a crucial tool in the study of the electroweak symmetry breaking mechanism. The
top-quark mass is an important input parameter in electroweak fits \cite{Flacher:2008zq},
and plays a role in the investigation of many Beyond the Standard Model (BSM) scenarios.
The dominant top-quark production mechanism at hadron colliders is the simultaneous
production of a top-anti-top pair. Several differential distributions related to the
production of top-quark pairs, such as the $t\bar{t}$ invariant-mass distribution and the
transverse-momentum distributions for the top quark, allow to investigate the existence of
possible heavy $s$-channel resonances \cite{Frederix:2007gi, Barger:2006hm, Baur:2008uv,
  Hill:1993hs} predicted by many BSM scenarios. Moreover, the rapidity distribution of the
top quark can be used to directly calculate the forward-backward asymmetry at the Tevatron
\cite{Aaltonen:2008hc,:2007qb, Aaltonen:2011kc}, an observable of much interest because of
its potential sensitivity to new physics.

So far, the best measurements of the top-quark mass, couplings, and production cross
sections have been performed using Tevatron data, where this particle was discovered in
1995 \cite{Abe:1995hr}. In the last 15 years a few thousand top-quark pair events were
studied by the CDF and D0 collaborations. The LHC is now running at a center-of-mass
energy of 7~TeV, and very recently the first measurements of the total pair-production
cross section was performed by the CMS \cite{Khachatryan:2010ez} and ATLAS
\cite{Aad:2010ey} collaborations. Hopes are that it will be possible to obtain an
integrated luminosity of up to 1\,fb$^{-1}$, which would produce roughly 150,000 
$t\bar{t}$ events before selection \cite{Bernreuther:2010cv}. With the planned increases
in the center-of-mass energy and luminosity at the LHC, it will eventually be possible to
observe millions of top quarks per year. The increase of the production rate will induce a
decrease of the experimental errors on the measured observables. For the total inclusive
cross section the relative experimental uncertainty is expected to become of the order of
5\%--10\% \cite{Bernreuther:2008ju}.
 
In view of the precise measurements at the LHC and at the Tevatron, it is crucial to
obtain theoretical predictions for the measured observables which are as accurate as
possible. The total pair-production cross section has been known at next-to-leading order
(NLO) for over two decades \cite{Nason:1987xz, Beenakker:1988bq, Beenakker:1990maa,
  Czakon:2008ii}. Some time later also differential distributions \cite{Nason:1989zy,
  Mangano:1991jk, Frixione:1995fj} and the forward-backward asymmetry \cite{Kuhn:1998jr,
  Kuhn:1998kw} were calculated to the same accuracy. As the NLO calculations suffer from
uncertainties larger than $10\%$, it would be desirable to extend them to
next-to-next-to-leading order (NNLO) in a fixed-order expansion in the strong coupling
constant. Full NNLO predictions require the calculation of two sets of corrections:
\textit{i) virtual corrections}, which can be split into genuine two-loop diagrams
\cite{Czakon:2007ej, Czakon:2007wk, Czakon:2008zk, Bonciani:2008az,
  Bonciani:2009nb,Bonciani:2010mn} and one-loop interference terms \cite{Korner:2008bn,
  Anastasiou:2008vd, Kniehl:2008fd}; \textit{ii) real radiation}, which involves one-loop
diagrams with the emission of one extra parton in the final state, and tree-level
diagrams with two extra partons in the final state \cite{Dittmaier:2007wz,
  Dittmaier:2008yq, Bevilacqua:2010ve, Melnikov:2010iu}. In spite of progress made by
several groups on different aspects of the NNLO calculations in the last few years,
especially in developing a new subtraction scheme and calculating the contributions from
double real radiation \cite{Czakon:2010td, Czakon:2011ve}, a significant amount of work is
still required to assemble all the elements.

Another way to improve on the fixed-order NLO calculation (and also the NNLO one, upon its
completion) is to supplement it with threshold resummation \cite{Sterman:1986aj,
  Catani:1989ne}. More precisely, one identifies a threshold parameter which vanishes in
the limit where real gluon emission is soft, expands the result to leading power in this
parameter, and uses renormalization-group (RG) methods to resum logarithmic corrections in this
parameter to all orders in the strong coupling constant. For the total cross section, one
such approach is to work in the threshold limit $\beta=\sqrt{1-4m_t^2/\hat{s}} \to 0$,
where $\hat{s}$ is the partonic center-of-mass energy squared, and $\beta$ is
approximately the velocity of the top (or anti-top) quark. In this production threshold
limit the top quarks are produced nearly at rest and there are logarithmic terms of the
form $\alpha_s^n\ln^m\beta$ (with $m\le 2n$), 
which can be resummed to all orders. This has been done to
leading-logarithmic (LL) order \cite{Laenen:1991af, Laenen:1993xr, Berger:1995xz,
  Berger:1996ad, Berger:1997gz, Catani:1996dj}, next-to-leading-logarithmic (NLL) order
\cite{Bonciani:1998vc}, and next-to-next-to-leading-logarithmic (NNLL) order
\cite{Moch:2008qy, Czakon:2008cx, Langenfeld:2009wd, Beneke:2009rj, Czakon:2009zw,
  Beneke:2009ye}. Note that in this case not only logarithmic corrections, but
  also Coulomb corrections involving inverse powers of $\beta$ occur. An approach to
  all-orders resummation of Coulomb terms can be found in \cite{Beneke:2010da}.

A drawback of this method is that it performs resummation of terms that become important in the region $\beta\to 0$, which however gives a very small contribution to the total cross section. At the Tevatron and LHC, typical values of $\beta$ are in the range between 0.4 and 0.9 \cite{Ahrens:2010zv}. 
An alternative approach is to perform the threshold expansion and resummation at the level
of differential distributions, which are interesting in their own right, and to obtain the
total cross section by integrating the results. One such method works with the top-pair
invariant-mass distribution $d\sigma/dM$, where $M$ is the invariant mass of the
$t\bar{t}$ pair. The threshold limit for this case of pair-invariant-mass (PIM) kinematics
is defined as $z=M^2/\hat{s} \to 1$, and the corresponding threshold logarithms are of the
form $\alpha_s^n\,[\ln^m(1-z)/(1-z)]_+$ (with $m\le 2n-1$). In this limit only soft gluons can be emitted, but
$\beta$ is a generic $\mathcal{O}(1)$ parameter, and the top-quark velocity need not be
small. Systematic resummation and fixed-order expansions of these logarithms has been
studied in Mellin moment space at NLL order \cite{Kidonakis:1996aq, Kidonakis:1997gm,
  Kidonakis:2000ui, Kidonakis:2001nj, Kidonakis:2003qe, Banfi:2004xa, Almeida:2008ug,
  Kidonakis:2008mu} and recently also in momentum space at NNLL order \cite{Ahrens:2009uz,
  Ahrens:2010zv}, using techniques from soft-collinear effective theory (SCET). In
general, it can be imagined that the approach based on differential distributions captures
more contributions than the approach based on the small-$\beta$ expansion, and therefore
gives more reliable predictions for the total cross section. In \cite{Ahrens:2010zv}, we
have argued that this is indeed the case. We also pointed out that in the effective-theory
calculation the argument of the soft logarithms involves the ratio $2E_s/\mu$, with $2E_s=M(1-z)/\sqrt{z}$
the energy of the soft radiation. This allowed us to identify a set of power-suppressed
terms proportional to $\ln z/(1-z)$, and a numerical analysis showed that incorporating
these terms in the fixed-order threshold expansion greatly improves the agreement between
the approximated results in PIM kinematics and the exact results, as was indeed expected
based on previous studies of Drell-Yan \cite{Becher:2007ty} and Higgs production
\cite{Ahrens:2008qu,Ahrens:2008nc} in SCET.

In addition to the invariant-mass distribution of the top-quark pair, the transverse-momen\-tum and rapidity distributions of the top quark (or anti-top quark) are also
interesting. In the case of distributions of the top quark, one collects the anti-top quark
and extra radiation into an inclusive hadronic state $X[\bar{t}]$ with total momentum
$p_X$, and defines the threshold limit as $p_X^2 \to m_t^2$. In this limit of
single-particle inclusive (1PI) kinematics, only soft radiation is allowed, but as in PIM
kinematics the parameter $\beta$ is an ${\cal O}(1)$ quantity. Soft gluon resummation for
this case has been developed in Mellin moment space \cite{Laenen:1998qw} and applied to
the top-quark transverse-momentum distribution at NLL order \cite{Kidonakis:2000ui} and
recently also at NNLL order \cite{Kidonakis:2010dk}, in the form of approximate NNLO
predictions. Starting from these distributions, it is possible to obtain the total cross
section by integrating over the kinematic variables. This approach has been taken in
\cite{Kidonakis:2001nj, Kidonakis:2003qe, Kidonakis:2008mu, Kidonakis:2010dk}, and a large
discrepancy from the results based on PIM kinematics was found. It was argued in
\cite{Kidonakis:2003qe, Kidonakis:2010dk} that the results based on PIM kinematics largely
underestimate the cross section, while those based on 1PI kinematics more reasonably
account for the higher-order corrections. In principle, in the threshold limit these two
kinematics encode the same soft gluon physics. Any differences between the two cases are
due to power-suppressed corrections. At realistic collider energies, however, subleading
terms are in general non-negligible, and one should study them carefully before drawing
any definite conclusions. In light of the improved behavior of the PIM expansion upon the
inclusion of the $\ln z/(1-z)$ terms observed in \cite{Ahrens:2010zv}, it is natural to
ask whether including the analogous set of terms in 1PI kinematics has a similar effect,
and if so, whether it can account for the discrepancies between the two types of
kinematics observed in previous work.

In this paper, we extend our effective-field theory approach for PIM kinematics to the
case of 1PI distributions, performing an NLO+NNLL resummation directly in momentum space,
and also deriving approximate NNLO formulas equivalent to those from
\cite{Kidonakis:2010dk}. We then apply these results to the transverse-momentum and
rapidity distributions of the top quark, as well as the forward-backward asymmetry at the
Tevatron. As in the case of PIM kinematics, in the effective-theory formulation the soft
threshold logarithms involve the ratio $2E_s/\mu$. This allows us to identify a set of
power-suppressed terms proportional to $\ln(1+s_4/m_t^2)/s_4$, where $s_4$ is defined in
(\ref{eq:mandelstampart}) below and approaches zero in the threshold limit. Through
detailed studies of differential distributions and the total cross section, we show that
although the power-suppressed terms in 1PI kinematics can be significant, including the
set of power corrections identified through the effective field-theory techniques
greatly improves the behavior of the 1PI threshold expansion in regions where the
differential cross section is large. In fact, we find that the results for the total cross
section at NLO+NNLL and approximate NNLO are very much compatible within the two
types of kinematics, as long as the extra terms appearing within the effective-theory
analyses are included.

The organization of the paper is as follows. In Section~\ref{sec:kin}, we review the
kinematics of top-quark pair production in hadronic collisions and define the threshold
regions in 1PI and PIM kinematics. Section~\ref{sec:SCET} contains the main new
theoretical results of our work. In that section we explain how to extend the SCET
formalism to 1PI kinematics and assemble all of the perturbative ingredients needed for
NLO+NNLL and approximate NNLO predictions. This includes a calculation of the soft
matching function at NLO in $\alpha_s$, which is a necessary ingredient for NNLL
resummation but was so far unknown. We also demonstrate how the extra subleading terms in the
threshold expansion appear naturally in our formalism. In Section~\ref{sec:numerical}, we
define the scheme needed to turn these theoretical results into numerical predictions.
This includes not only a review of input parameters such at the top-quark mass and parton
distribution functions (PDFs), but also a definition of the scheme used in evaluating the
formulas at threshold. We then turn to numerical results in Section~\ref{sec:threshold},
performing a very detailed comparison of the 1PI results with exact results at NLO for the
differential distributions, and also with the PIM results at NLO and NNLO for the total
cross section. Finally, in Section~\ref{sec:pheno} we give concrete results for the
transverse-momentum and rapidity distributions, the forward-backward asymmetry at the
Tevatron, and the total cross section, both at NLO+NNLL and approximate NNLO. We
conclude with a summary of our findings in Section~\ref{sec:conclusions}.

\section{Kinematics}
\label{sec:kin}

We consider the scattering process
\begin{align}
  \label{eq:process}
  N_1(P_1) + N_2(P_2) \to t(p_3) + \bar{t}(p_4) + X \, ,
\end{align}
where $N_1$ and $N_2$ indicate the incoming protons (LHC) or proton and anti-proton
(Tevatron), while $X$ represents an inclusive hadronic final state. In the Born
approximation, two different production channels contribute to the scattering process
(\ref{eq:process}): the quark-anti-quark annihilation and gluon-gluon fusion channels. At
higher orders in the strong coupling constant, in addition to these two channels, there
are also contributions from other partonic channels such as quark-gluon scattering. These
additional channels are power suppressed in the partonic threshold region which we will
define below, therefore we will not discuss them in this work. Nevertheless we will
include them in the numerical results when matching with fixed-order calculations.

The partonic processes which we will analyze in detail are thus
\begin{align}
  \label{eq:partprocess}
  q(p_1) + \bar{q}(p_2) &\to t(p_3) + \bar{t}(p_4) + \hat{X}(k) \, , \nonumber
  \\
  g(p_1) + g(p_2) &\to t(p_3) + \bar{t}(p_4) + \hat{X}(k) \, .
\end{align}
Note that the hadronic state $\hat{X}$ in the above equations is different than the state
$X$ in (\ref{eq:process}): $\hat{X}$ contains only the products of the hard-scattering,
while $X$ contains also the beam remnants from the initial hadrons. The relations between
the hadronic and partonic momenta are $p_1=x_1P_1$ and $p_2=x_2P_2$. At the hadronic
level, we define the kinematic invariants as
\begin{align}
  \label{eq:mandelstam}
  s &= (P_1+P_2)^2 \, , \quad t_1 = (P_1-p_3)^2-m_t^2 \, , \quad u_1=(P_2-p_3)^2 -m_t^2 \,
  ,
\end{align}
while the corresponding quantities at the partonic level are given by
\begin{gather}
  \hat{s} = x_1 x_2 s \, , \quad \hat{t}_1 = x_1 t_1 \, , \quad \hat{u}_1= x_2 u_1 \, ,
  \nonumber
  \\
  M^2 = (p_3+p_4)^2 \, , \quad s_4 = \hat{s} + \hat{t}_1 + \hat{u}_1 = (p_4+k)^2 - m_t^2
  \, . \label{eq:mandelstampart}
\end{gather}
Momentum conservation implies that at Born level (for $k=0$) we have $\hat{s}=M^2$ and $s_4 = 0$.

Given the kinematic variables above, it is possible to define different threshold regions
depending on the observables of interest. For example, in the case of the invariant-mass
distribution of the top-quark pair, it is natural to define the threshold region as
$\hat{s} \to M^2$. This case, which is conventionally called pair invariant-mass (PIM)
kinematics, has been analyzed in \cite{Kidonakis:1997gm, Kidonakis:2001nj,
  Kidonakis:2008mu, Kidonakis:2003qe, Ahrens:2010zv}. On the other hand, in the case of
single-particle inclusive (1PI) observables such as the transverse-momentum or rapidity
distribution of a single top quark, it is natural to define the threshold region as $s_4
\to 0$. This threshold limit was first studied in \cite{Laenen:1998qw} and is the subject
of this work. It should be stressed that, in both the PIM and 1PI threshold regions, the
top and anti-top quarks are not forced to be nearly at rest. Instead, the top-quark
velocity $\beta=\sqrt{1-4m_t^2/\hat{s}}$ is considered a generic ${\cal O}(1)$ parameter,
neither close to zero nor close to unity, which is in fact the region of phase space where
the differential cross section is largest \cite{Ahrens:2010zv}. This differs from the situation encountered in
many calculations of the total top-quark pair production cross section, which are carried
out in the limit of vanishing top-quark velocity $\beta\to 0$ \cite{Bonciani:1998vc,
  Kidonakis:2000ui, Kidonakis:2001nj, Kidonakis:2003qe, Moch:2008qy, Kidonakis:2008mu,
  Czakon:2008cx}. We will refer to the kinematic region where $\beta \to 0$ as the
production threshold limit. Note that at very large transverse momentum the top-quark velocity $\beta$ approaches 
1. To deal with such highly boosted top quarks would require a different type of
effective theory, which has been developed in \cite{Fleming:2007qr} for $e^+e^-$
collisions, but has not yet been extended to hadron colliders.

The common link between these different threshold limits is that they in one way or
another force the emitted real radiation to be soft. One uses this fact to define a
partonic expansion parameter which vanishes in the exact threshold limit, where the energy
of the soft radiation goes to zero, and calculates the partonic hard-scattering kernels as
a threshold expansion in that parameter. However, unless restrictive kinematic cuts are
applied, the convolution integral between the partonic hard-scattering kernels and PDFs is sensitive to kinematic regions where the partonic
threshold parameter is not small. In that case, power corrections to the threshold
expansion can become important, and these have a different structure in each of the three
limits mentioned above. In \cite{Ahrens:2010zv}, we analyzed several advantages of
calculations of the total cross section carried out in the partonic threshold limit for
PIM kinematics with respect to calculations based on the production threshold limit.
Similar considerations apply to the partonic threshold limit for 1PI kinematics, so when
we discuss the total cross section in the present work we focus instead on a comparison
between 1PI and PIM.

Several phenomenologically interesting observables are related to the detection of a
single particle rather than a pair, and 1PI kinematics is suitable to describe such
observables. In this paper we focus our attention on the top-quark transverse-momentum and
rapidity distributions, as well as on the total cross section. The double-differential
cross section in the transverse momentum $p_T$ and the rapidity $y$ can be written in the
factorized form
\begin{align}
  \label{eq:pt,y}
  \frac{d\sigma}{dp_Tdy} &= \frac{16\pi p_T}{3s} \sum_{i,j} \int_{x_1^{\text{min}}}^1
  \frac{dx_1}{x_1} \int_{x_2^{\text{min}}}^{1} \frac{dx_2}{x_2} \, f_{i/N_1}(x_1,\mu_f) \,
  f_{j/N_2}(x_2,\mu_f) \, C_{ij}(s_4,\hat{s},\hat{t}_1,\hat{u}_1,m_t,\mu_f) \, ,
\end{align}
where the $f_{i/N}$ are universal non-perturbative PDFs for the
parton $i$ in the hadron $N$, and the hard-scattering kernels $C_{ij}$ are related to the
partonic cross section and can be calculated perturbatively as series in the strong
coupling constant. We define their expansion coefficients according to
\begin{align}
  \label{eq:cexp}
  C_{ij} = \alpha_s^2 \left[ C_{ij}^{(0)} + \frac{\alpha_s}{4\pi} C_{ij}^{(1)} + \left(
      \frac{\alpha_s}{4\pi} \right)^2 C_{ij}^{(2)} + \ldots \right] .
\end{align}
The hadronic Mandelstam variables are related to $p_T$ and $y$ in the laboratory frame via
\begin{align}
  t_1 = -\sqrt{s} \, m_\perp \, e^{-y} \, , \quad u_1 = -\sqrt{s} \, m_\perp e^{y} \, ,
\end{align}
where $m_\perp=\sqrt{p_T^2+m_t^2}$. Together with (\ref{eq:mandelstampart}), the kinematic
variables entering $C_{ij}$ can be expressed as functions of $p_T$, $y$, $x_1$ and $x_2$.
The lower limits of integrations in (\ref{eq:pt,y}) are
\begin{align}
  x_1^{\text{min}} = \frac{-u_1}{s+t_1} \, , \quad x_2^{\text{min}} =
  \frac{-x_1t_1}{x_1s+u_1} \, . \nonumber
\end{align}
Finally, the total cross section can be calculated by integrating the double-differential
distribution over the ranges
\begin{align}
  \label{eq:ylimits}
  0 \leq |y| \leq \frac{1}{2} \ln\frac{1+ \sqrt{1-4m_\perp^2/s}}{1-
      \sqrt{1-4m_\perp^2/s}} \, , \quad 0 \leq p_T \leq \sqrt{\frac{s}{4}-m_t^2} \,
  .
\end{align}

In this paper we will mainly discuss the distributions of the top quark, but our results
can also be applied to the transverse momentum $\bar{p}_T$ and rapidity $\bar{y}$
distributions of the anti-top quark after appropriate replacements. At the Tevatron,
charge-conjugation invariance of the strong interactions implies that within QCD we have the simple
relation
\begin{align}
  \label{eq:cpi}
  \frac{d\sigma}{d\bar{p}_Td\bar{y}} = \frac{d\sigma}{dp_Tdy} \Big|_{p_T\to\bar{p}_T, \,
    y=\to-\bar{y}} \, ,
\end{align}
which we will use later on in interpreting the charge asymmetry in terms of a
forward-backward asymmetry.

In the threshold limit $s_4 \to 0$ only the emission of soft radiation is allowed. In this
limit calculations are greatly simplified, since one effectively deals with a two-body
final state. The hard-scattering kernels can be factorized into a hard function
$\bm{H}_{ij}$ and a soft function $\bm{S}_{ij}$ as
\begin{align}
  \label{eq:fac}
  C_{ij}(s_4,\hat{s},\hat{t}_1,\hat{u}_1,m_t,\mu_f) = \Tr \left[
    \bm{H}_{ij}(\hat{s}',\hat{t}_1',\hat{u}_1',m_t,\mu_f) \,
    \bm{S}_{ij}(s_4,\hat{s}',\hat{t}_1',\hat{u}'_1,m_t,\mu_f) \right] + \mathcal{O}(s_4)
  \, .
\end{align}
The notation above is meant to emphasize that there are ambiguities in the choice of
$\hat{s}'$, $\hat{t}_1'$ and $\hat{u}_1'$, which can in general differ from the exact
Mandelstam variables $\hat{s}$, $\hat{t}_1$ and $\hat{u}_1$ by power corrections that
vanish at $s_4=0$. For instance, given an explicit result for the hard and soft functions,
one can always rewrite it using $\hat{s}'+\hat{t}_1'+\hat{u}_1'=0$ or
$\hat{s}'+\hat{t}_1'+\hat{u}_1'=s_4$. Although the difference is suppressed by positive
powers of $s_4$, the two choices give different numerical results upon integration. In
Section~\ref{sec:numerical}, we will explain in detail our method for dealing with this
ambiguity.

The use of boldface in (\ref{eq:fac}) indicates that the hard and soft functions are
matrices in color space. The hard function originates from virtual corrections and is the
same as that in the PIM case \cite{Ahrens:2010zv}, once the variables $\hat{s}'$,
$\hat{t}_1'$, $\hat{u}_1'$ are expressed in terms of $M$ and $\cos\theta$. The soft
function captures contributions arising from the emission of soft real radiation. It
depends on the details of the phase-space integrals and is different in 1PI and PIM
kinematics. The 1PI soft function contains singular distributions in $s_4$, which are of
the form
\begin{align}
  \label{eq:plus}
  P_n(s_4) \equiv \left[ \frac{1}{s_4} \ln^n\frac{s_4}{m_t^2} \right]_+ 
  ,
\end{align}
where the plus-distributions are defined by
\begin{align}
  \int_0^{m_t^2} ds_4 \left[ \frac{1}{s_4} \ln^n \frac{s_4}{m_t^2} 
  \right]_+ g(s_4) = \int_0^{m_t^2} ds_4 \, \frac{1}{s_4} \ln^n\!\bigg( \frac{s_4}{m_t^2} \bigg)
  \left[ g(s_4) - g(0) \right] .
\end{align}
With this definition
\begin{align}
  \int_0^{s_4^{\text{max}}} \left[ \frac{1}{s_4} \ln^n \frac{s_4}{m_t^2} 
  \right]_+ g(s_4) = \int_0^{s_4^{\text{max}}} ds_4 \, \frac{1}{s_4} \ln^n\!\bigg( 
    \frac{s_4}{m_t^2} \bigg) \left[ g(s_4) - g(0) \right] + \frac{g(0)}{n+1} \ln^{n+1}
  \!\bigg( \frac{s_4^{\text{max}}}{m_t^2} \bigg) \,.
\end{align}
Near threshold, these singular distributions lead to a bad convergence of the perturbation
series. More generally, they give rise to the dominant higher-order corrections to the
hadronic cross section if the product of PDFs in (\ref{eq:pt,y}) falls off very quickly
away from values of $x_1,x_2$ where $s_4\to 0$. In either case, resumming such terms to all
orders can lead to improved theoretical predictions. This is the topic of the next
section.

\section{Factorization and resummation in SCET}
\label{sec:SCET}

In \cite{Ahrens:2010zv}, factorization and resummation for the case of PIM kinematics was
studied in detail. Most of that discussion can be carried over directly to the case of 1PI
kinematics. In fact, the differences between 1PI and PIM kinematics arise solely from the
structure of real emission in the two cases and therefore affect only the soft function,
which must be modified in two essential ways. First, the phase-space integrals for real
emission in the soft limit change, so the fixed-order expansion of the soft function is
different from its PIM counterpart and must be calculated from scratch. Second, the RG
equation for the soft function, derived using the RG invariance of the cross section along
with the evolution equations of the hard function and PDFs, also differs slightly from its
expression in PIM kinematics. In what follows we focus on how to deal with these two differences with
respect to PIM kinematics and otherwise just quote results from \cite{Ahrens:2010zv} for
the pieces which remain unchanged. In particular, we derive the explicit one-loop soft
function and the RG equation needed for NNLL resummation in 1PI
kinematics, and present results for approximate NNLO formulas in fixed-order perturbation
theory. We also discuss the structure of power corrections to the leading-order term in
the threshold expansion and explain how a certain set of subleading corrections in $s_4$
appears naturally within the SCET formalism.

\subsection{Soft function in 1PI kinematics}
\label{sec:soft}

In general, the soft function is related to the vacuum expectation value of a soft Wilson-loop operator. To calculate it explicitly, we first generalize the derivation of the
differential cross section at partonic threshold given for PIM kinematics in
\cite{Ahrens:2010zv} to the 1PI case.

In the limit where extra gluon radiation is soft, the differential cross section
can be factorized as\footnote{The functions $\bm{H}$ and $\bm{W}$ are summed over the channel indices
  $ij$. In order to keep the notation as simple as possible, in the remainder of this
  section we suppress the sum and indices.}
\begin{align}
  \label{eq:Forward}
  d\hat\sigma &= \frac{1}{2\hat{s}} \frac{d^3{\vec{p}}_3}{(2\pi)^32E_3} \int
  \frac{d^3\vec{p}_4}{(2\pi)^32E_4} \int d^4x \, e^{i(p_1+p_2-p_3-p_4) \cdot x} \nonumber
  \\
  &\quad\times \frac{128\pi^2}{3} \Tr \big[
  \bm{H}(\hat{s}',\hat{t}_1',\hat{u}_1',m_t,\mu_f) \,
  \bm{W}(x,\hat{s}',\hat{t}_1',\hat{u}_1',m_t,\mu_f) \big] \, ,
\end{align}
where $\bm{W}$ is the expectation value of the Wilson-loop operator in position space.
Since the integrand depends on $\vec{p}_4$ only through $E_4=\sqrt{|\vec{p}_4|^2+m_t^2}$
and in the exponent, its calculation is simplified by going to the rest frame of the
inclusive final state $\bar t+\hat{X}$ in (\ref{eq:partprocess}), which consists of the
anti-top-quark plus additional soft radiation. In this frame $|\vec{p}_4|={\cal
  O}(s_4/m_t)$, and we can drop the dependence on it in $E_4 \sim m_t$. The integral over
$\vec{p}_4$ then produces a factor of $(2\pi)^3 \delta^{(3)}(\vec x)$, and the exponent
depends only on $E_s \equiv (p_1+p_2-p_3-p_4)^0 = s_4/(2\sqrt{s_4+m_t^2})$, which is the
energy of the soft radiation in this particular frame. Using the $\delta$-function to
perform the integral over $d^3\vec x$, and converting the result to a differential cross
section in $\hat{t}_1$ and $\hat{u}_1$, we obtain
\begin{align}
  \label{eq:fac2}
  \frac{d\hat\sigma}{d\hat{t}_1 d\hat{u}_1} &= \frac{8\pi}{3\hat{s}^2} \frac{1}{m_t} \int
  \frac{dx^0}{4\pi} \, \exp \bigg( \frac{ix^0 s_4}{2\sqrt{s_4+m_t^2}} \bigg) \nonumber
  \\
  &\quad \times \Tr \big[ \bm{H}(\hat{s}',\hat{t}_1',\hat{u}_1',m_t,\mu_f) \,
  \bm{W}\!\left((x^0,\vec{x}=0),\hat{s}',\hat{t}_1',\hat{u}_1',m_t,\mu_f\right) \big] \, .
\end{align}
We now introduce the momentum-space soft function according to
\begin{align}
  \hat{\bm{W}}(\omega,\hat{s}',\hat{t}'_1,\hat{u}'_1,m_t,\mu) = \int \frac{dx^0}{4\pi}
  \exp \bigg( \frac{i\omega x^0}{2} \bigg)\,
  \bm{W}\!\left((x^0,\vec{x}=0),\hat{s}',\hat{t}_1',\hat{u}_1',m_t,\mu\right) .
\end{align}
The soft function entering the factorization formula is then given by
\begin{align}
  \label{eq:Sdef}
  \bm{S}(s_4,\hat{s}',\hat{t}'_1,\hat{u}'_1,m_t,\mu) = \frac{1}{m_t} \hat{\bm{W}} \bigg(
    \frac{s_4}{\sqrt{s_4+m_t^2}},\hat{s}',\hat{t}'_1,\hat{u}'_1,m_t,\mu \bigg) \,.
\end{align}
Using the results above, one recovers the factorization formula (\ref{eq:fac}) for the
hard-scattering kernels by noting that
\begin{align}
  C_{ij}(s_4,\hat{s},\hat{t}_1,\hat{u}_1,m_t,\mu_f) = \frac{3\hat{s}^2}{8\pi}
  \frac{d\hat\sigma_{ij}}{d\hat{t}_1 d\hat{u}_1} \, .
\end{align}

It is instructive to compare this derivation with that given in \cite{Ahrens:2010zv} for
PIM kinematics. In both cases, the calculations are simplified by working in a frame where
the soft function depends only on the time component $x^0$, or in momentum space, on the
energy of the extra soft radiation (although of course the final results are Lorentz
invariant and do not depend on the frame). In PIM kinematics this is the partonic center-of-mass
frame, while in 1PI kinematics it is the center-of-mass frame of the unobserved partonic final state,
consisting of the anti-top-quark plus additional soft radiation. The difference between these
two cases comes from which combinations of the momenta are counted as ``small'': in PIM kinematics it
is $(p_3+p_4) \cdot k$ against $(p_3+p_4)^2=M^2$, while in 1PI kinematics it is $p_4 \cdot k$ against
$p_4^2=m_t^2$. This difference has important implications for the structure of power
corrections in the two types of kinematics. Such power corrections come both from
time-ordered products involving the subleading SCET Lagrangian and operators, which before
phase-space integrations are the same in both cases, and from the approximations in the
phase-space integrals, e.g. $E_4 = \sqrt{|\vec{p}_4|^2+m_t^2} \sim m_t$ in 1PI and
$E_3+E_4 = \sqrt{|\vec{p}_3+\vec{p}_4|^2+M^2} \sim M$ in PIM kinematics. The corrections from the
latter source can be quite different: for 1PI kinematics they involve the expansion parameter
$|\vec{p}_4|/m_t = s_4/(2m_t\sqrt{m_t^2+s_4})$, while for PIM kinematics they involve the
expansion parameter $|\vec{p}_3+\vec{p}_4|/M =\sqrt{\hat{s}}(1-z)/(2M) =
(1-z)/(2\sqrt{z})$.

A complete analysis of power corrections is beyond the scope of this paper. However, in the
case of Drell-Yan \cite{Becher:2007ty} and Higgs production \cite{Ahrens:2008qu,Ahrens:2008nc} near
threshold it was found that by keeping the exact dependence $\omega=2E_g$ in the SCET
soft functions analogous to (\ref{eq:Sdef}) one can reproduce a set of logarithmic power
corrections involving $\ln z/(1-z)$, which are indeed present in the analytic results for
the fixed-order expansions of the hard-scattering kernels. Keeping such terms improved
agreement of the threshold-expanded hard-scattering kernels with the exact results in QCD.
In \cite{Ahrens:2010zv}, we followed this procedure in PIM kinematics, using the exact
form $\omega=2E_g=M(1-z)/\sqrt{z}$ instead of $\omega=M(1-z)$ in the first argument of the
soft function. Results in this ``$\scetpim$'' scheme include the same type of logarithmic
corrections found in Drell-Yan and Higgs production, and the numerical results for the
threshold expansion at NLO in this scheme are indeed improved significantly compared to the traditional
PIM scheme \cite{Kidonakis:2001nj, Kidonakis:2003qe, Kidonakis:2008mu}, which does not
include such corrections. In 1PI kinematics, the equivalent procedure is to use
$\omega=2E_g=s_4/\sqrt{m_t^2+s_4}$, as we have already indicated explicitly in
(\ref{eq:Sdef}). We will refer to numerical results obtained with this choice as being
calculated in the ``$\scetopi$'' scheme. Since the factorization formula (\ref{eq:fac2})
is derived in the limit $s_4 \ll m_t^2$, it would be equally valid to use $\omega=s_4/m_t$ in
the first argument of the soft function (\ref{eq:Sdef}). This is in fact the choice that
has been made in previous calculations in 1PI kinematics \cite{Laenen:1998qw}, and later
on we will refer to this as the ``1PI'' scheme. When expanded in fixed-order perturbation theory, the
two schemes differ through terms involving $\ln(1+s_4/m_t^2)/s_4$, which are power
suppressed in the limit $s_4\to 0$. However, in our analysis in
Sections~\ref{sec:threshold} and \ref{sec:pheno} we will see that these power-suppressed
effects can be numerically important, and that the agreement with the exact numerical
results at NLO is improved in the $\scetopi$ scheme. Furthermore, although in this case we
do not have explicit analytic results to compare with, we note that the logarithms of
$\ln(1+s_4/m_t^2)$ appear naturally in the fixed-order NLO calculations of hard gluon
corrections through terms of the form $\ln(2E_g/\mu)$, see for instance Eqs.~(4.16) and
(4.17) of \cite{Beenakker:1988bq}.

We now present the calculation of the soft function in 1PI kinematics at one-loop order. The
results can be written as
\begin{align}
  \bm{W}^{(1)}_{\text{bare}}(\epsilon,x^0,\mu) = \sum_{i,j} \, \bm{w}_{ij} \,
  \mathcal{I}'_{ij}(\epsilon,x^0,\mu) \, ,
\end{align}
where the matrices $\bm{w}_{ij}$ are related to products of color generators and can be
found in \cite{Ahrens:2010zv}. The integrals $\mathcal{I}'_{ij}$ are defined as
\begin{align}
  \label{eq:softintegrals}
  \mathcal{I}'_{ij}(\epsilon,x^0,\mu) = -\frac{(4\pi\mu^2)^\epsilon}{\pi^{2-\epsilon}} \,
  v_i \cdot v_j \int d^dk \, \frac{e^{-ik^0x^0}}{v_i \cdot k \, v_j \cdot k} \, (2\pi) \,
  \delta(k^2) \, \theta(k^0) \, .
\end{align}
Evaluating these objects in the rest frame of the heavy anti-top quark, we find for the
non-vanishing integrals
\begin{align}
  \mathcal{I}'_{12} &= - \left[ \frac{2}{\epsilon^2} + \frac{2}{\epsilon} \left( L_0 -
      \ln\frac{\hat{s}'m_t^2}{\hat{t}'_1\hat{u}'_1} \right) + \left( L_0 -
      \ln\frac{\hat{s}'m_t^2}{\hat{t}'_1\hat{u}'_1} \right)^2 + \frac{\pi^2}{6} + 2\Li_2
    \left( 1 - \frac{\hat{s}'m_t^2}{\hat{t}'_1\hat{u}'_1} \right) \!\right] , \nonumber
  \\
  \mathcal{I}'_{33} &= \frac{2}{\epsilon} + 2L_0 - \frac{2(1+\beta_t^2)}{\beta_t} \ln x_s
  \, , \nonumber
  \\
  \mathcal{I}'_{44} &= \frac{2}{\epsilon} + 2L_0 + 4 \, , \nonumber
  \\
  \mathcal{I}'_{14} = \mathcal{I}'_{24} &= -\frac{1}{\epsilon^2} - \frac{1}{\epsilon} L_0
  - \frac{1}{2} L_0^2 - \frac{\pi^2}{12} \, ,
  \\
  \mathcal{I}'_{13} &= - \Biggl[ \frac{1}{\epsilon^2} + \frac{1}{\epsilon} \left(L_0 - 2
    \ln{\frac{\hat{t}'_1}{\hat{u}'_1}} \right) + \frac{1}{2} \left( L_0 -
    2\ln\frac{\hat{t}'_1}{\hat{u}'_1} \right)^2 + \frac{\pi^2}{12} \nonumber
  \\
  &\hspace{1.1cm} + 2\Li_2 \left( 1 - \frac{\hat{t}'_1}{\hat{u}'_1x_s} \right) + 2\Li_2
  \left( 1 - \frac{\hat{t}'_1x_s}{\hat{u}'_1} \right) \Biggr] \, , \nonumber
  \\
  \mathcal{I}'_{23} &= \mathcal{I}'_{13} \, (\hat{t}'_1 \leftrightarrow \hat{u}'_1) \, ,
  \nonumber
  \\
  \mathcal{I}'_{34} &= \frac{1+\beta_t^2}{2\beta_t} \left[- \frac{2}{\epsilon} \ln x_s
    -2L_0\ln x_s + 2\ln^2x_s - 4\ln x_s \ln(1-x_s^2) - 2\Li_2(x_s^2) + \frac{\pi^2}{3}
  \right] , \nonumber
\end{align}
where $\beta_t=\sqrt{1-4m_t^2/\hat{s}'}$, $x_s=(1-\beta_t)/(1+\beta_t)$ and
\begin{align}
  L_0 = \ln\bigg( -\frac{\mu^2 (x^0)^2 e^{2 \gamma_E}}{4} \bigg) \, .
\end{align}
The renormalized function is obtained by subtracting the $1/\epsilon^n$ poles in the bare function. When
performing resummation it is more convenient to introduce the Laplace transform of this
object, which is defined as
\begin{align}
  \label{eq:stildedef}
  \tilde{\bm{s}}(L,\hat{s}',\hat{t}'_1,\hat{u}'_1,m_t,\mu) &= \int_0^\infty d\omega \exp
  \bigg( -\frac{ \omega}{e^{\gamma_E}\mu e^{L/2}} \bigg)\,
  \hat{\bm{W}}(\omega,\hat{s}',\hat{t}'_1,\hat{u}'_1,m_t,\mu) \, , \nonumber
  \\
  &= \bm{W}\!\left( \!\bigg( x^0=-\frac{2i}{e^{\gamma_E}\mu e^{L/2}},\vec{x}=0 \bigg)
    ,\hat{s}',\hat{t}'_1,\hat{u}'_1,m_t,\mu \right) .
\end{align}
From the second line it is clear that the Laplace-transformed function $\tilde{\bm{s}}$ is
determined directly from the position-space soft function through the replacement $L_0\to
-L$ \cite{Becher:2007ty}.

The above expression for the one-loop soft function is new, but we have been able to perform two
important checks. First, we have verified that our results are consistent
with the results for real emission in the soft limit given in \cite{Beenakker:1988bq} for
the $gg$ channel and in \cite{Beenakker:1990maa} for the $q\bar q$ channel. Second, we have
made sure that the divergence structure of the one-loop soft function is consistent with the
RG equation derived in the following section.

\subsection{RG equations and resummation}

The physical cross section should be independent of the factorization scale, which implies
\begin{align}
  \label{eq:RGinv}
  0 &= \int_{x_1^{\text{min}}}^1 \frac{dx_1}{x_1} \int_{x_2^{\text{min}}}^1
  \frac{dx_2}{x_2}\, \bigg\{ \left[ \frac{d}{d\ln\mu_f} \left[f_{i/N_1}(x_1,\mu_f) \,
      f_{j/N_2}(x_2,\mu_f) \right] \right] \,
  C_{ij}(s_4,\hat{s},\hat{t}_1,\hat{u}_1,m_t,\mu_f) \nonumber
  \\
  &\hspace{4.3cm}\mbox{} + f_{i/N_1}(x_1,\mu_f) \, f_{j/N_2}(x_2,\mu_f) \, \frac{d}{d\ln\mu_f}
  C_{ij}(s_4,\hat{s},\hat{t}_1,\hat{u}_1,m_t,\mu_f) \bigg\} \, .
 \end{align}
 Compared to the PIM case, the terms arising from the derivatives acting on the PDFs are
 slightly different. To understand their structure, first consider the term where the
 derivative acts on $f_{j/N_2}(x_2,\mu_f)$. 
 Under the dynamical assumption of steeply falling PDFs, the DGLAP evolution equations can be simplified by keeping only the leading terms for $x\to 1$ in the
 Altarelli-Parisi splitting functions, in which case the evolution becomes diagonal and reads
\begin{align}
  \frac{d}{d\ln\mu} f_{j/N_2}(y,\mu) = \int_y^1 \frac{dx}{x} \, P_{jj}(x) \,
  f_{j/N_2}(y/x,\mu) \, ,
\end{align}
with $P_{jj}(x)$ given by 
\begin{align}
  \label{eq:SplittingF}
  P_{jj}(x) = \frac{2\Gamma^j_{\text{cusp}}(\alpha_s)}{(1-x)_+} + 2\gamma^\phi_j(\alpha_s)
  \, \delta(1-x) \, .
\end{align}
We then obtain (up to power-suppressed terms)
\begin{align}
  \int_{x_1^{\text{min}}}^1 \frac{dx_1}{x_1}f_{i/N_1}(x_1,\mu_f) \left[
    \int_{x_2^{\text{min}}}^1 \frac{dx_2}{x_2} \int_{x_2}^1 \frac{d\xi}{\xi} \,
    f_{j/N_2}(x_2/\xi,\mu_f) P_{jj}(\xi) \,
    C_{ij}(s_4,\hat{s},\hat{t}_1,\hat{u}_1,m_t,\mu_f) \right].
 \end{align}
 To derive the evolution equation for the soft function, we arrange the integrations such
 that the Altarelli-Parisi kernel acts on the $C_{ij}$ rather than on the PDF. After some
 manipulations, the term in the square brackets above can be written as
\begin{align}
  \int_{x_2^{\text{min}}}^1 \frac{dx_2}{x_2} \, f_{j/N_2}(x_2,\mu_f) \int_{x_2^{\rm
      min}/x_2}^1 \frac{d\xi}{\xi} P_{jj}(\xi) \, C_{ij}(s_4',\xi \hat{s},\hat{t}_1,\xi
  \hat{u}_1,m_t,\mu_f)
\end{align}
with $s_4'=\hat{t}_1 + \xi(\hat{s}+\hat{u}_1)$. Changing the integration variable from
$\xi$ to $s_4'$, taking the threshold limit $s_4\to 0$, and using the identity
\begin{align}
  \int_0^{s_4}ds_4'\,\frac{f(s_4')}{(-\hat{t}_1) [(s_4-s_4')/(-\hat{t}_1)]_+}=
  \int_0^{s_4}ds_4'\,\frac{f(s_4')-f(s_4)}{s_4-s_4'} +f(s_4)\ln\frac{s_4}{-\hat{t}_1} \,,
\end{align}
we can convert this term into a form from which the RG equation for the soft function is
more easily derived. The analogous procedure is then used for the term where the
derivative acts on the other PDF. Finally, we use the RG equation for the hard function
\begin{align}
  \label{eq:Hev}
  \frac{d}{d\ln\mu} \bm{H}(\hat{s}',\hat{t}'_1,\hat{u}'_1,m_t,\mu) &=
  \bm{\Gamma}_H(\hat{s}',\hat{t}'_1,\hat{u}'_1,\mu) \,
  \bm{H}(\hat{s}',\hat{t}'_1,\hat{u}'_1,m_t,\mu) \nonumber
  \\
  &\quad\mbox{}+ \bm{H}(\hat{s}',\hat{t}'_1,\hat{u}'_1,m_t,\mu) \,
  \bm{\Gamma}_H^\dagger(\hat{s}',\hat{t}'_1,\hat{u}'_1,m_t,\mu) \, ,
\end{align}
with
\begin{align}
  \label{eq:gammaH}
  \bm{\Gamma}_H(\hat{s}',\hat{t}'_1,\hat{u}'_1,m_t,\mu) = \Gamma_{\text{cusp}}(\alpha_s)
  \left( \ln\frac{\hat{s}'}{\mu^2} - i\pi \right) \bm{1} +
  \bm{\gamma}^h(\hat{s}',\hat{t}'_1,\hat{u}'_1,m_t,\alpha_s) \, ,
\end{align}
where $\Gamma_{\text{cusp}}$ is equal to $C_F \, \gamma_{\text{cusp}}$ for the $q\bar{q}$ and
$C_A \, \gamma_{\text{cusp}}$ for the $gg$ channel, and the matrices $\bm{\gamma}^h$ can be found in
\cite{Ahrens:2010zv}.

Assembling the different pieces and enforcing (\ref{eq:RGinv}), the evolution equation for
the momentum-space soft function reads
\begin{align}
  \frac{d}{d\ln\mu} \bm{S}(s_4,\hat{s}',\hat{t}'_1,\hat{u}'_1,m_t,\mu) &= - \left[
    2\Gamma_{\text{cusp}}(\alpha_s) \ln\frac{s_4}{m_t\mu} +
    \bm{\gamma}^{s\dagger}(\hat{s}',\hat{t}'_1,\hat{u}'_1,m_t,\alpha_s) \right]
  \bm{S}(s_4,\hat{s}',\hat{t}'_1,\hat{u}'_1,m_t,\mu) \nonumber
  \\
  &\hspace{-4em} - \bm{S}(s_4,\hat{s}',\hat{t}'_1,\hat{u}'_1,m_t,\mu) \left[
    2\Gamma_{\text{cusp}}(\alpha_s) \ln\frac{s_4}{m_t\mu} +
    \bm{\gamma}^{s}(\hat{s}',\hat{t}'_1,\hat{u}'_1,m_t,\alpha_s) \right] \nonumber
  \\
  &\hspace{-4em} - 4\Gamma_{\text{cusp}}(\alpha_s) \int_0^{s_4} ds_4' \,
  \frac{\bm{S}(s_4',\hat{s}',\hat{t}'_1,\hat{u}'_1,m_t,\mu)-
    \bm{S}(s_4,\hat{s}',\hat{t}'_1,\hat{u}'_1,m_t,\mu)} {s_4-s_4'} \, ,
\end{align}
where we have defined
\begin{align}\label{eq33}
  \bm{\gamma}^s(\hat{s}',\hat{t}'_1,\hat{u}'_1,m_t, \alpha_s) =
  \bm{\gamma}^h(\hat{s}',\hat{t}'_1,\hat{u}'_1,m_t,\alpha_s) +
  \left(2\gamma^{\phi}(\alpha_s)+ \Gamma_{\text{cusp}}(\alpha_s)\ln\frac{\hat{s}'
      m_t^2}{\hat{t}'_1\hat{u}'_1} \right) \bm{1} \, .
\end{align}
This evolution equation is of the same form as for the PIM case, but the soft anomalous
dimension is modified by the logarithmic term in (\ref{eq33}), which can be traced back to the different
form of the collinear evolution terms after arranging the integrations as appropriate for
1PI kinematics (this extra term vanishes in the production threshold limit $\beta\to 0$,
where PIM and 1PI kinematics agree). Therefore, we can use the expression for the resummed
soft function derived in \cite{Ahrens:2010zv}, taking into account the changes in the
anomalous dimension and soft matching function, and the fact that now
$\omega=s_4/\sqrt{s_4+m_t^2}$ sets the mass scale in the soft logarithms. The resummed
soft function is then given by
\begin{align}
  \bm{S}(s_4,\hat{s}',\hat{t}_1',\hat{u}'_1,m_t,\mu_f) &= \exp \left[ -4S(\mu_s,\mu_f) +
    2a_{\Gamma}(\mu_s,\mu_f) \ln\frac{\hat{s}'m_t^2}{\hat{t}_1'\hat{u}'_1} +
    4a_{\gamma^\phi}(\mu_s,\mu_f) \right] \nonumber
  \\
  &\hspace{-2em} \times \bm{u}^\dagger(\hat{s}',\hat{t}_1',\hat{u}'_1,m_t,\mu_f,\mu_s) \,
  \tilde{\bm{s}}(\partial_\eta,\hat{s}',\hat{t}_1',\hat{u}'_1,m_t,\mu_s) \,
  \bm{u}(\hat{s}',\hat{t}_1',\hat{u}'_1,m_t,\mu_f,\mu_s) \nonumber
  \\
  &\hspace{-2em} \times \frac{1}{s_4}\,
  \bigg( \frac{s_4}{\sqrt{m_t^2+s_4}\mu_s}\bigg)^{2\eta}\,
  \frac{e^{-2\gamma_E\eta}}{\Gamma(2\eta)}\,\,\Bigg|_{\eta=2a_\Gamma(\mu_s,\mu_f)} ,
\end{align}
and combining this with the solution for the hard function, the final result for the
resummed hard-scattering kernels is
\begin{align}
  \label{eq:resC}
  C(s_4,\hat{s}',\hat{t}_1',\hat{u}'_1,m_t,\mu_f) &= \exp \left[ 2a_{\Gamma}(\mu_s,\mu_f)
    \ln\frac{m_t^2\mu_s^2}{\hat{t}_1'\hat{u}'_1} + 4a_{\gamma^{\phi}}(\mu_s,\mu_f) \right]
  \nonumber
  \\
  &\hspace{-2em} \times \Tr \Bigg[ \bm{U}(\hat{s}',\hat{t}_1',\hat{u}'_1,m_t,\mu_h,\mu_s)
  \, \bm{H}(\hat{s}',\hat{t}_1',\hat{u}'_1,m_t,\mu_h) \,
  \bm{U}^\dagger(\hat{s}',\hat{t}_1',\hat{u}'_1,m_t,\mu_h,\mu_s) \nonumber
  \\
  &\hspace{-2em} \times
  \tilde{\bm{s}}(\partial_\eta,\hat{s}',\hat{t}_1',\hat{u}'_1,m_t,\mu_s) \Bigg]
  \frac{1}{s_4}\, \bigg(\frac{s_4}{\sqrt{m_t^2+s_4}\mu_s}\bigg)^{2\eta}\,
  \frac{e^{-2\gamma_E\eta}}{\Gamma(2\eta)}\,\,\Bigg|_{\eta=2a_\Gamma(\mu_s,\mu_f)} .
\end{align}
The result (\ref{eq:resC}) has already been given in \cite{Ahrens:2010zv}, where the explicit 
definitions of the RG exponents $a_\Gamma$ and $a_{\gamma^\phi}$ and the evolution factors
$\bm U$ and $\bm u$ can be found. The notation is such that one must first take the
derivatives with respect to $\eta$ appearing in the first argument of the
Laplace-transformed soft function $\tilde{\bm{s}}$, defined in (\ref{eq:stildedef}),
and then set $\eta=2a_\Gamma(\mu_s,\mu_f)$. For values of the scale where $\eta<0$, one
must use analytic continuation to interpret the formula in terms of plus-distributions.

Our result for the resummed hard-scattering kernels is equivalent to the Mellin-space
resummation formula from \cite{Laenen:1998qw} when expanded to any fixed order in
$\alpha_s$, if we approximate $\sqrt{m_t^2+s_4}\sim m_t$. However, the scale choices used
in Mellin-space resummation are typically such that the one encounters a Landau-pole
ambiguity in the evaluation of the all-orders formula, upon inverting the Mellin transform
and integrating over $s_4$. One way of dealing with this, as done in \cite{Kidonakis:2001nj,
  Kidonakis:2003qe, Kidonakis:2008mu, Kidonakis:2010dk}, is to instead use the resummation
formula only to construct approximate fixed-order expansions at NNLO. Another way is to
instead view $\mu_s$ as a function of the observables $p_T$ and $y$, and choose it in such
a way that the perturbative expansion of the soft function at $\mu_s$ is well behaved.
With such a choice of $\mu_s$, one can still evaluate the all-orders resummation formula,
but without encountering Landau-pole ambiguities. We will describe how to construct the
fixed-order expansion to NNLO in the next section, and then compare the two methods in
more detail in Section~\ref{sec:comparison}.

\subsection{Approximate NNLO results}

In addition to the resummed formulas derived above, it is sometimes useful to construct
approximate NNLO formulas in a fixed-order expansion. In our formalism, this is achieved
by evaluating the expressions
\begin{align}
  \label{eq:FixedOrder}
  C(s_4,\hat{s}',\hat{t}_1',\hat{u}'_1,m_t,\mu) &= \tilde{c}(\partial_\eta,
  \hat{s}',\hat{t}_1',\hat{u}'_1,m_t,\mu)\,\frac{e^{-2\gamma_E \eta}}{\Gamma(2\eta)}\,
  \frac{1}{s_4}\,\bigg(\frac{s_4}{\sqrt{m_t^2+s_4}\mu_s}\bigg)^{2\eta}\,\Bigg|_{\eta=0} \,
  ,
\end{align}
where
\begin{align}
  \label{eq:ctilde}
  \tilde{c}(L,\hat{s}',\hat{t}_1',\hat{u}'_1,m_t,\mu) = \Tr \Big[
  \bm{H}(\hat{s}',\hat{t}_1',\hat{u}'_1,m_t,\mu)\,
  \tilde{\bm{s}}(L,\hat{s}',\hat{t}_1',\hat{u}'_1,m_t ,\mu) \Big] \, .
\end{align}
Using the ingredients of our NNLL resummation formula, it is possible to use
RG methods to determine at NNLO the coefficients of all powers of $L$
in $\tilde{c}$, as well as the $\mu$-dependent part of the constant piece. To do so, we
use the same methods as in \cite{Ahrens:2009uz}. We then convert the derivatives with
respect to the auxiliary parameter $\eta$ into distributions in $s_4$ defined in
(\ref{eq:plus}), which can be easily done by using the following replacement rules:
\begin{align}
  1 &\longrightarrow \delta(s_4) \, , \nonumber
  \\
  L &\longrightarrow 2 P_0(s_4) - \delta(s_4) \, L_m \, , \nonumber
  \\
  L^2 &\longrightarrow 8 P_1(s_4) - 4 L_m P_0(s_4) + \delta(s_4) \left( L_m^2 -
    \frac{2\pi^2}{3} \right) - \frac{4L_4}{s_4} \, , \nonumber
  \\
  L^3 &\longrightarrow 24 P_2(s_4) - 24L_m P_1(s_4) + \left( 6L_m^2 - 4\pi^2 \right)
  P_0(s_4) + \delta(s_4) \left( -L_m^3 + 2\pi^2 L_m + 16\zeta_3 \right) \nonumber
  \\
  &\quad - \frac{6L_4}{s_4} \left[ -L_4 + 2\ln\frac{s_4^2}{m_t^2\mu^2} \right] ,
  \nonumber
  \\
  L^4 &\longrightarrow 64 P_3(s_4) - 96L_m P_2(s_4) + \left( 48L_m^2 - 32\pi^2 \right)
  P_1(s_4) + \left( -8L_m^3 + 16\pi^2 L_m + 128\zeta_3 \right) P_0(s_4) \nonumber
  \\
  &\quad + \delta(s_4) \left( L_m^4 - 4\pi^2 L_m^2 - 64\zeta_3 L_m + \frac{4\pi^4}{15}
  \right) \nonumber
  \\
  \label{eq:conversion}
  &\quad - \frac{8L_4}{s_4} \left[ L_4^2 - 3 L_4 \ln\frac{s_4^2}{m_t^2\mu^2} + 3
    \ln^2\frac{s_4^2}{m_t^2\mu^2} - 2\pi^2 \right] ,
\end{align}
where $L_m=\ln(\mu^2/m_t^2)$ and $L_4=\ln(1+s_4/m_t^2)$.

The final result for the hard-scattering kernels at NNLO can be written as
\begin{align}
  \label{eq:C2}
  C^{(2)}(s_4,\hat{s}',\hat{t}'_1,\hat{u}'_1,m_t,\mu) = D_3 \, P_3(s_4) + D_2 \, P_2(s_4)
  + D_1 \, P_1(s_4) + D_0 \, P_0(s_4) + C_0\,\delta(s_4) + R(s_4) \, ,
\end{align}
where the coefficients $D_0,\ldots,D_3$ and $C_0$ are functions of the variables
$\hat{s}'$, $\hat{t}'_1$, $\hat{u}'_1$, $m_t$ and $\mu$. The explicit results are quite
lengthy and are contained in a computer program 
which can be downloaded together with the arXiv 
version of
this paper. The analytic expressions for $D_i$ ($i=1,2,3$) were first derived in
\cite{Kidonakis:2003qe} starting from resummed formulas in Mellin moment space. We have
compared with those results and found agreement. In a recent paper the coefficient $D_0$ 
was also determined \cite{Kidonakis:2010dk}, but its explicit form was not reported there. 
We are thus unable to compare this term. The regular piece
$R(s_4)$ collects terms involving $L_4$, which arise from choosing
$\omega=s_4/\sqrt{m_t^2+s_4}$ in the argument of the soft function. As noted in Section
\ref{sec:soft}, dropping $R(s_4)$ recovers the 1PI scheme used in earlier work, for
instance \cite{Kidonakis:2010dk}. The $C_0$ term, on the other hand, is ambiguous since
only its scale-dependent part is exactly determined, and one needs to specify which
contributions are included there. One contribution to $C_0$ comes from the conversion of
powers of $L$ in $\tilde{c}^{(2)}$ according to (\ref{eq:conversion}), which is determined
exactly, therefore the ambiguity comes from the constant term of $\tilde{c}^{(2)}$, which
is
\begin{align}
  \label{eq:NNLOscheme}
  \tilde{c}^{(2)}(0) = \Tr \Big[ \bm{H}^{(1)} \, \tilde{\bm{s}}^{(1)}(0) + \bm{H}^{(0)} \,
  \tilde{\bm{s}}^{(2)}(0) + \bm{H}^{(2)} \, \tilde{\bm{s}}^{(0)}(0) \Big] \, ,
\end{align}
where we have suppressed the dependence on $\hat{s}'$, $\hat{t}_1'$, $\hat{u}_1'$, $m_t$
and $\mu$ for convenience. In the three terms above, the first term is known exactly and
is therefore included, while the constant term of the two-loop soft function in the second
term is unknown. As for the two-loop hard function in the third term, one can determine
its scale-dependent part and include it in the formula, as was done for the PIM case in
\cite{Ahrens:2010zv}. In that case there is a freedom to write the result in terms of
$\ln(\mu_0/\mu)$, where the scale $\mu_0$ can be chosen as $\sqrt{\hat{s}}$ or $m_t$.
However, by including these extra $\mu$-dependent terms one runs the risk of artificially
reducing the scale dependence, rendering it an ineffective means of estimating theoretical
uncertainties. Therefore, we will not include these terms here and instead drop the
contributions of the two-loop hard function. Later, when we compare the numerical results
for the total cross section derived using PIM and 1PI kinematics, the PIM numbers will also be
computed in the equivalent way, i.e., by dropping the two-loop hard function completely.

\subsection{Resummation vs. NNLO expansions}
\label{sec:comparison}

In our numerical studies later on we will typically give results from both resummed
perturbation theory and the approximate NNLO formulas. Although the perturbative
information used in these formulas is the same -- the NLO matching functions and the NNLO
anomalous dimensions -- the implementation and philosophy is different.

In resummed perturbation theory, one views the soft function as depending on the
distributions
\begin{align}
  \label{eq:plusprime}
  P_n'(s_4) = \left[ \frac{1}{s_4} \ln^n \bigg( \frac{s_4}{\sqrt{s_4+m_t^2}\mu_s} \bigg)
  \right]_+ ,
\end{align}
as well as $\delta(s_4)$. The logarithmic corrections are in general considered large
compared to the $\delta$-function term (and of course subleading terms in $s_4$), but it
is assumed that with a proper choice of $\mu_s$ they can be treated on the same footing,
so that the soft function at this scale can be reliably calculated in fixed-order
perturbation theory. This would obviously be the case if we chose
$\mu_s=s_4/\sqrt{s_4+m_t^2}$, since then the logarithmic corrections would vanish, but in
that case the running coupling in the soft corrections would, for some value of $s_4$, 
be evaluated at the Landau pole, which would spoil the clear separation between 
perturbative and non-perturbative physics accomplished by using the effective field-theory 
formalism. Therefore, as usually done in momentum-space
resummation \cite{Becher:2006nr, Becher:2006mr, Becher:2007ty, Ahrens:2008qu, Ahrens:2008nc,
  Ahrens:2010zv}, we will view the soft scale as a function of the observables $p_T$ and
$y$, and choose it based on the convergence of the physical cross section. In particular,
we study corrections from the soft function to the cross section as a function of $\mu_s$,
and choose the numerical value of the scale as the point where the correction is
minimized. The formulas then sum logarithms of the numerical ratio $\mu_s/\mu_f$, where $\mu_s$
is the dynamically generated soft scale. The same reasoning applies to the choice of the hard
and factorization scales, and an advantage of the resummation formalism is that the three
scales can be varied independently as a way of estimating perturbative uncertainties.

This approach should be contrasted with that based on approximate NNLO
formulas. In using such an approximation, one assumes that the
logarithmic corrections from the $P_n'$ distributions account for the
bulk of the NNLO corrections in fixed order at an arbitrary
factorization scale $\mu_f$, and at the same time that the corrections
at NNNLO and beyond, evaluated at that scale, are small enough that
the perturbative series is well behaved. From this point of view, the
resummation formalism is just a useful tool for constructing the
approximate fixed-order expansion, and no physical significance is
given to the soft scale $\mu_s$, on which the final answer does not
depend.

It is worth emphasizing that the fixed-order expansion of the NLO+NNLL formulas to NNLO
in $\alpha_s$ is not exactly equivalent to the approximate NNLO formulas from the previous section.
The direct expansion of the NLO+NNLL formulas to NNLO contains explicit dependence
on the scale $\mu_s$ and also a different pattern of plus-distributions
compared to the approximate NNLO formulas. For instance, the approximate NNLO formula
contains $P_3$ distributions, but the expansion of the NLO+NNLL formula to NNLO in fixed
order contains at most $P_2$ distributions, as required by the Altarelli-Parisi equations.
This aspect of the calculation is discussed in greater detail in the Appendix.

The optimal method for including the higher-order perturbative effects is not entirely
clear without further information. The reason is that it is not possible to tell whether
the logarithmic corrections from the partonic threshold region can be considered large, or
how to minimize them with a proper choice of $\mu_s$, until {\it after} the integration
over $s_4$. After that integration, they give large perturbative corrections to the
differential cross section if the PDFs fall off very quickly away from the
region where $s_4\to 0$, an effect referred to as ``dynamical threshold enhancement'' 
\cite{Becher:2007ty}. Since
the PDFs are not known analytically, it is only possible to assess the extent to which the
corrections in the partonic threshold region are dynamically enhanced through a numerical
study. We thus return to a discussion of the relative merits of resummation vs.
approximate NNLO expansions after our numerical studies in Sections~\ref{sec:threshold} 
and~\ref{sec:pheno}.

\section{Numerical implementation}
\label{sec:numerical}

We now describe the numerical implementation of our results. A central issue
is that there are power-suppressed ambiguities in the evaluation of (\ref{eq:pt,y}) and
(\ref{eq:fac}) away from the exact threshold limit, and we describe in detail how we deal
with these ambiguities in the following. We also use this section to summarize some of the
inputs used throughout the rest of the paper, in particular the PDF sets, the strong
coupling constant, and the top-quark mass.

As pointed out in Section~\ref{sec:kin}, there are power-suppressed ambiguities in the
choice of the variables $\hat{s}'$, $\hat{t}'_1$ and $\hat{u}'_1$ of the hard and soft functions. Apart
from when it appears in the $\delta$-function or plus-distributions, in the perturbative
calculation of the hard and soft functions one can set $s_4=0$ everywhere and use
$\hat{s}'+\hat{t}'_1+\hat{u}'_1=0$ to rewrite the hard-scattering kernels in many
different forms. While these are all formally equivalent in the threshold limit $s_4\to
0$, they change the functional dependence of the hard-scattering kernels on $x_1$ and
$x_2$, so the integration in (\ref{eq:pt,y}) gives different results for the pieces
multiplying plus-distributions in $s_4$. Moreover, since one typically trades either $x_1$
or $x_2$ in favor of $s_4$ as an integration variable, another obvious choice is to use
$\hat{s}'+\hat{t}'_1+\hat{u}'_1=s_4$ before integration, again leading to numerically
different answers which are nonetheless equivalent in the threshold limit $s_4\to 0$.
 
Our method of fixing this ambiguity is as follows.  First, we enforce
$\hat{s}'+\hat{t}'_1+\hat{u}'_1=0$ in the hard-scattering kernels, and
use this to eliminate either $\hat{t}_1'$ or $\hat{u}_1'$ as an
independent variable.  We then define the two cross sections
\begin{align}
  \frac{d\sigma^t}{dp_Tdy} &= \frac{16\pi p_T}{3s} \sum_{i,j} \int_{-u_1/(s+t_1)}^1
  \frac{dx_1}{x_1} \int_0^{x_1(s+t_1)+u_1} \frac{ds_4}{s_4-x_1t_1} \nonumber
  \\
  &\quad \times f_{i/N_1}(x_1,\mu_f) \, f_{j/N_2}(x_2(s_4),\mu_f) \,
  C_{ij}(s_4,\hat{s}',\hat{t}_1',-\hat{s}'-\hat{t}'_1,m_t,\mu_f) \, , \label{eq:sigmat}
  \\
  \frac{d\sigma^u}{dp_Tdy} &= \frac{16\pi p_T}{3s} \sum_{i,j} \int_{-t_1/(s+u_1)}^1
  \frac{dx_2}{x_2} \int_0^{x_2(s+u_1)+t_1} \frac{ds_4}{s_4-x_2u_1} \nonumber
  \\
  &\quad \times f_{i/N_1}(x_1(s_4),\mu_f) \, f_{j/N_2}(x_2,\mu_f) \,
  C_{ij}(s_4,\hat{s}',-\hat{s}'-\hat{u}'_1,\hat{u}'_1,m_t,\mu_f) \, . \label{eq:sigmau}
\end{align}
We have changed variables from $x_2$ or $x_1$ to $s_4$ in the two equations,
respectively, so that
\begin{align}
  \label{eq:xi}
  x_1(s_4) = \frac{s_4-x_2u_1}{x_2s+t_1} \, , \quad x_2(s_4) = \frac{s_4-x_1t_1}{x_1s+u_1}
  \, .
\end{align}
Finally, we drop all dependence on $s_4$ in the hard-scattering kernels by using
\begin{align}
  \hat{t}'_1 = \hat{t}_1 = x_1 t_1 \, , \quad \hat{s}' = x_1 x_2(0) s
\end{align}
in (\ref{eq:sigmat}), and 
\begin{align}
  \hat{u}'_1 = \hat{u}_1 = x_2 u_1 \, , \quad \hat{s}' = x_1(0) x_2 s
\end{align}
in (\ref{eq:sigmau}). It is easy to see that with this choice $\sigma^t$ and $\sigma^u$
are not necessarily the same, although the difference is power suppressed. We shall take
the average of the two as the final result for the differential cross section:
\begin{align}
  \label{eq:tot}
  \frac{d\sigma}{dp_Tdy} = \frac{1}{2} \left[ \frac{d\sigma^t}{dp_Tdy} +
    \frac{d\sigma^u}{dp_Tdy} \right] .
\end{align}
In this way, the rapidity distribution in the gluon channel is invariant under $y \to -y$,
as it should be, and the relation (\ref{eq:cpi}) is preserved.

The scheme above specifies our procedure for the numerical evaluation of the threshold
formulas for the differential distribution in $p_T$ and $y$. In the next section we will
also study the differential distribution in $\beta$. Rather than define yet another
scheme, we will calculate this through an exact change of variables and integration orders
in (\ref{eq:tot}). The differential cross section in $\beta$ then takes the form
\begin{align}
  \label{eq:tot2}
  \frac{d\sigma}{d\beta} &= \frac{4\pi\beta}{3sm_t^2} \sum_{i,j} \ff_{ij} \left(
    \frac{\hat{s}}{s}, \mu_f \right) \nonumber
  \\
  &\quad \times \frac{1}{2} \bigg[ \int_{-\hat{s}(1+\beta)/2}^{-\hat{s}(1-\beta)/2} d\hat{t}_1
  \int_0^{\hat{s}+\hat{t}_1+\hat{s}m_t^2/\hat{t}_1} ds_4 \,
  C_{ij}(s_4,\hat{s}_t',\hat{t}_1',-\hat{s}_t'-\hat{t}_1' ,m_t,\mu_f) \nonumber
  \\
  &\qquad\mbox{} + \int_{-\hat{s}(1+\beta)/2}^{-\hat{s}(1-\beta)/2} d\hat{u}_1
  \int_0^{\hat{s}+\hat{u}_1+\hat{s}m_t^2/\hat{u}_1} ds_4 \,
  C_{ij}(s_4,\hat{s}_u',-\hat{s}_u'-\hat{u}_1' , \hat{u}_1', m_t,\mu_f) \bigg] \, ,
\end{align}
where $\hat{s}=4m_t^2/(1-\beta^2)$, and
\begin{gather}
  \hat{s}_t' = \hat{s} \, \frac{-\hat{t}_1}{s_4-\hat{t}_1} \, , \quad \hat{t}_1'=\hat{t}_1
\end{gather}
in the first term in the bracket, while
\begin{gather}
  \hat{s}_u' = \hat{s} \, \frac{-\hat{u}_1}{s_4-\hat{u}_1} \, , \quad \hat{u}_1'=\hat{u}_1
\end{gather}
in the second. We have also defined the parton luminosity function
\begin{align}
  \ff_{ij}(y,\mu_f) = \int_y^1 \frac{dx}{x} \, f_{i/N_1}(x,\mu_f) \, f_{j/N_2}(y/x,\mu_f)
  \, .
\end{align}
Since the change of variables is carried out exactly, (\ref{eq:tot}) and (\ref{eq:tot2})
give the same result for the total cross section when integrated over.

We should emphasize again that the procedure above is by no means unique. In
\cite{Kidonakis:2001nj}, for instance, the scheme was instead specified at the level of
the $\beta$ distribution. In particular, the hard-scattering kernels as a function of
$\hat{t}_1', \hat{u}_1', \hat{s}'$ were written in a form specified in the Appendix of
that paper, $u_1'$ was eliminated as an integration variable using
$\hat{s}'+\hat{t}'_1+\hat{u}'_1=s_4$, and then the integration over $s_4$ was carried out
followed by that over $\hat{t}_1'$.\footnote{We are grateful to Sven Moch for the
  clarification of this point, and for providing the numerical code used in the comparison
  in Section~\ref{subsec:1PIcompare}.} We will discuss the numerical differences which
result from using such an equivalent procedure when we compare with previous literature in
Section~\ref{subsec:1PIcompare}.

Having specified our procedure for evaluating the formulas in the threshold region, we
next clarify how to match the results with fixed-order perturbation theory at NLO. The
exact results contain the perturbative corrections to our formula which vanish in the
limit $s_4 \to 0$, and to obtain solid phenomenological results it is important to include
them. In resummed perturbation theory, we achieve NLO+NNLL accuracy by evaluating
differential cross sections according to
\begin{align}
  \label{eq:FixedMatching}
  d\sigma^{\text{NLO+NNLL}} &\equiv d\sigma^{\text{NNLL}} \Big|_{\mu_h,\mu_s,\mu_f} +
  d\sigma^{\text{NLO, subleading}} \Big|_{\mu_f} \nonumber
  \\
  &\equiv d\sigma^{\text{NNLL}} \Big|_{\mu_h,\mu_s,\mu_f} + \left( d\sigma^{\text{NLO}}
    \Big|_{\mu_f} - d\sigma^{\text{NLO, leading}} \Big|_{\mu_f} \right) ,
\end{align}
where $d\sigma^{\text{NLO}}$ is the exact result in fixed order, and
$d\sigma^{\text{NLO,leading}}|_{\mu_f} \equiv d\sigma^{\text{NNLL}}|_{\mu_h=\mu_s=\mu_f}$
captures the leading singular terms in the threshold limit $s_4 \to 0$ at NLO. To obtain
approximate NNLO results in fixed order, we simply compute 
\begin{equation}
  \label{eq:NNLOapprox}
  d\sigma^{\text{NNLO, approx}} = d\sigma^{\text{NLO}} + d\sigma^{\text{(2), approx}} \, ,
\end{equation}
where $d\sigma^{\text{(2), approx}}$ is the approximate NNLO correction to the differential cross
section obtained using the coefficient function (\ref{eq:C2}).

The size of the power corrections contained in parentheses in the second term of
(\ref{eq:FixedMatching}) depends on the region of phase space in which the differential
cross sections are evaluated, and also on the interplay of the hard-scattering kernels
with the PDFs. The power corrections are expected to be small when $\hat{s}\to
4m_\perp^2$, since in that case $s_4\to 0$. However, experiments do not typically
reconstruct $\hat{s}$ as an observable. 
 For more interesting 
differential distributions the limit $s_4\to 0$ can be enforced via a restriction to 
the machine threshold, for instance by requiring that $m_\perp\to \sqrt{s}/2$ for the $p_T$ spectrum, 
but in this case the differential cross section would be extremely small.
Away from such special kinematic regions the
leading terms in the partonic threshold limit are dominant only if a dynamical enhancement
occurs, because the product of PDFs appearing in the cross section falls off sharply away
from the region where $s_4 \to 0$. Some systematic studies of the conditions under which
threshold enhancement is effective have been made in \cite{Becher:2007ty, Bauer:2010jv}, 
but in the end this is largely a numerical question, which must be examined on a
case-by-case basis. Given its importance, we address the issue of threshold enhancement
and power corrections to the threshold expansion in some detail in the next section.

We end this section by summarizing some of the inputs needed for numerical predictions,
which we shall use in the rest of the paper unless otherwise indicated. The results depend
on the input parameters $m_t$, $\alpha_s$, and the PDFs. We shall use the MSTW2008 PDF
sets \cite{Martin:2009bu}, and change the PDF set according to the order at which the
perturbative hard-scattering kernel is evaluated, as indicated in Table~\ref{tab:PDForder}.
The running couplings $\alpha_s(\mu)$ are taken in the $\overline{\rm{MS}}$ scheme with
five active flavors, using one-loop running at LO, two-loop running at NLO, and three-loop
running at NNLO. By default we take $m_t=173.1$~GeV in the pole scheme. Using a more
physical quark mass such as that in the $\overline{\mbox{MS}}$ scheme can lead to a better
convergence of the perturbative series \cite{Langenfeld:2009wd, Ahrens:2010zv}, but for
the purposes of this work we shall leave that issue aside.

Finally, in the following we do not consider the theoretical uncertainty on the top-pair
cross section induced by the error on $\alpha_s(M_Z)$, which is an input parameter.
However, a recent study employing CTEQ PDFs shows that this
additional theoretical uncertainty can be as big as $4 \%$ for the total top-quark pair
production cross section at the LHC \cite{Lai:2010nw}. Such an uncertainty is therefore
not negligible in comparison to the residual scale uncertainty in NNLO and NNLL
calculations, and it will need to be considered when comparing data and theoretical
predictions. An analysis of the effects of the $\alpha_s$ uncertainty on several cross
sections of interest at the Tevatron and at the LHC was also carried out by the MSTW
collaboration \cite{Martin:2009bu}. However, the top-quark pair production is not among
the processes considered in that work.

\begin{table}
\vspace{4mm}
\begin{center}
\begin{tabular}{|l|c|c|c|}
  \hline
  Order & PDF set & $\alpha_s(M_Z)$ \\ 
  \hline
  LO & MSTW2008LO & 0.139 \\ 
  NLO, NLL & MSTW2008NLO & 0.120 \\ 
  NNLO approx, NLO+NNLL & MSTW2008NNLO & 0.117 \\ 
  \hline
\end{tabular}
\end{center}
\vspace{-2mm}
\caption{\label{tab:PDForder}
  Order of the PDFs \cite{Martin:2009bu} and the corresponding values of the strong
  coupling used for the different perturbative approximations.}
\end{table}

\section{Threshold enhancement and power corrections}
\label{sec:threshold}

We now proceed to study threshold enhancement and the numerical importance of power
corrections to the factorization formula (\ref{eq:fac}). The goal is to examine under
which conditions the higher-order corrections dominating in the limit $s_4\to0$ can be
expected to give a good approximation to the full result. Given its importance, we approach
this question from several different angles. In Section~\ref{subsec:1PIdists} we compare
the leading terms in the 1PI and $\scetopi$ threshold expansions for the $p_T$ and
rapidity distributions with the exact results at NLO in QCD. We then gain more insight
into the patterns observed there by studying the differential cross section with respect
to $\beta=\sqrt{1-4m_t^2/\hat{s}}$, which is obtained from the parton luminosity functions
multiplied by the total partonic cross section in Section~\ref{subsec:betadists}. We make
general comments on the pattern of power corrections in 1PI and PIM kinematics, and also
investigate the extent by which corrections stemming from logarithmic plus-distributions
are enhanced compared to other terms.

\subsection{Transverse-momentum and rapidity distributions at NLO}
\label{subsec:1PIdists}

\begin{figure}
\begin{center}
\begin{tabular}{ll}
\includegraphics{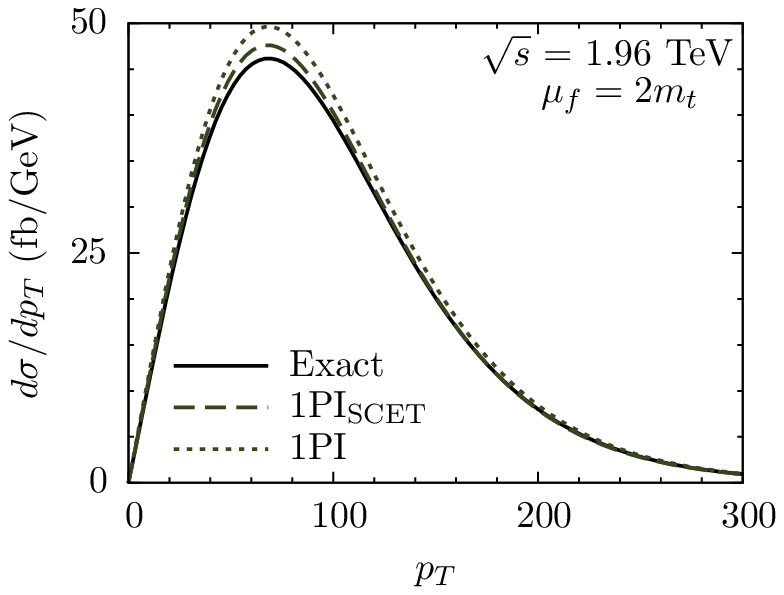}
&
\includegraphics{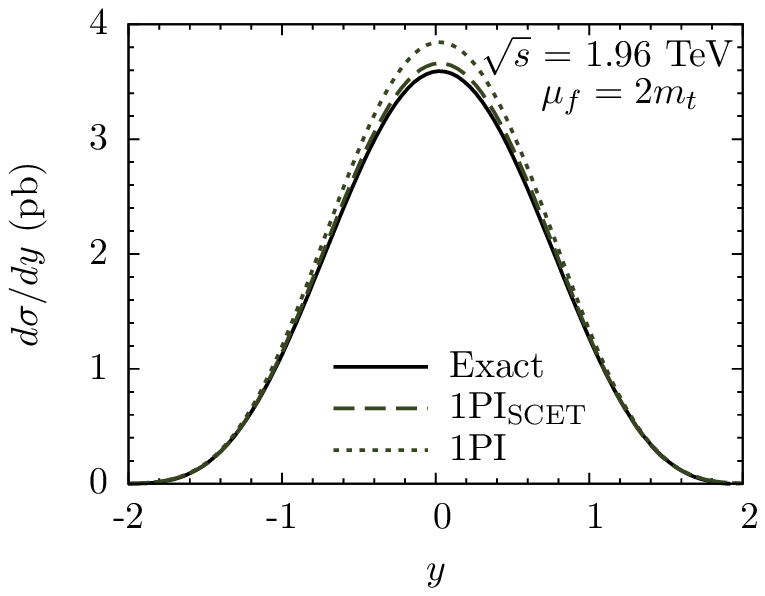}
\\
\includegraphics{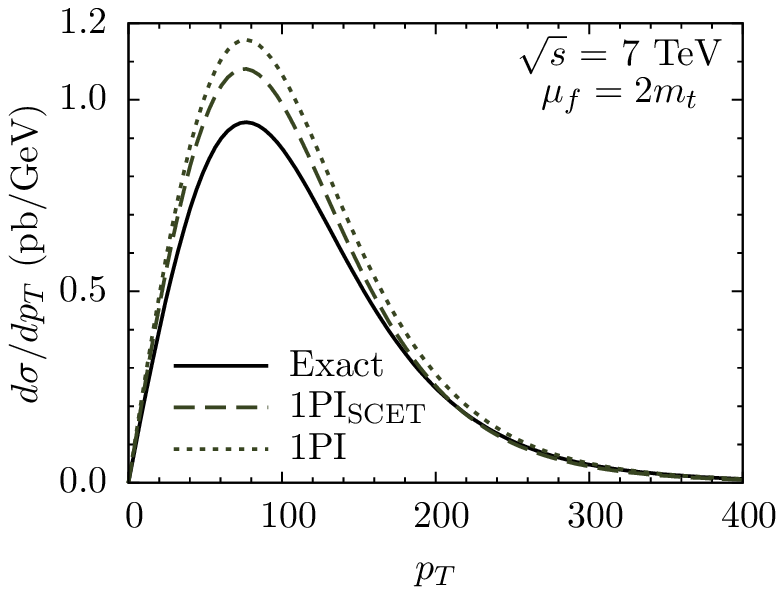}
&
\includegraphics{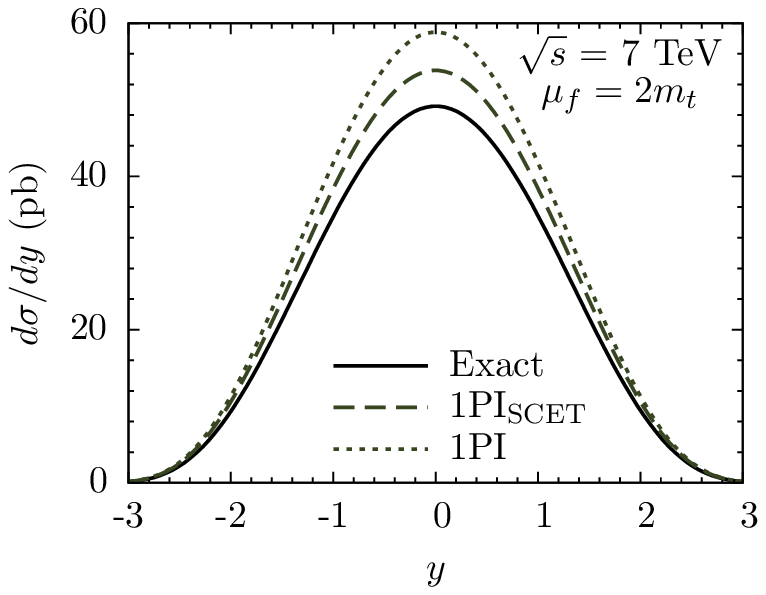}
\\
\includegraphics{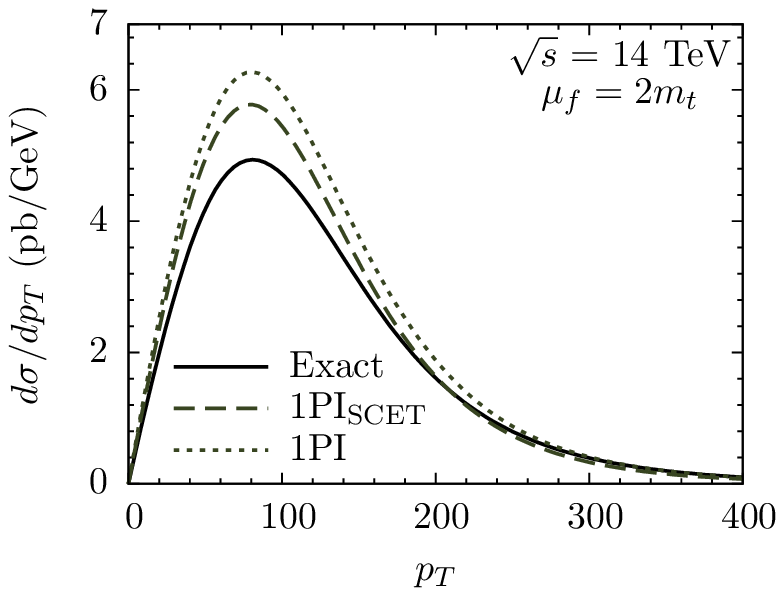}
&
\includegraphics{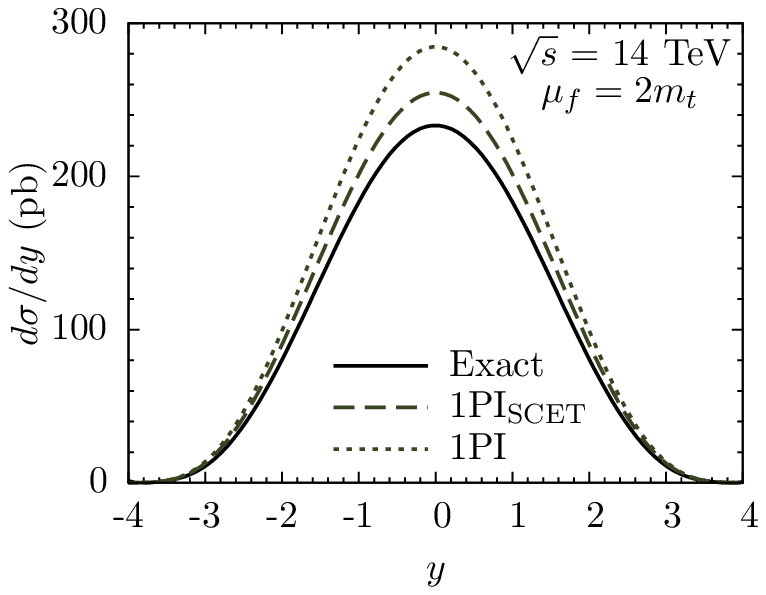}
\end{tabular}
\end{center}
\vspace{-1ex}
\caption{\label{fig:subl_terms} The transverse-momentum and rapidity distributions at
  NLO at the Tevatron with $\sqrt{s}=1.96$~TeV and at the LHC with $\sqrt{s}=7$ and 14~TeV.}
\end{figure}

\begin{figure}
\begin{center}
\begin{tabular}{ll}
\includegraphics{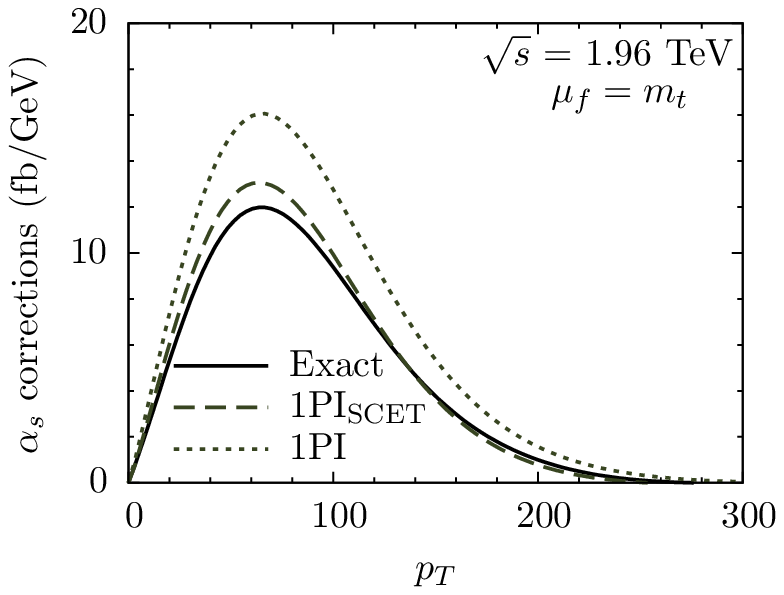}
&
\includegraphics{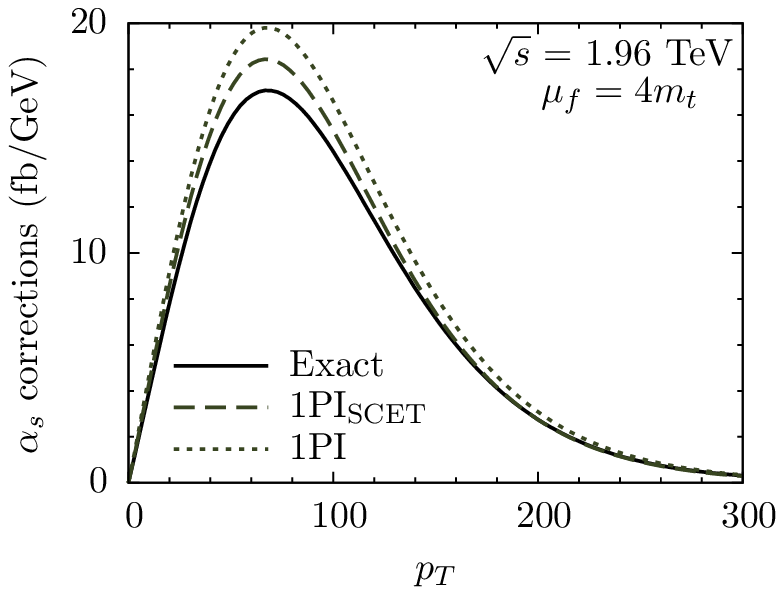}
\\
\includegraphics{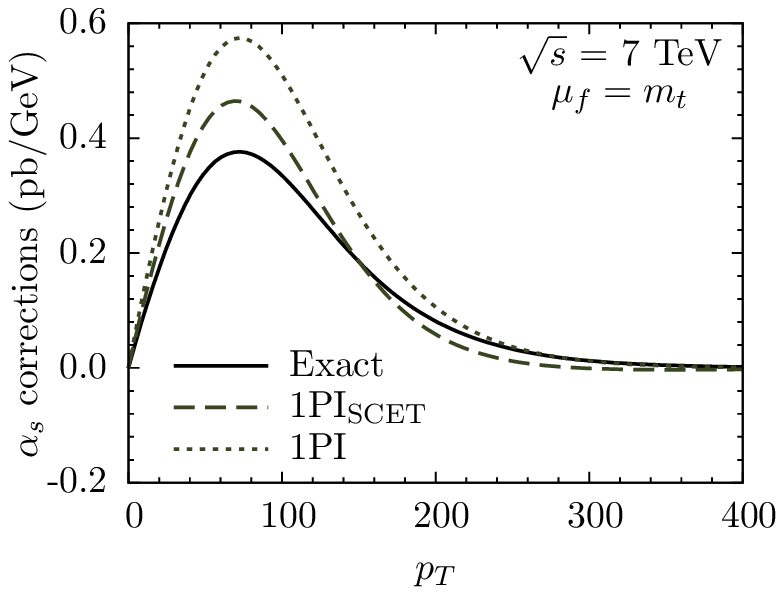}
&
\includegraphics{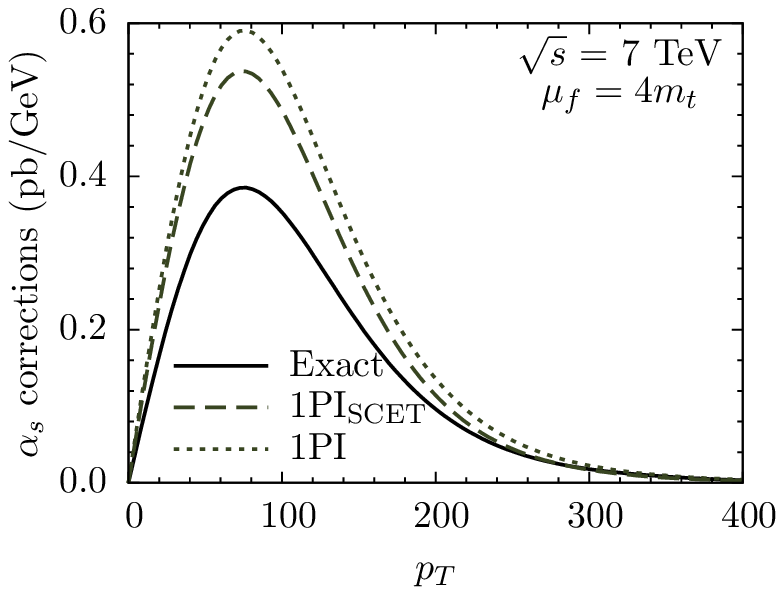}
\\
\includegraphics{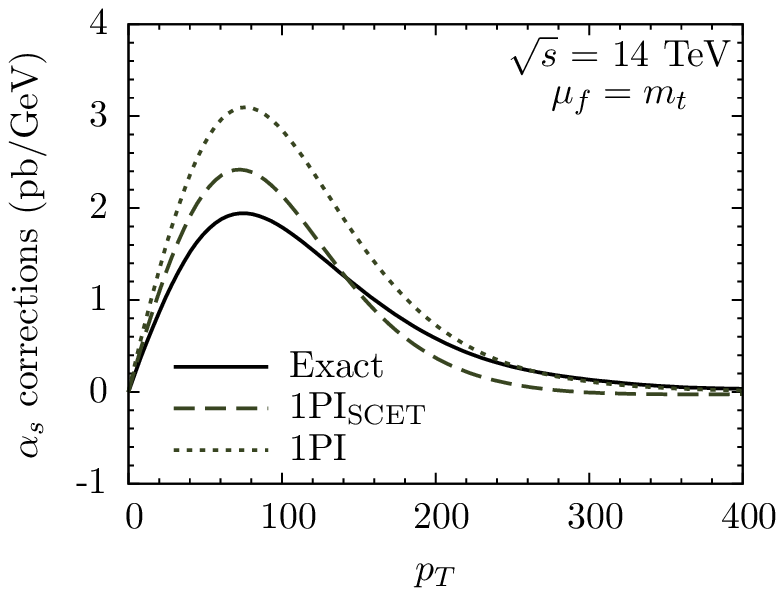}
&
\includegraphics{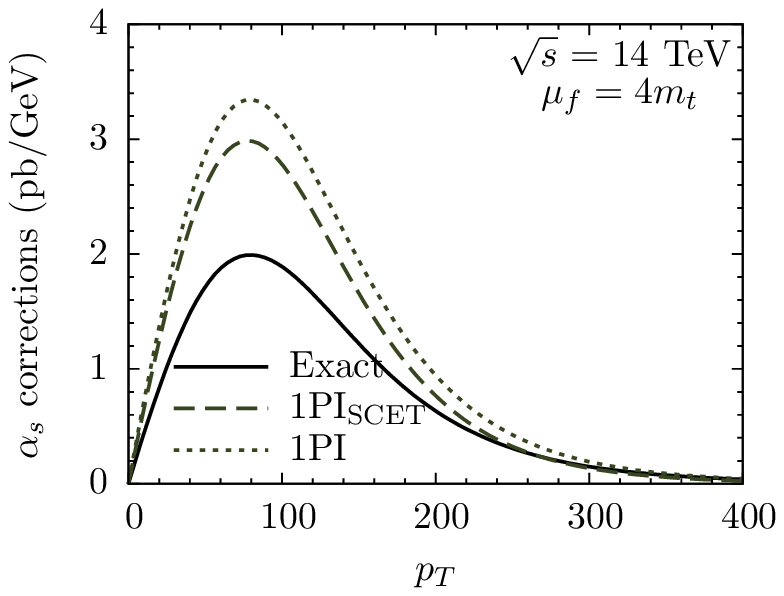}
\end{tabular}
\end{center}
\vspace{-1ex}
\caption{\label{fig:subl_cors} The NLO corrections to $d\sigma/dp_T$.}
\end{figure}

Our first test of the threshold expansion in 1PI kinematics is to compare its predictions
for the transverse-momentum and rapidity distributions with the exact ones at NLO in QCD.
In other words, we study whether the leading singular pieces in the 1PI or $\scetopi$
threshold expansions provide a good approximation to the exact NLO hard-scattering
kernels, so that the power corrections contained in the parentheses of the second term in
(\ref{eq:FixedMatching}) are small. The results of this comparison can be found in
Figure~\ref{fig:subl_terms}, where the transverse-momentum and rapidity distributions at
the Tevatron and the LHC with $\sqrt{s}=7$ and 14~TeV are displayed, for the choice
$\mu_f=2m_t$. To compute the exact QCD corrections we rely on the Monte Carlo programs
MCFM \cite{Campbell:2000bg} and an internal NLO version of MadGraph/MadEvent\footnote{We
  are grateful to Rikkert Frederix for providing the code.} \cite{Alwall:2007st,
  Frederix:2009yq}, whereas to implement the leading pieces of the threshold expansion in
$\scetopi$ and 1PI we have used the procedure described in Section~\ref{sec:numerical}.
All inputs, including PDFs and $m_t$, can also be found in that section. It is obvious
from the figures that the $\scetopi$ approximation does better than the 1PI approximation
in reproducing the exact QCD results. Since the three curves differ only through the NLO
corrections to the hard-scattering kernels, we can compare them much better if we isolate
these pieces. We have done so for the $p_T$ distribution in Figure~\ref{fig:subl_cors}, in
this case for two different values of $\mu_f$. We then see that at the Tevatron the
$\scetopi$ approximation works remarkably well over the full range of $\mu_f$. At the LHC,
the $\scetopi$ approximation reasonably reproduces the correction at $\mu_f = m_t$, but
does relatively poorly in reproducing the correct $\mu_f$ dependence. The 1PI
approximation significantly overestimates the true result at all three values of $\mu_f$,
both at the Tevatron and at the LHC.

Given that the numerical differences between the $\scetopi$, 1PI, and exact results are
due solely to subleading terms as $s_4\to 0$, we conclude that power corrections to the
pure threshold expansion can be sizeable at NLO. At the Tevatron, where the $q\bar q$
channel gives the largest contributions, the extra terms related to $L_4$ included in
$\scetopi$ account for the dominant power corrections. They are also important at the LHC,
where the $gg$ channel dominates, but so are other corrections, which cannot be obtained
in our formalism.
 
We have focused on the region of $p_T$ where the differential cross section is
largest. In principle, our results can also be used to predict the high-$p_T$ tail of
the distribution. In the limit $m_\perp\to \sqrt{s}/2$, for instance, $s_4\to 0$, so
threshold expansion is bound to work well when compared to the exact NLO result.
However, in such kinematic regions the differential cross section is so small that it is essentially
unobservable, and the top-quark is so highly boosted that $m_t\ll \sqrt{\hat{s}}$, so
the power counting in the effective theory would need to be modified. A more interesting
region would be up to around $400$~GeV at the Tevatron, and up to around a TeV at the
LHC. We will include the higher-$p_T$ region for the Tevatron in the phenomenological
studies. For the LHC, however, $s_4$ can be on average rather large at such values of
$p_T$, so power corrections to the $gg$ and $qq$ channels can become significant, and the
$qg$ channel can also give non-negligible contributions. Given these problems, we will not
study the high-$p_T$ distributions at the LHC. 
  
This simple NLO study is instructive, but in the end the real issue is how well the
threshold approximation is expected to work at NNLO. On the one hand, at NNLO one
encounters plus-distributions enhanced by up to three powers of logarithms, so it is not
unreasonable to expect that the power corrections are of less relative importance than at
NLO. On the other hand, at NLO the coefficients multiplying both the $P_n$ distributions and
$\delta$-function terms are known exactly, while at NNLO only those multiplying the $P_n$ distributions
are available. It is therefore difficult to anticipate the behavior of the threshold
expansion at NNLO based only on its behavior at NLO. However, we can gain additional
insights through the studies of the $\beta$ distribution in the next subsection.

\subsection{The $\bm\beta$ distribution and total cross section}
\label{subsec:betadists}

The differential cross section with respect to $\beta=\sqrt{1-4m_t^2/\hat{s}}$ can be
expressed as
\begin{align}
  \label{eq:dbeta}
  \frac{d\sigma}{d\beta} = \frac{1}{s}\frac{8\beta}{(1-\beta^2)^2} \sum_{ij} \,
  \ff_{ij}\bigg(\frac{\hat{s}}{s},\mu_f\bigg)\,\alpha_s^2 \,f_{ij} \bigg( \frac{4
    m_t^2}{\hat{s}} ,\mu_f \bigg) \,.
\end{align}
The coefficient functions $f_{ij}$ are proportional to the total partonic cross sections, and are obtained from
our results by comparison with (\ref{eq:tot2}). We define expansion coefficients for these
functions as
\begin{align}
  \label{eq:fexp}
  f_{ij} &= f_{ij}^{(0)} + 4\pi \alpha_s f_{ij}^{(1)} + \left(4\pi\alpha_s\right)^2 \left[
    f_{ij}^{(2,0)} + f_{ij}^{(2,1)} \ln\frac{\mu_f^2}{m_t^2} + f_{ij}^{(2,2)}
    \ln^2\frac{\mu_f^2}{m_t^2} \right] + \dots \,.
\end{align}
This differential cross section is not measured experimentally, but from the theoretical
perspective it is convenient because answers for the partonic cross section are known
analytically to NLO \cite{Czakon:2008ii}, and for the functions $f_{ij}^{(2,1)}$ and
$f_{ij}^{(2,2)}$ multiplying the scale-dependent logarithms to NNLO
\cite{Langenfeld:2009wd}. Moreover, this cross section is calculable in both 1PI and PIM kinematics, so it
gives us a way of directly comparing the threshold expansion and power corrections for
these two cases. The results must agree in the limit $\beta\to 0$, since in that case
gluon emission is soft, but beyond that they receive a different set of power corrections,
so the agreement of the two approximations with each other is one way of testing whether
these power corrections are under control. 

\begin{figure}
\begin{center}
\begin{tabular}{ll}
\includegraphics{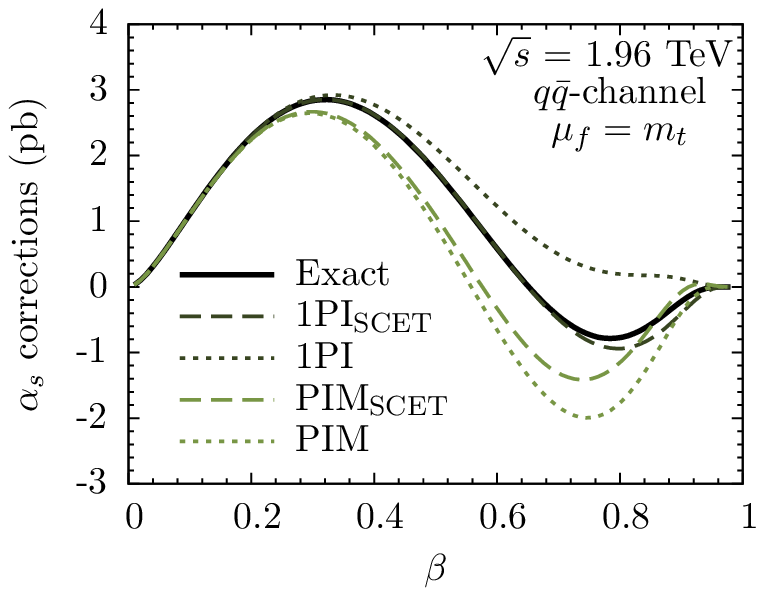} & 
\includegraphics{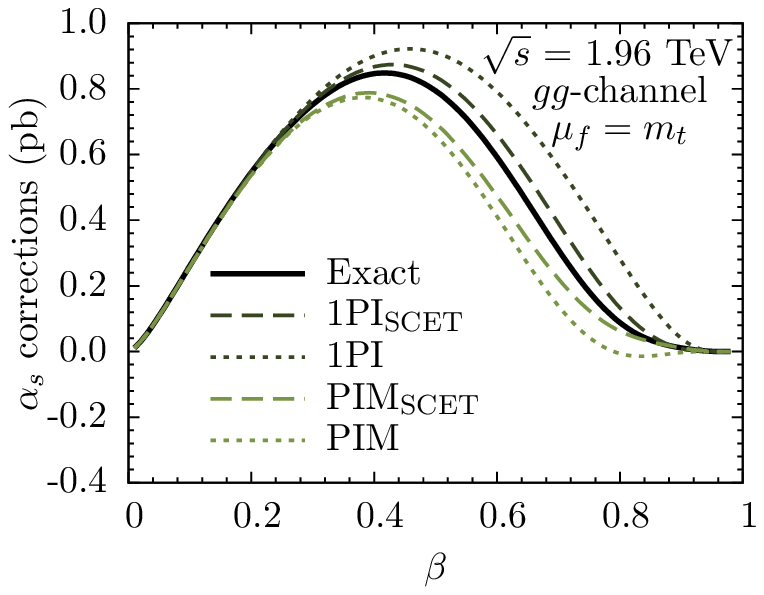}
\\
\includegraphics{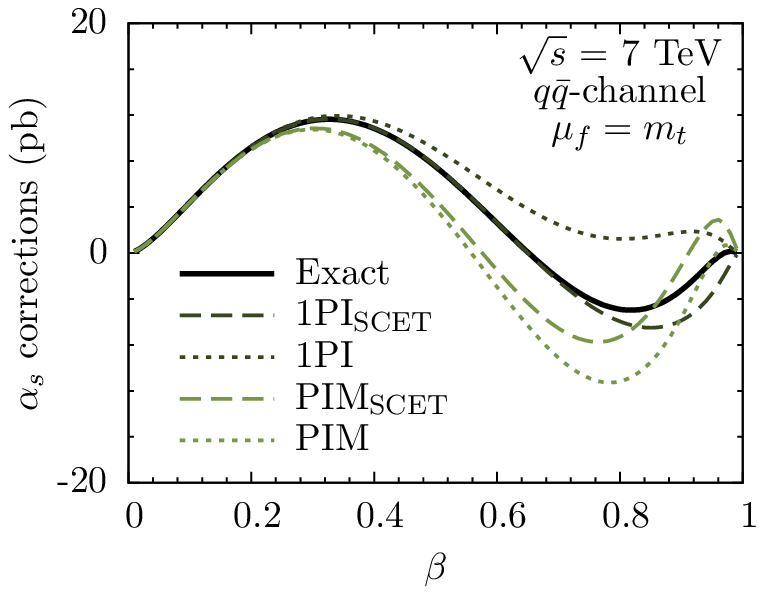} & 
\includegraphics{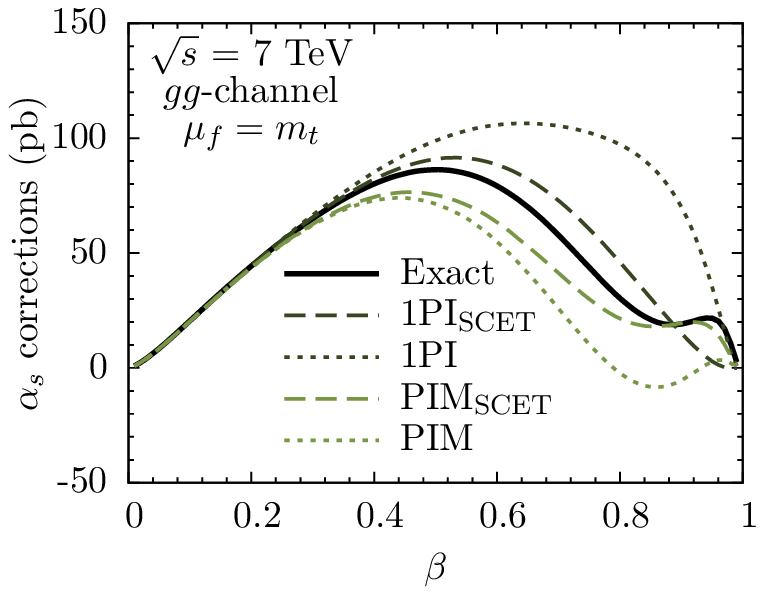} 
\\
\includegraphics{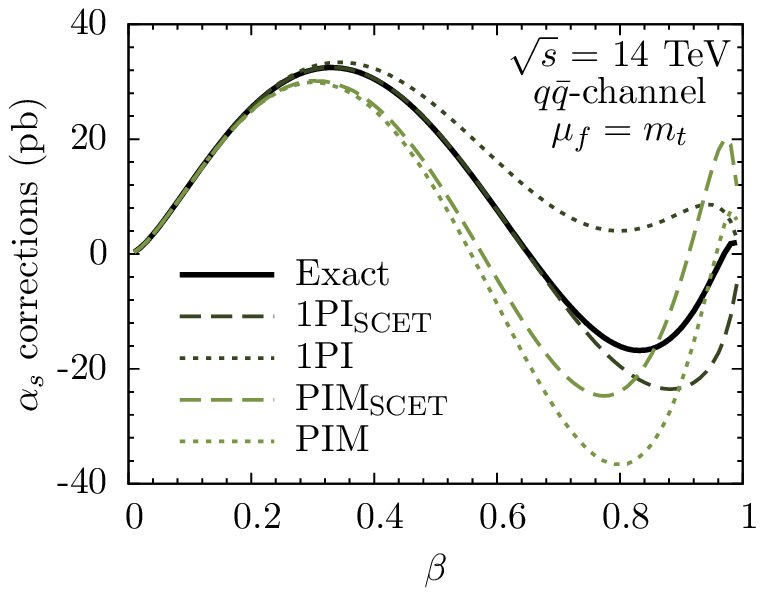} & 
\includegraphics{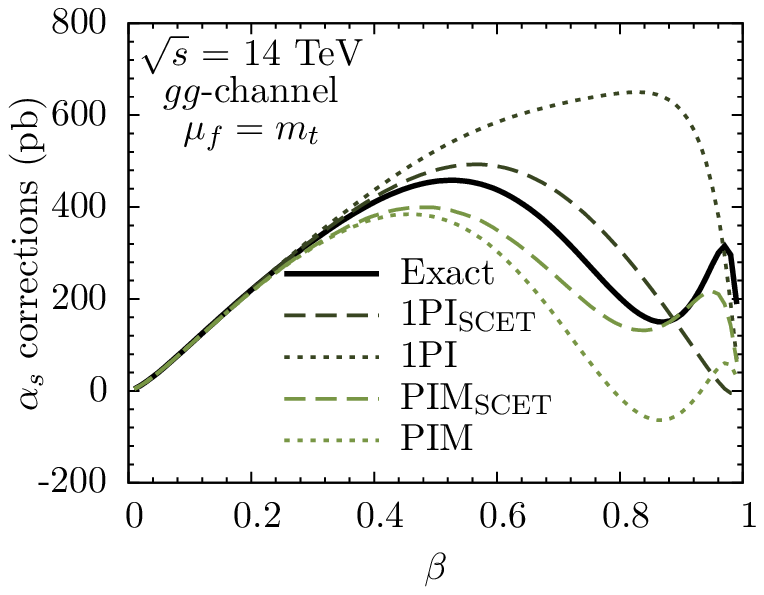} 
\end{tabular}
\end{center}
\vspace{-1ex}
\caption{\label{fig:beta-plot} The $\alpha_s$ corrections to $d\sigma/d\beta$ for the
  different approximations mentioned in the text, with $\mu_f=m_t$.}
\end{figure}

We begin with a study of the NLO corrections, similar to that performed for the $p_T$
distribution in the previous section. In this case, however, we can also include the
results from PIM kinematics, which we calculate using the expressions in
\cite{Ahrens:2010zv}. In Figure~\ref{fig:beta-plot} we compare results for the $\alpha_s$
correction to (\ref{eq:dbeta}) obtained within the different expansions. In contrast to
the $p_T$ spectrum, we compare the quark and gluon channels separately and look at the
corrections in a kinematic range from the production threshold at $\beta=0$ to the machine
threshold at $\beta_{\rm max}=\sqrt{1-4m_t^2/s}$. As anticipated from the results of the
previous section, the $\scetopi$ approximation works quite well in the $q\bar q$ channel,
but somewhat worse in the $gg$ channel, especially at the LHC. The 1PI results
overestimate the exact corrections in all cases. As for the results in PIM kinematics,
they are lower than the exact results in all cases, but the power-suppressed terms
included in $\scetpim$ bring the leading terms in the threshold expansion closer to the
full result. The $\scetpim$ approximation is slightly worse than $\scetopi$ in the $q\bar
q$ channel, and slightly better in the $gg$ channel, but the differences are not major.

\begin{figure}[t]
\begin{center}
\begin{tabular}{cc}
\includegraphics{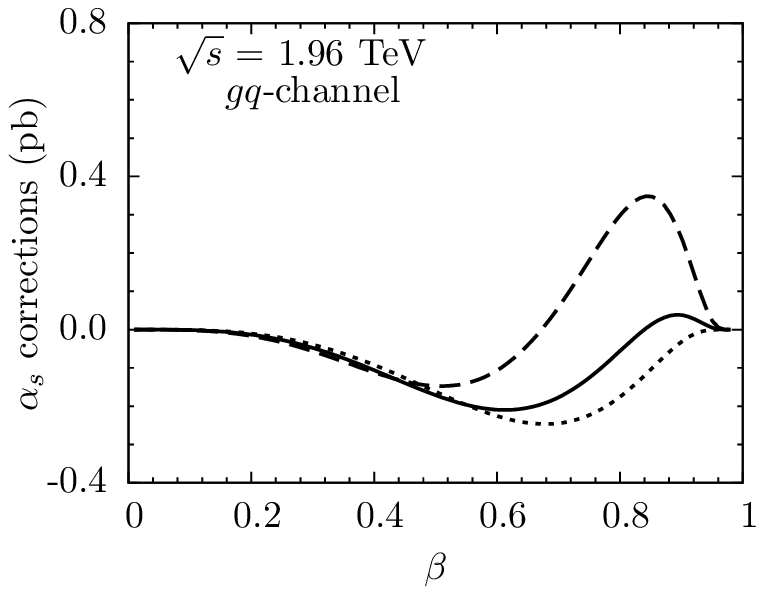}
&
\includegraphics{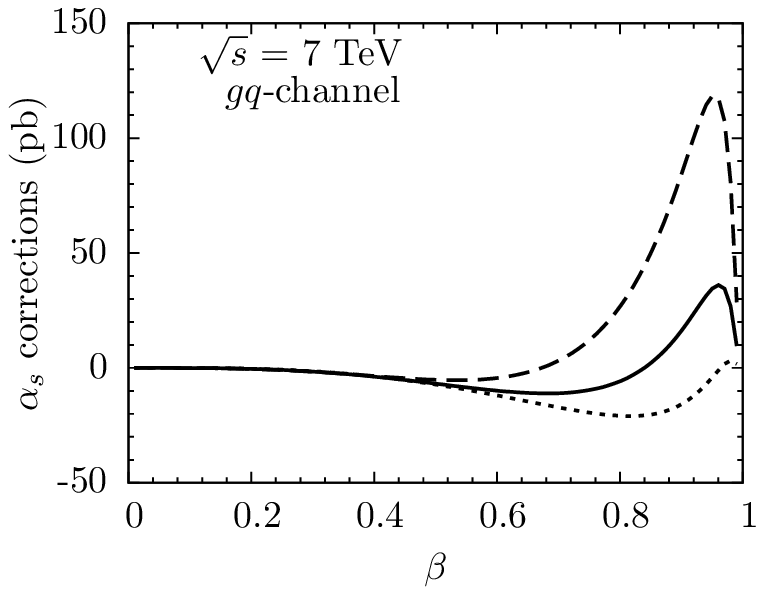} \\
\end{tabular}
\begin{tabular}{c}
\includegraphics{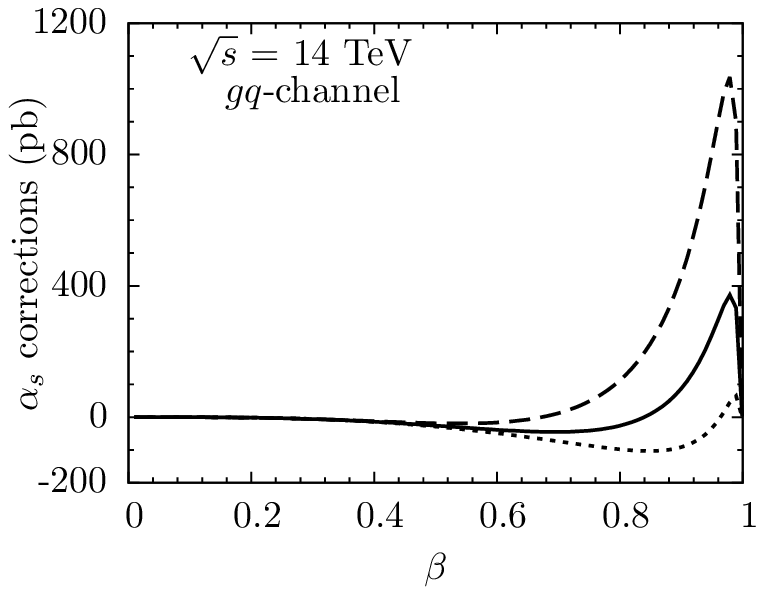} 
\end{tabular}
\end{center}
\vspace{-1ex}
\caption{\label{fig:beta-plot-gq} The $\alpha_s$ corrections to $d\sigma/d\beta$ from the
  $gq$ channel, for $\mu_f = m_t$ (solid), $\mu_f = m_t/2$ (dashed) and $\mu_f = 2m_t$
  (dotted).}
\end{figure}

As mentioned above, all of the approximations have the same leading-order expansion in the
limit $\beta\to 0$. The power-suppressed effects accounting for the differences between
the curves start to become noticeable at $\beta\sim 0.2$. After that point, there is more
phase space for hard gluon emission and the size of the power corrections increases.
Evidently, the subleading terms included in the $\scetopi$ and $\scetpim$ approximations
account for these power-suppressed terms in part, although not completely. This is
especially noticeable in 1PI kinematics, where the power corrections are generically more
important than in PIM kinematics, a point we will return to below. The power
corrections become progressively more important at higher values of the collider energy,
since then the luminosities are larger at high $\beta$ and the differences in the
partonic cross sections in that region are magnified. This is most easily seen by
comparing the LHC results at the two different collider energies, where one observes
larger gaps between the approximations at 14~TeV than at 7~TeV. A careful examination of
the results at the LHC with $\sqrt{s}=14$~GeV also shows a feature not obvious in the
other cases: at very high values of $\beta$ close to the endpoint, the exact correction
in the $gg$ channel remains a positive number. In fact, the exact NLO correction to the
partonic cross section in the $gg$ channel tends to a positive constant at very high
$\hat{s}$ \cite{Nason:1987xz}, while that for the $q\bar q$ channel, and also the
threshold approximations to the corrections in PIM and 1PI kinematics in both channels,
approach zero at high $\hat{s}$. This feature is only visible at the highest collider
energy, because otherwise the high-$\beta$ cross section is completely damped by the
luminosities. For the same reason, the $qg$ channel, for which the partonic cross
section also tends to a positive constant at high $\hat{s}$, can become important at
high $\beta$. We show the $\alpha_s$ correction from this channel at the LHC in
Figure~\ref{fig:beta-plot-gq}, for three different values of $\mu_f$ (at the Tevatron,
the contribution from this channel is still very small.) For higher values of $\beta$,
and especially at lower values of $\mu_f$, it can be as important as the $gg$ channel,
even though it is suppressed in the limit $s_4\to 0$. We discuss this effect at the
level of the total cross section in more detail in Section~\ref{sec:pheno}.

%

%

%

%

\begin{figure}
\begin{center}
\begin{tabular}{ll}
  \includegraphics{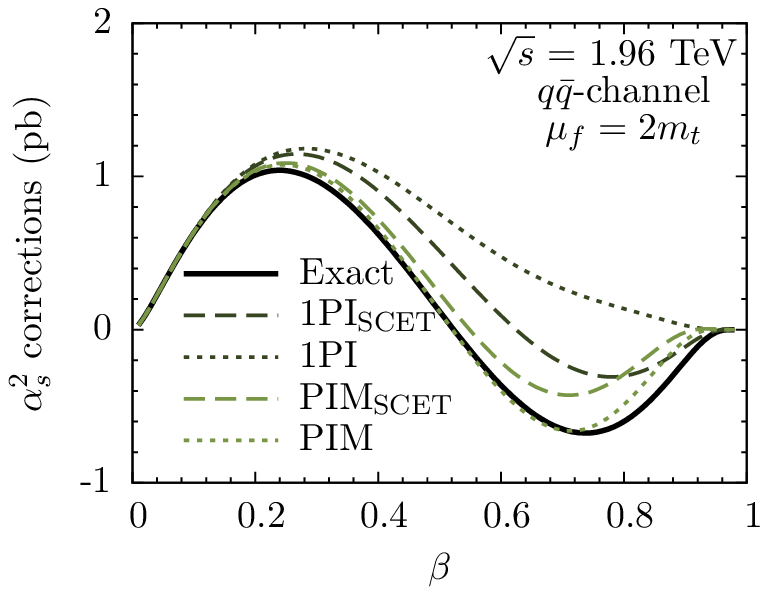}
  &
  \includegraphics{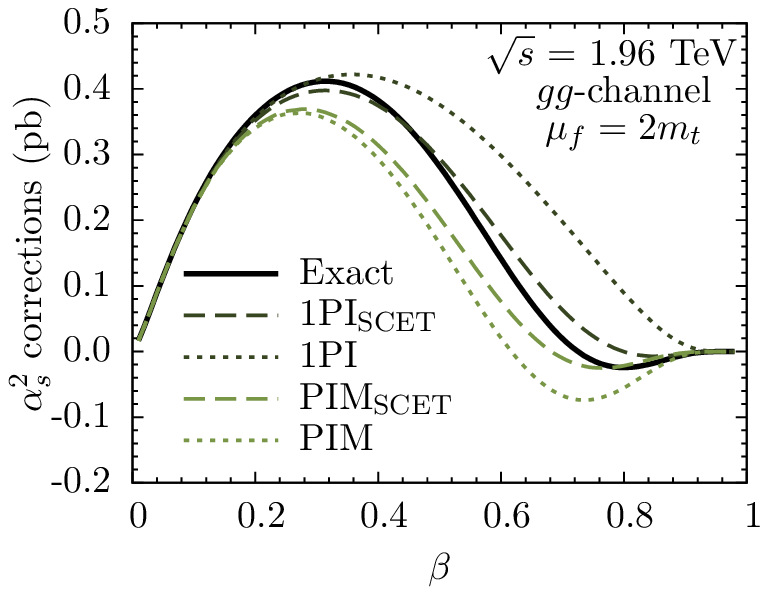}
  \\
  \includegraphics{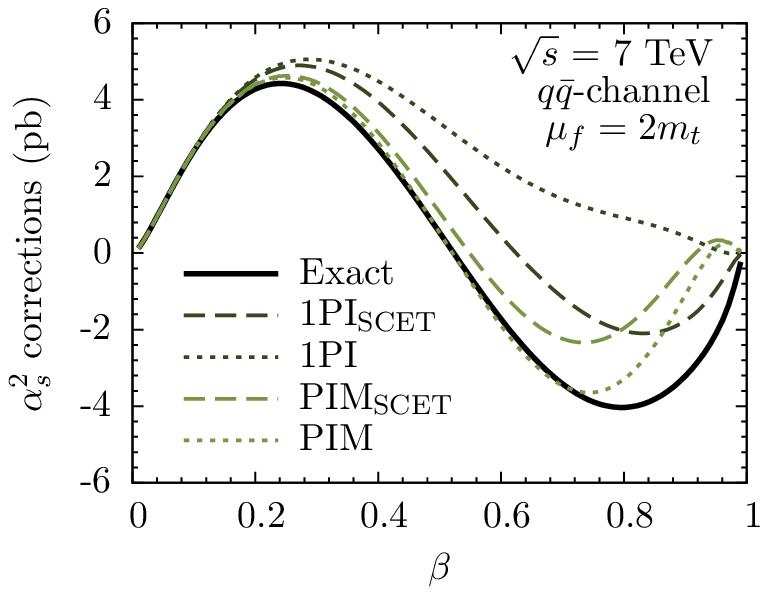}
  &
  \includegraphics{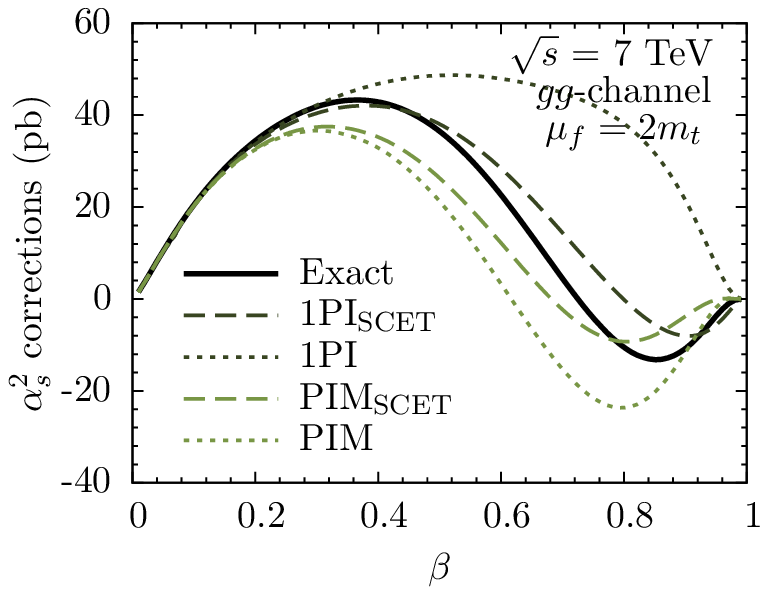}
  \\
  \includegraphics{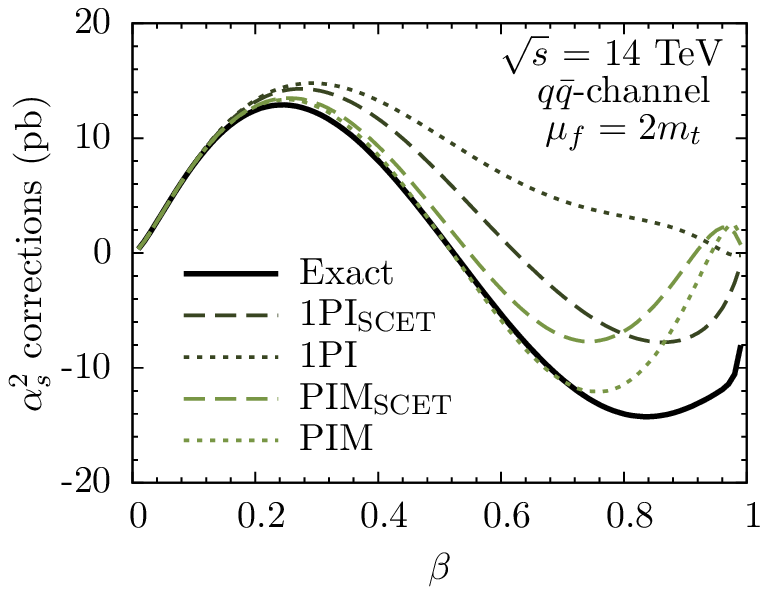}
  &
  \includegraphics{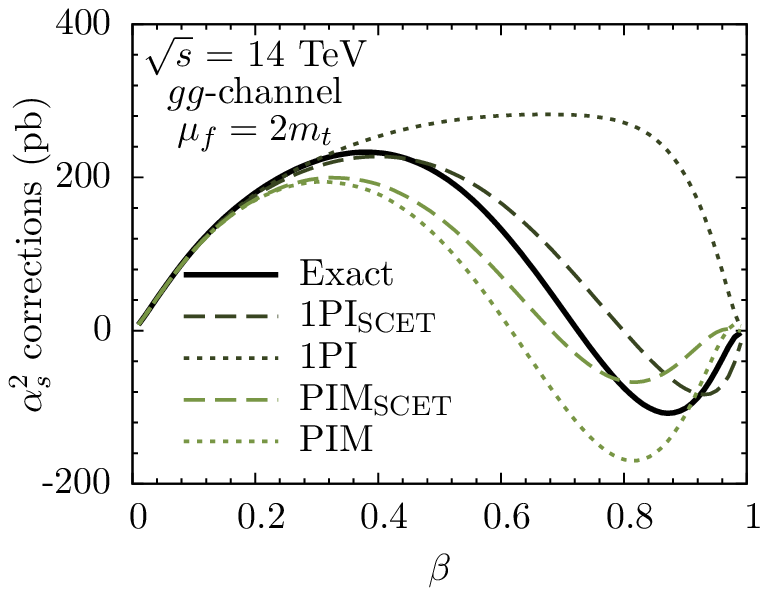}
\end{tabular}
\end{center}
\vspace{-1ex}
\caption{\label{fig:beta-plot-nnlomu} The $\alpha_s^2$ corrections to $d\sigma/d\beta$
  obtained by dropping the scale-independent piece $f^{(2,0)}$ in (\ref{eq:fexp}), for
  $\mu_f=2m_t$. The exact result is the black line, the dashed red line $\scetopi$, the
  dotted red line 1PI, the dashed blue line $\scetpim$, and the dotted blue line PIM.}
\end{figure}

\begin{figure}
\begin{center}
\begin{tabular}{ll}
  \includegraphics{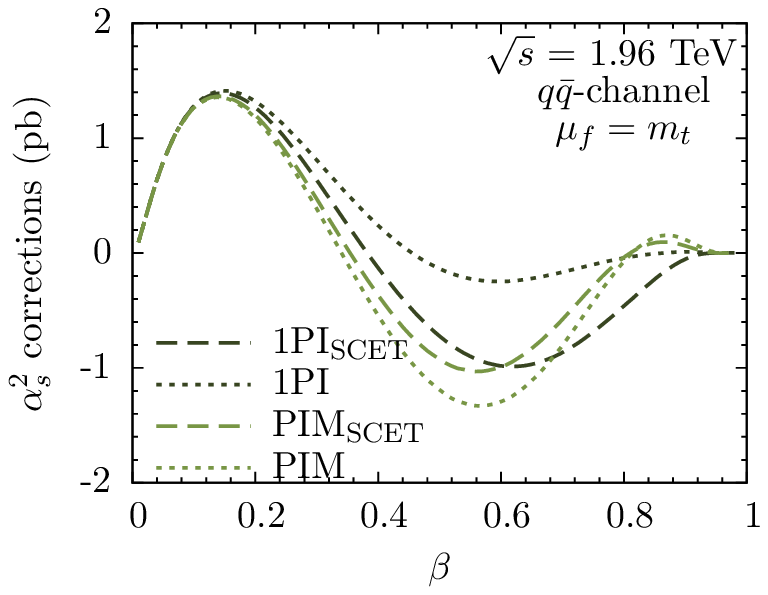}
  & 
  \includegraphics{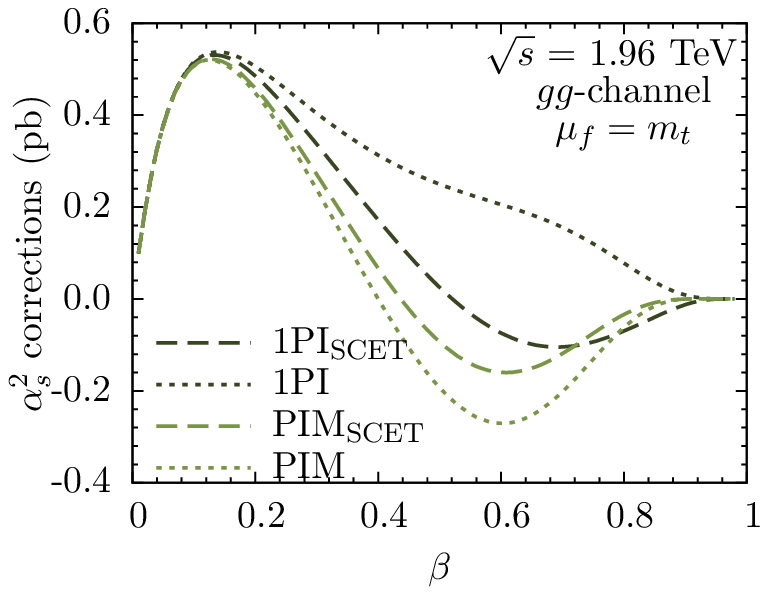}
  \\
  \includegraphics{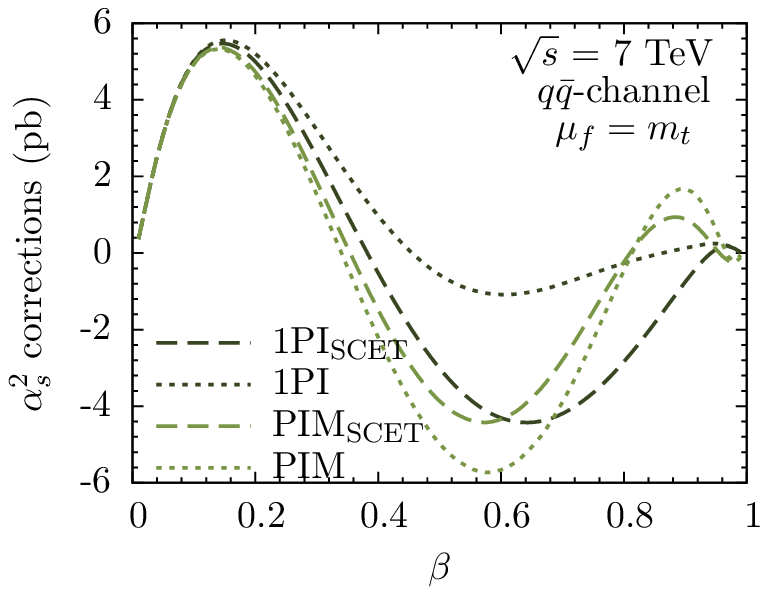}
  & 
  \includegraphics{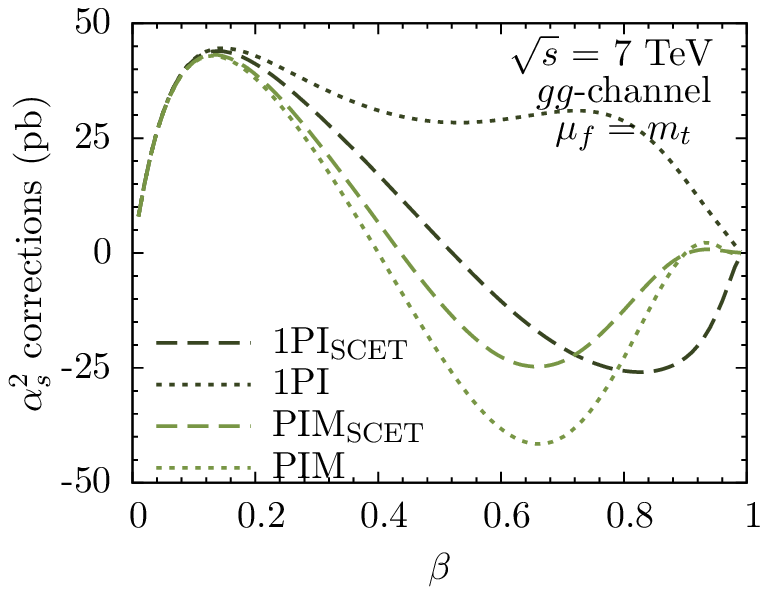} 
  \\
  \includegraphics{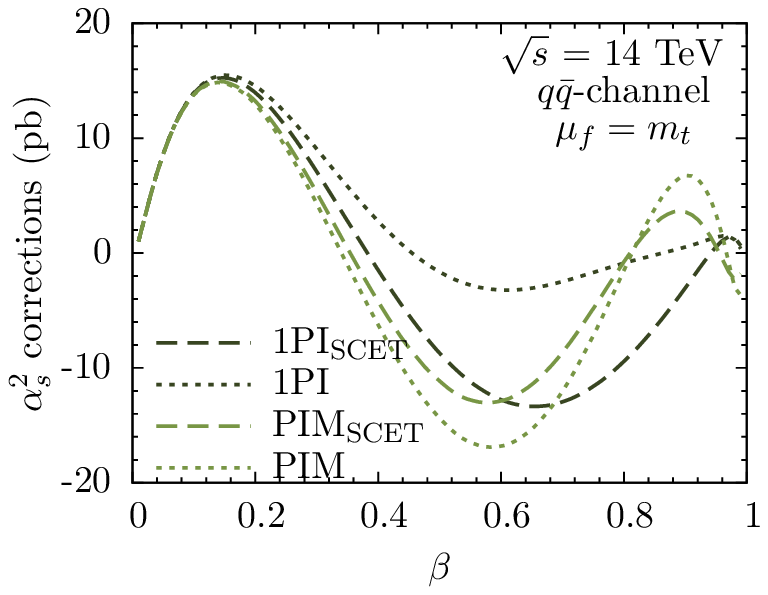}
  & 
  \includegraphics{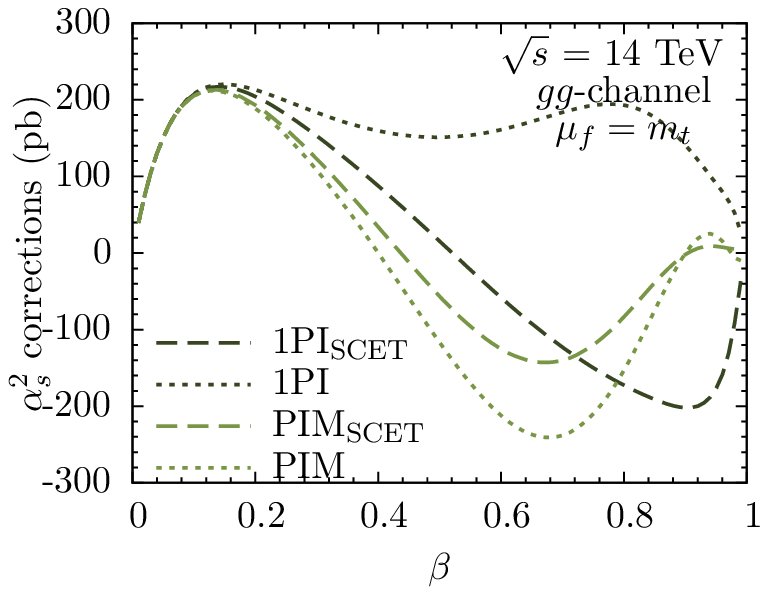} 
\end{tabular}
\end{center}
\vspace{-1ex}
\caption{\label{fig:beta-plot-nnlo} The $\alpha_s^2$ corrections to $d\sigma/d\beta$
  arising from the scale-independent piece, for $\mu_f=m_t$.}
\end{figure}

In contrast to the transverse-momentum and rapidity distributions, for which little is
known beyond NLO, for the $\beta$ distribution we can also perform some comparisons with
other results at NNLO. First of all, results for the $\beta$ distribution at the same
level of NNLO approximation, but in PIM kinematics, were obtained in \cite{Ahrens:2010zv},
and the agreement of these results with each other gives some information about the size
of power-suppressed terms. In addition, the NNLO corrections proportional to the
scale-dependent logarithmic terms in the $\beta$ spectrum (\ref{eq:dbeta}) are known
exactly \cite{Langenfeld:2009wd}. This gives us an opportunity to also compare both types
of kinematics with an exact result beyond NLO.

The NNLO corrections proportional to the scale-dependent logarithmic terms in the $\beta$
distribution (\ref{eq:dbeta}) within different approximations are shown in
Figure~\ref{fig:beta-plot-nnlomu}. The results are obtained by dropping the
scale-independent coefficient $f^{(2,0)}$ from (\ref{eq:fexp}). Since these contributions
would vanish for $\mu_f=m_t$, we have chosen in this case $\mu_f=2m_t$ instead. We have
evaluated the exact results in QCD using the formulas from \cite{Langenfeld:2009wd}, and
the $\scetpim$ and PIM results using those from \cite{Ahrens:2010zv}. To obtain the
threshold expansions for this comparison, we have differed from our normal scheme for
including the NNLO corrections described at the end of Section~\ref{sec:SCET} by
including the $\mu$-dependent pieces of the two-loop hard function, and by rewriting all
$\mu$-dependent logarithms in the form $\ln(m_t^2/\mu^2)$. Concerning the agreement between
the 1PI approximations and the exact results, one sees the same qualitative behavior as at
NLO. The $\scetopi$ results are consistently a better approximation than the 1PI results,
especially at the LHC, where the power corrections are large at higher values of $\beta$.
As for the PIM results, the $\scetpim$ approximation fares slightly better than the PIM
results in the gluon channel, but slightly worse in the quark channel. In any case, the
differences between the $\scetpim$ and PIM results are much smaller than those between the 
$\scetopi$ and 1PI results, which can be taken as an indication that the power corrections 
are smaller in PIM than in 1PI kinematics. However, once the extra corrections unique to 
the $\scetopi$ scheme are taken
into account, the results in these two types of kinematics are very much compatible with
one another and provide a good approximation to the exact results.

The NNLO corrections from the scale-independent pieces $f^{(2,0)}$ to the $\beta$ spectrum
(\ref{eq:dbeta}) within the different PIM and 1PI approximations are shown in
Figure~\ref{fig:beta-plot-nnlo}, for the choice $\mu_f=m_t$. For these pieces it is not
possible to make a comparison with an exact result, but we can make a couple of comments
based on the agreement of the different approximations with each other. As before, the
difference between the two PIM schemes is small compared to that between the two 1PI
schemes, indicating that the power corrections in PIM kinematics are smaller, and the
difference between the $\scetopi$ and $\scetpim$ results are much reduced compared to the
difference between the 1PI and PIM results. In general, the NNLO corrections in the 1PI
approximation are much larger than any of the others.

\begin{figure}
\begin{center}
\includegraphics{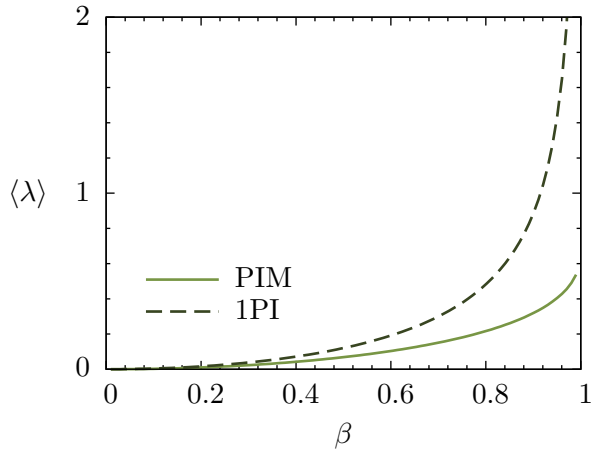}
\end{center}
\vspace{-1ex}
\caption{\label{fig:means4} Mean value of the ``small'' parameter $\lambda$, where
  $\lambda=(1-z)/(2\sqrt{z})$ in PIM and $\lambda=s_4/(2m_t\sqrt{m_t^2+s_4})$ in 1PI
  kinematics.}
\end{figure}

The explicit results from these studies all point to the fact that 1PI kinematics is more
susceptible to power-suppressed effects than PIM kinematics. We can gain more insight into
this observation through a very simple analysis. As discussed in Section \ref{sec:SCET},
the leading power corrections in 1PI kinematics are related to the partonic expansion parameter
$\lambda=E_s/m_t$, where $2E_s\sim s_4/\sqrt{m_t^2+s_4}$ is the energy of extra soft
radiation in the partonic scattering process. For the case of PIM kinematics, the equivalent
parameter is $\lambda = E^{\rm PIM}_s/M$, where $2E^{\rm PIM}_s = M(1-z)/\sqrt{z}$. We can
quantify in part the relative size of these parameters as a function of $\beta$ by
evaluating the mean value
\begin{align}
  \Braket{\lambda}_{\rm 1PI} = \left. \int_{t_1^{\text{min}}}^{t_1^{\text{max}}}
    d\hat{t}_1 \int_0^{s_4^{\text{max}}} ds_4
    \left(\frac{s_4}{2m_t\sqrt{m_t^2+s_4}}\right) \right/
  \int_{t_1^{\text{min}}}^{t_1^{\text{max}}} d\hat{t}_1 \int_0^{s_4^{\text{max}}} ds_4
\end{align}
in 1PI kinematics, where the appropriate integration range can be read off from (\ref{eq:tot2}), and
the analogous expression
\begin{align}
  \Braket{\lambda}_{\rm PIM} = \left. \int_{4m_t^2/\hat{s}}^1 dz
    \left(\frac{1-z}{2\sqrt{z}}\right) \right/ \int_{4m_t^2/\hat{s}}^1 dz
\end{align}
in PIM kinematics. The results are shown in Figure~\ref{fig:means4}. In the 1PI scheme, there is a sharp growth
in the average value of the ``small'' parameter $\lambda$ with increasing $\beta$. The
expansion parameter in the PIM scheme also increases as a function of $\beta$, but not as quickly.
Note that this behavior of the partonic expansion parameter $\lambda$ does not translate
directly into correspondingly large corrections to the threshold expansion for the
physical cross sections. The singular distributions in $s_4$ or $(1-z)$ still enhance the
region where $\lambda$ is small, even if the integration range covers regions where it is
not, and the regions of $\beta$ closer to the endpoint are damped by the parton
luminosities. However, given the results of the figure, it is not surprising that the
power corrections are generally larger in 1PI than in PIM kinematics, and that they are especially
important at the LHC, where the parton luminosities are larger at higher values of
$\beta$.

So far, we have focused on the agreement of the corrections in 1PI or PIM kinematics with
the exact QCD results or with each other. We can gain more information by looking at the
contributions of the individual terms in the decomposition (\ref{eq:C2}) in the 1PI scheme and also
at the analogous contributions in the PIM scheme. The assumption of dynamical threshold enhancement
is that contributions from regions of phase space where the partonic expansion parameter
$\lambda$ is large, as shown for example in Figure~\ref{fig:means4}, are suppressed due to the
properties of the PDFs, so one can expect to see a hierarchy between the different terms
in the expansion. In particular, one would expect that the plus-distributions contribute
more than the $\delta$-function and of course the power-suppressed contributions contained in
$R$.

The exact structure of contributions to the total cross section from the different terms
are shown in Table~\ref{tab:DistCors}, for the choice $\mu_f=m_t$. Generally speaking, the
$P_i$ distributions are indeed enhanced compared to the other terms. In fact, the NNLO
contributions from $P_3$ can be as large as the NLO contributions from $P_1$. However, in
both 1PI and PIM kinematics, there are large cancellations between the different terms, so
that the total NNLO correction turns out to be small. In PIM kinematics, this happens at the level of the
distributions, and to a lesser extent between the $\delta$-function and $R$ terms, which
seem to be generically smaller than the other terms. In 1PI kinematics, the $\delta$-function terms
and especially the power-suppressed terms in $R$ are relatively larger than in PIM. We
emphasize that the results for the coefficients of the $P_i$ distributions are exact, so a full NNLO calculation will change
only the $\delta$-function and $R$ pieces, moreover in such a way that the cross section
from both types of kinematics agrees exactly. The numbers above suggest that these terms
are relatively small in PIM kinematics, because of threshold enhancement, so in order to
preserve the good agreement between the two types of kinematics they would also need to be
small in 1PI kinematics. In that case the terms already included in our calculation are
the dominant ones at NNLO, although this can only be confirmed through the full NNLO
results.

\begin{table}[t!]
  \begin{center}
    \begin{tabular}{|l|c|l|c|c|c|c|c|c|c|c|}
      \hline
      &  &  & $P_3$ &  $P_2$  & $P_1$ & $P_0$ & $\delta$ & $R$ & sum
      \\
      \hline  
      \multirow{8}{*}{\rotatebox{90}{Tevatron}} &
      \multirow{4}{*}{\rotatebox{90}{$\scetopi$}} &  $\delta \sigma^{(1)}_{q\bar q}$
      & & &  1.1 & 0.30 & $-0.06$ & $-0.38$ & 0.94  \\  
      & &  $\delta \sigma^{(2)}_{q\bar q}$
      & 0.57 & 0.39 & $-0.09$ & $-0.19$ & $-0.37$ & $-0.31$ & $0.01$  \\ 
      \cline{3-10}
      & &  $\delta \sigma^{(1)}_{gg}$ 
      & &  &  0.32 & 0.08 & $-0.11$ & $-0.06$ & 0.45   \\  
      & &  $\delta \sigma^{(2)}_{gg}$
      & 0.39 & 0.18 & $-0.07$ & $-0.14$ & $-0.12$ & $-0.12$ & 0.14   \\   
      \cline{2-10}
      & \multirow{4}{*}{\rotatebox{90}{$\scetpim$}} &  $\delta \sigma^{(1)}_{q\bar q}$
      & &  &  1.2 & $-0.67$ & $-0.11$ & 0.20 & 0.64   \\    
      & &  $\delta \sigma^{(2)}_{q\bar q}$  
      & 0.64 & $-0.24$ & $-0.26$ & $-0.16$ & $-0.08$ & 0.11 & 0.02   \\  
      \cline{3-10}
      & &  $\delta \sigma^{(1)}_{gg}$  
      & &  &  0.48 & $-0.24$ & 0.12 & 0.03 & 0.38   \\ 
      & &  $\delta \sigma^{(2)}_{gg}$ 
      & 0.60 & $-0.31$ & $-0.11$ & $-0.11$ & $-0.08$ & 0.04 & 0.10   \\ 
      \hline
      \multirow{8}{*}{\rotatebox{90}{LHC14}}  &
      \multirow{4}{*}{\rotatebox{90}{$\scetopi$}} &  $\delta \sigma^{(1)}_{q\bar q}$
      &  &  &  18 & 1 & $-2$ & $-9$ & 7  \\  
      & &  $\delta \sigma^{(2)}_{q\bar q}$
      & 8.1 & 2.9 & $-1.6$ & $-0.6$ & $-5.4$ & $-4.6$ & $-1.1$  \\  
      \cline{3-10}
      & &  $\delta \sigma^{(1)}_{gg}$
      &  &  & 280 & 14 & 124 & $-125$ & 292   \\  
      & &  $\delta \sigma^{(2)}_{gg}$
      & 296 & 83 & $-20$ & $-35$ & $-160$ & $-149$ & 16   \\   
      \cline{2-10}
      & \multirow{4}{*}{\rotatebox{90}{$\scetpim$}} &  $\delta \sigma^{(1)}_{q\bar q}$
      &  &  &  11 & $-9$ & $-1$ & 5 & 6   \\  
      & &  $\delta \sigma^{(2)}_{q\bar q}$ 
      & 5.5 & $-2.8$ & $-1.7$ & $-1.0$ & $-1.3$ & 1.1 & $-0.1$   \\  
      \cline{3-10}
      & &  $\delta \sigma^{(1)}_{gg}$  
      &  &  &  250 & $-189$ & 120 & 60 & 240   \\
      & &  $\delta \sigma^{(2)}_{gg}$
      & 287 & $-194$ & $-41$ & $-60$ & $-10$ & 37 & 19   \\ 
      \hline
    \end{tabular}
  \end{center}
  \vspace{-2mm}
  \caption{\label{tab:DistCors} 
Corrections in pb from the different types of distributions at 
NLO and NNLO, for $\mu_f=m_t$. 
}
\end{table}

\section{Phenomenology}
\label{sec:pheno}

\begin{figure}
\begin{center}
\begin{tabular}{ll}
\includegraphics{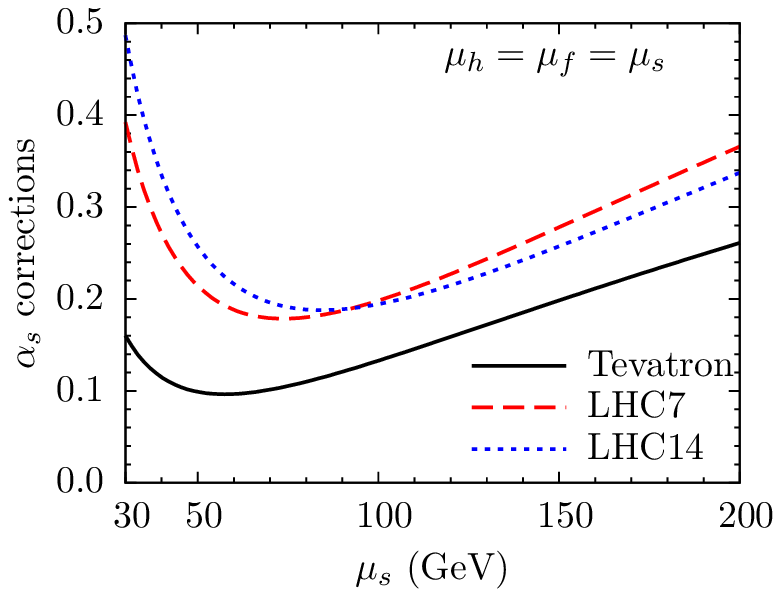}
& 
\includegraphics{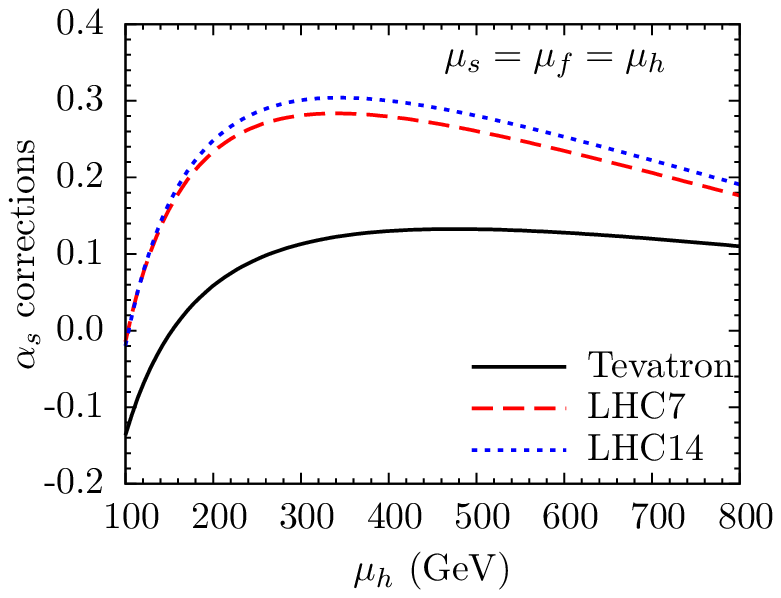}
\end{tabular}
\end{center}
\vspace{-1ex}
\caption{\label{fig:SoftScales} The one-loop correction from the soft function (left) and
  the hard function (right) to the total cross section, divided by the leading-order
  result. The solid black line is for the Tevatron, the dashed red line for LHC7, and the
  dotted blue line for LHC14.}
\end{figure}

\begin{figure}
\begin{center}
\begin{tabular}{ll}
\includegraphics{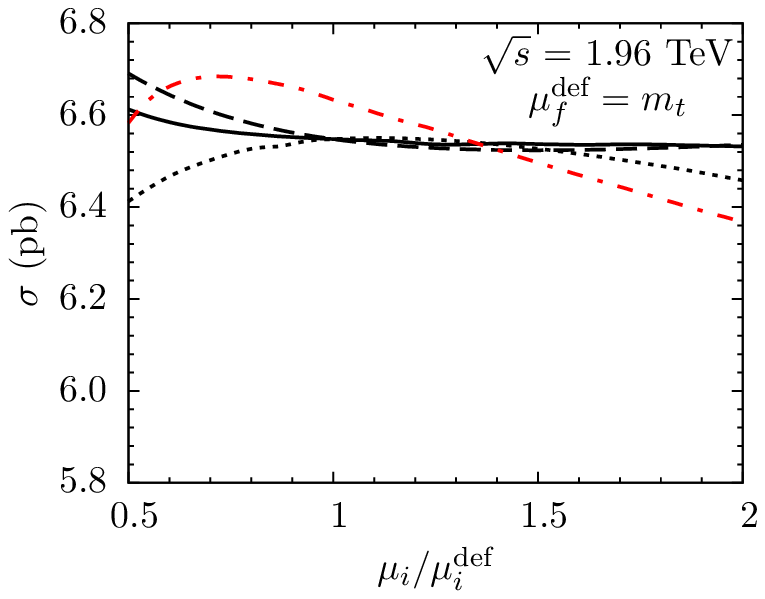}
& 
\includegraphics{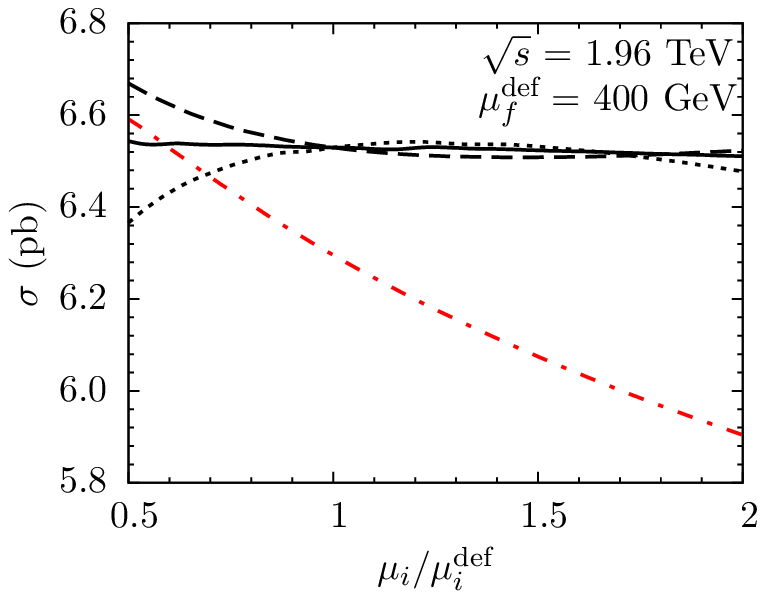}
\\
\includegraphics{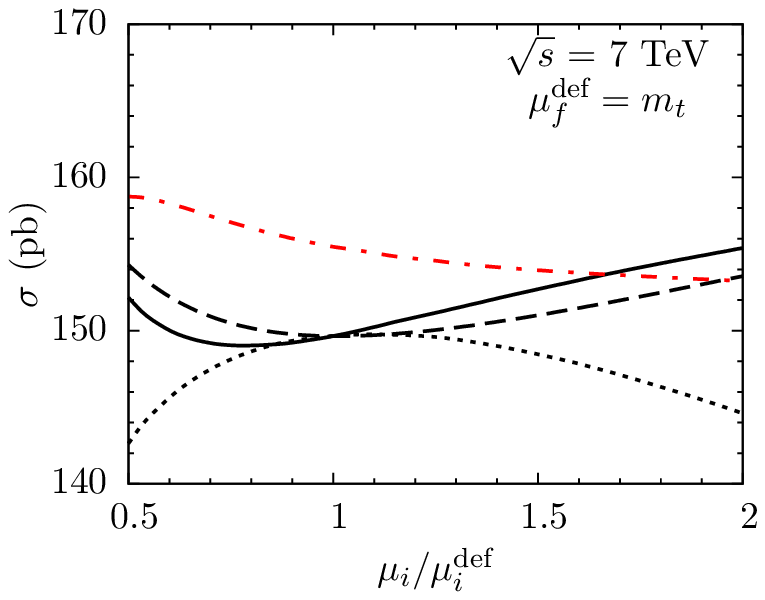}
& 
\includegraphics{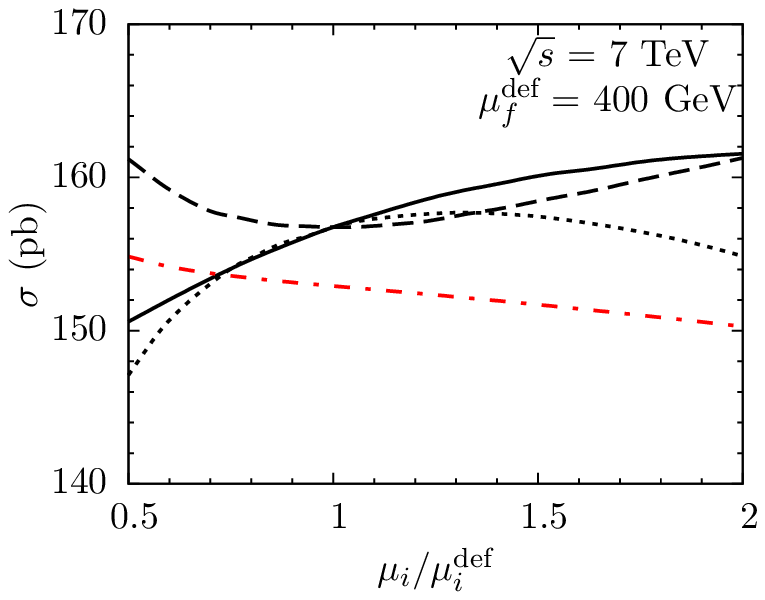}
\\
\includegraphics{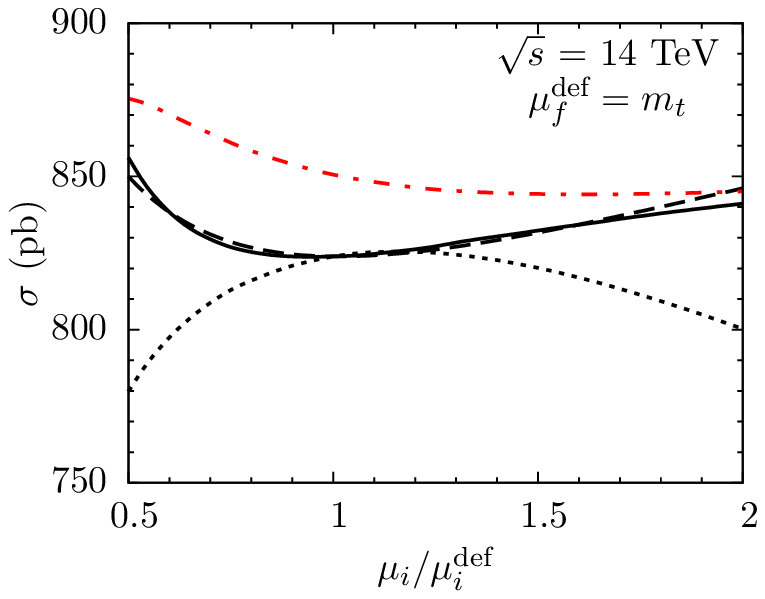}
& 
\includegraphics{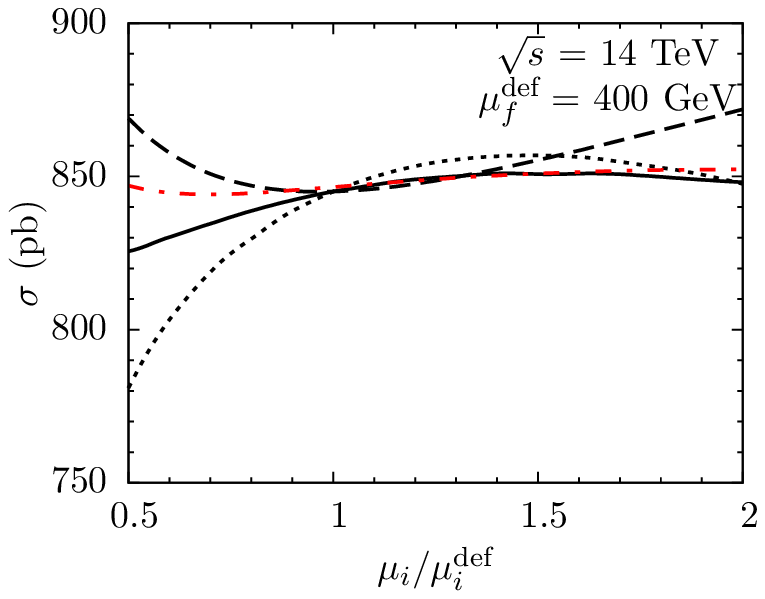}
\end{tabular}
\end{center}
\vspace{-1ex}
\caption{\label{fig:cs_with_muvar} Dependence of the cross section at NLO+NNLL and
  approximate NNLO in $\scetopi$ on the scales $\mu_f,\mu_s$, and $\mu_h$. The dot-dashed
  red lines show the dependence of the cross section on $\mu_f$ in approximate NNLO. The
  dependence of the cross section on the scales at NLO+NNLL order is represented by the
  solid black lines ($\mu_f$), the dashed black lines ($\mu_s$), and the dotted black
  lines ($\mu_h$).}
\end{figure}

We now have all the ingredients in place to present detailed phenomenological results for the $p_T$ and rapidity
distributions, the forward-backward asymmetry at the Tevatron, and the total cross
section.
Before doing so, we review some of the necessary inputs. We have already described our
scheme for evaluating the approximate NNLO corrections in Section~\ref{sec:SCET}, and that
for evaluating the formulas at threshold in Section~\ref{sec:numerical}, where our
treatment of PDFs and the top-quark mass was also summarized. In addition, we must specify
the choice of the matching scales $\mu_s$ and $\mu_h$, and of the factorization scale $\mu_f$.

We begin by discussing our method for determining an appropriate choice of the soft scale
$\mu_s$. From theoretical arguments, we expect that the perturbative expansion of the soft
function should be well-behaved at a scale characteristic of the energy of the real soft
radiation, which is generally smaller than the hard scales $m_t$ and $\sqrt{\hat{s}}$. As
explained in Section~\ref{sec:comparison}, a general analysis would determine the soft
scale by requiring that the corrections from the soft function to the double-differential
spectrum are well behaved, after integration over the partonic variables. We have
performed such an analysis for the $p_T$ distribution and found reasonable results at the
Tevatron. This is also the case for the LHC at lower values of $p_T$, where the
differential cross section is large. As long as we study a relatively modest range in
$p_T$, an equally valid procedure for determining the soft scale is to study the
corrections to the total cross section as a function of $\mu_s$. This automatically
samples the regions of phase-space where the double-differential cross section is largest.
We show the results of such an analysis in Figure~\ref{fig:SoftScales}. To isolate the
$\alpha_s$ correction from the soft function shown there, we pick out the piece of the
NNLL approximation to the hard scattering kernels arising from $\widetilde{\bm{s}}^{(1)}$,
evaluate the total cross section using only this piece, and divide the result by that at
NLL, working in the $\scetopi$ scheme. We furthermore make the scale choice
$\mu_f=\mu_h=\mu_s$, which amounts to looking at the correction at NLO in fixed-order. As
seen from the figure, a well-defined minimum in the soft correction appears for $\mu_s\sim
60$~GeV at the Tevatron, $\mu_s\sim 80$~GeV at the LHC with $\sqrt{s}=7$~GeV, and
$\mu_s\sim 90$~GeV at the LHC with $\sqrt{s}=14$~GeV. We will use these as the default
choices of $\mu_s$ in the rest of this section, both for the total cross section and for
differential distributions.

One can apply this same procedure to determine an appropriate choice of the hard scale.
Since the hard function is the same in PIM and 1PI kinematics, we first recall the
analysis of \cite{Ahrens:2010zv}, where it was argued that $\mu_h = M$ is a reasonable
default value. Translated to 1PI kinematics, this would imply the choice
$\mu_h=\sqrt{\hat{s}}$. The actual result for the $\alpha_s$ correction to the total cross
section arising from the hard function as a function of $\mu_h$ is shown in the right panel of Figure~\ref{fig:SoftScales}. We isolate this correction as we did for the soft
function, except for this time we pick out the piece of the NNLL cross section
proportional to $\bm{H}^{(1)}$, and examine the result as a function of $\mu_h$. As for
the analysis with the soft function, we make the choice $\mu_f=\mu_h=\mu_s$. At lower
values of $\mu_h$ the correction becomes negative and depends strongly on the scale. To
avoid sensitivity to that region, we will choose $\mu_h=400$~GeV by default, which is
close to the average value of $\sqrt{\hat{s}}$ for the total cross section, and in any
case will be varied by a factor of two in the error analysis.

For the factorization scale, we will consider the two different choices $\mu_f=m_t$ and
$\mu_f=400$~GeV used in \cite{Ahrens:2010zv}. 
Of course we could also just choose a
single, intermediate value of $\mu_f$ and vary it in a larger range, but in resummed
perturbation theory it is useful to have independent variations of the matching scales
$\mu_h$ and $\mu_s$ for the two different values of $\mu_f$. For the differential
distributions in the following subsection, on the other hand, we use $\mu_f=2m_t$ as the
central value. A more refined analysis could use other choices, such as $\mu_f =
m_\perp$ for the $p_T$ distribution, but since we do not study tails of the distributions
we prefer to stick to a single value which is roughly intermediate between the two values
used for the total cross section.

We should mention that another method often used to argue for a particular scale choice is
to look for areas where the scale dependence of the observable is flat. As part of our
analysis below, we show in Figure~\ref{fig:cs_with_muvar} the dependence of the total
cross section on the scales $\mu_s$, $\mu_h$, and $\mu_f$, at NLO+NNLL and approximate
NNLO. We note that the scale-dependence of the cross section at NLO+NNLL order is
indeed flat close to our default values of $\mu_s$ and $\mu_h$, and also close to
$\mu_f=m_t$. The approximate NNLO results, on the other hand, do not seem to favor a
particular choice of $\mu_f$ based on this criteria.

\subsection{Rapidity and transverse-momentum distributions}

\begin{figure}
\begin{center}
\begin{tabular}{lr}
\psfrag{y}[][][1][90]{$d\sigma/dy$ [fb]}
\psfrag{x}[]{$y$}
\psfrag{z}[][][0.85]{$\sqrt{s}=1.96$\,TeV}
\includegraphics[width=0.43\textwidth]{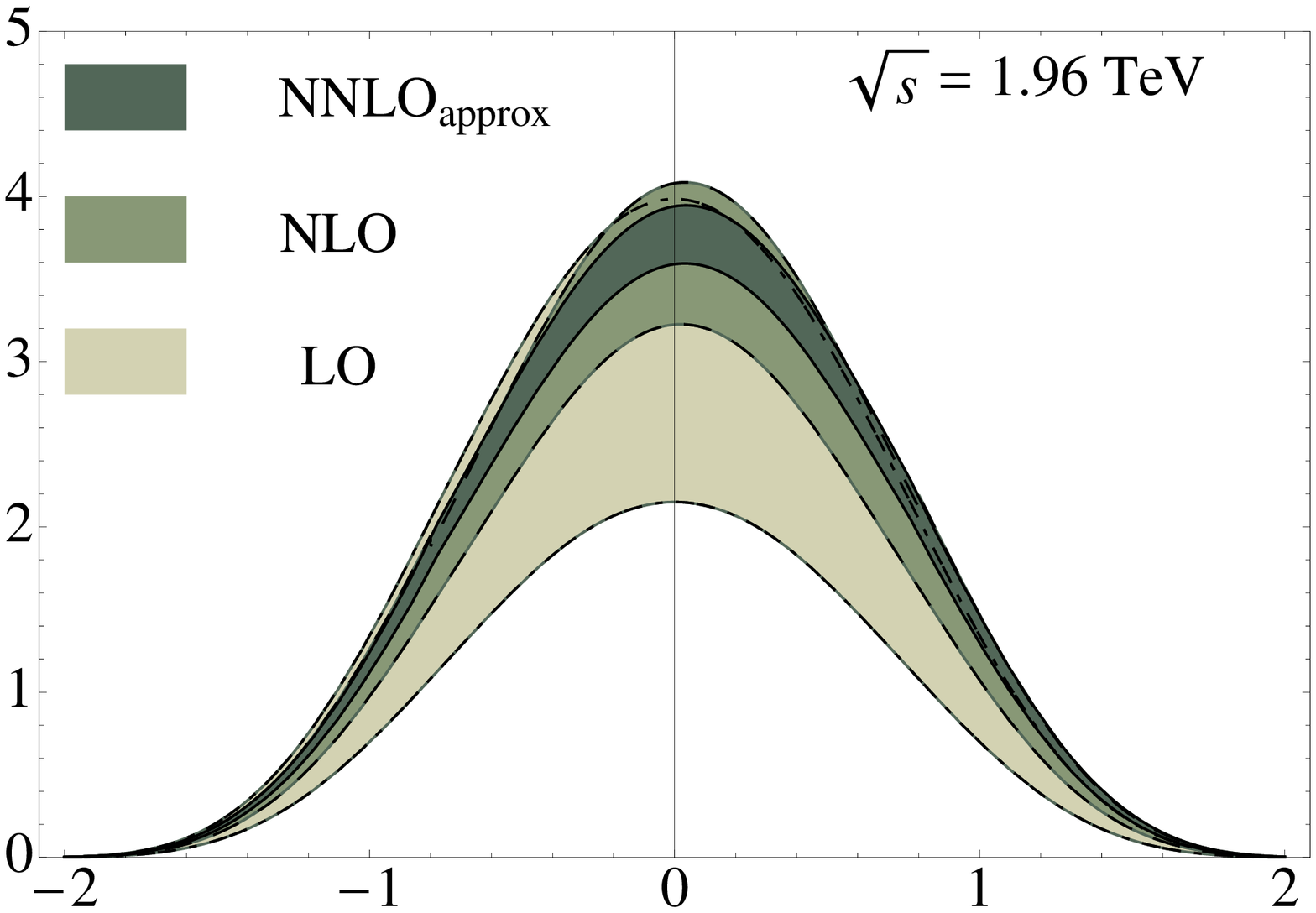}
&
\psfrag{y}[][][1][90]{$d\sigma/dy$ [fb]}
\psfrag{x}[]{$y$ }
\psfrag{z}[][][0.85]{$\sqrt{s}=1.96$\,TeV}
\includegraphics[width=0.43\textwidth]{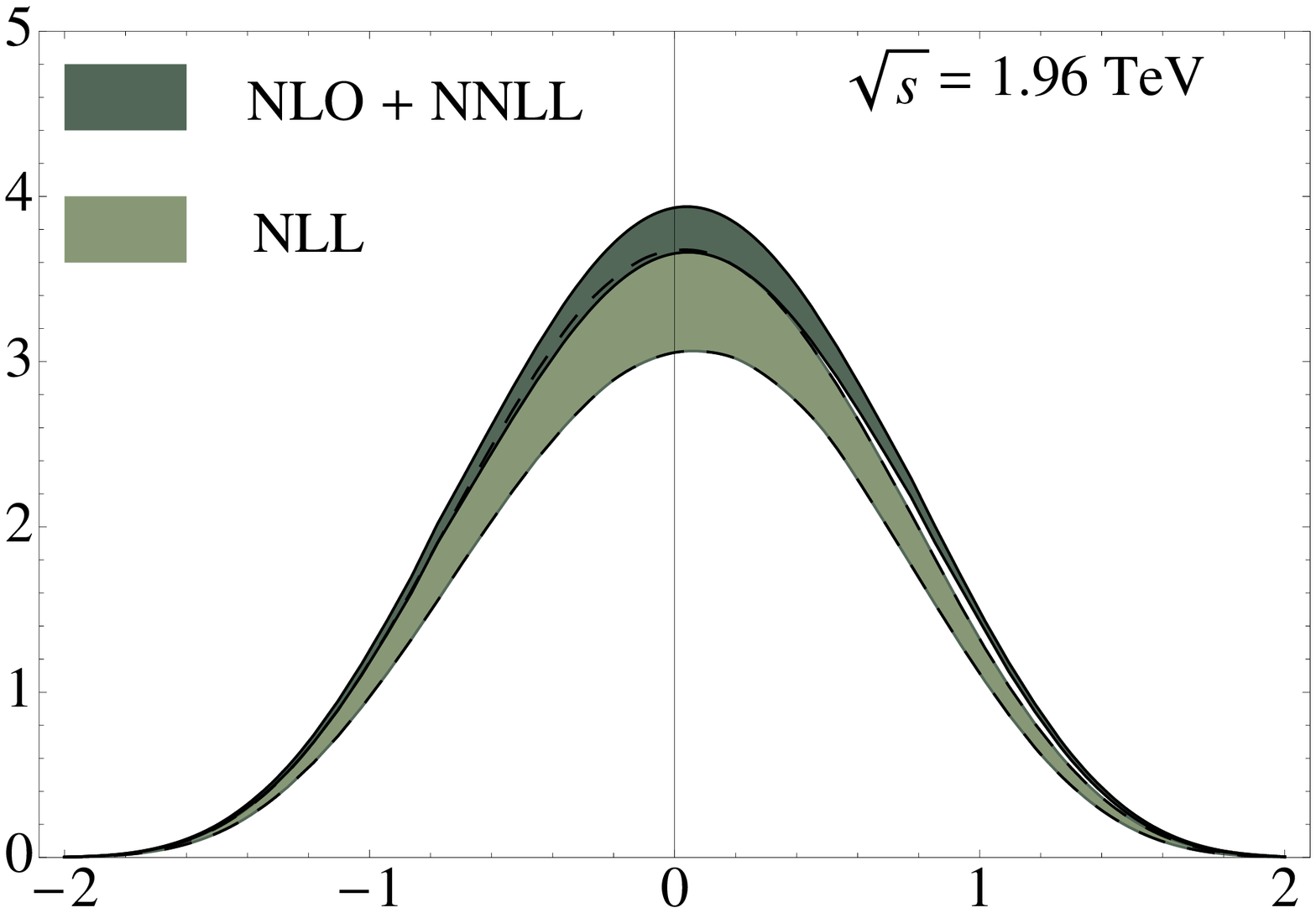}
\\[3mm]
\psfrag{y}[][][1][90]{$d\sigma/dy$ [pb]}
\psfrag{x}[]{$y$ }
\psfrag{z}[][][0.85]{$\sqrt{s}=7$\,TeV}
\includegraphics[width=0.43\textwidth]{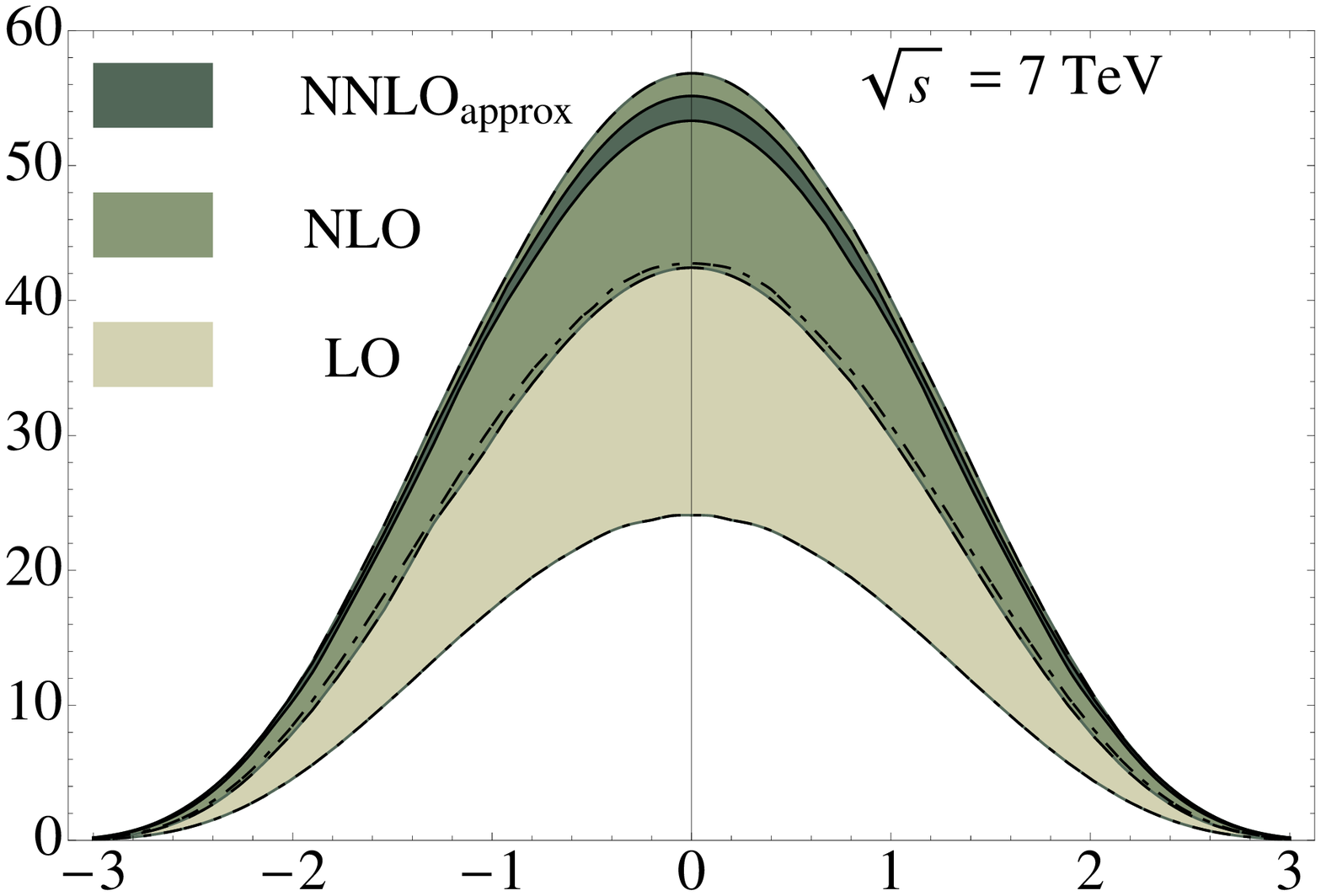}
&
\psfrag{y}[][][1][90]{$d\sigma/dy$ [pb]}
\psfrag{x}[]{$y$ }
\psfrag{z}[][][0.85]{$\sqrt{s}=7$\,TeV}
\includegraphics[width=0.43\textwidth]{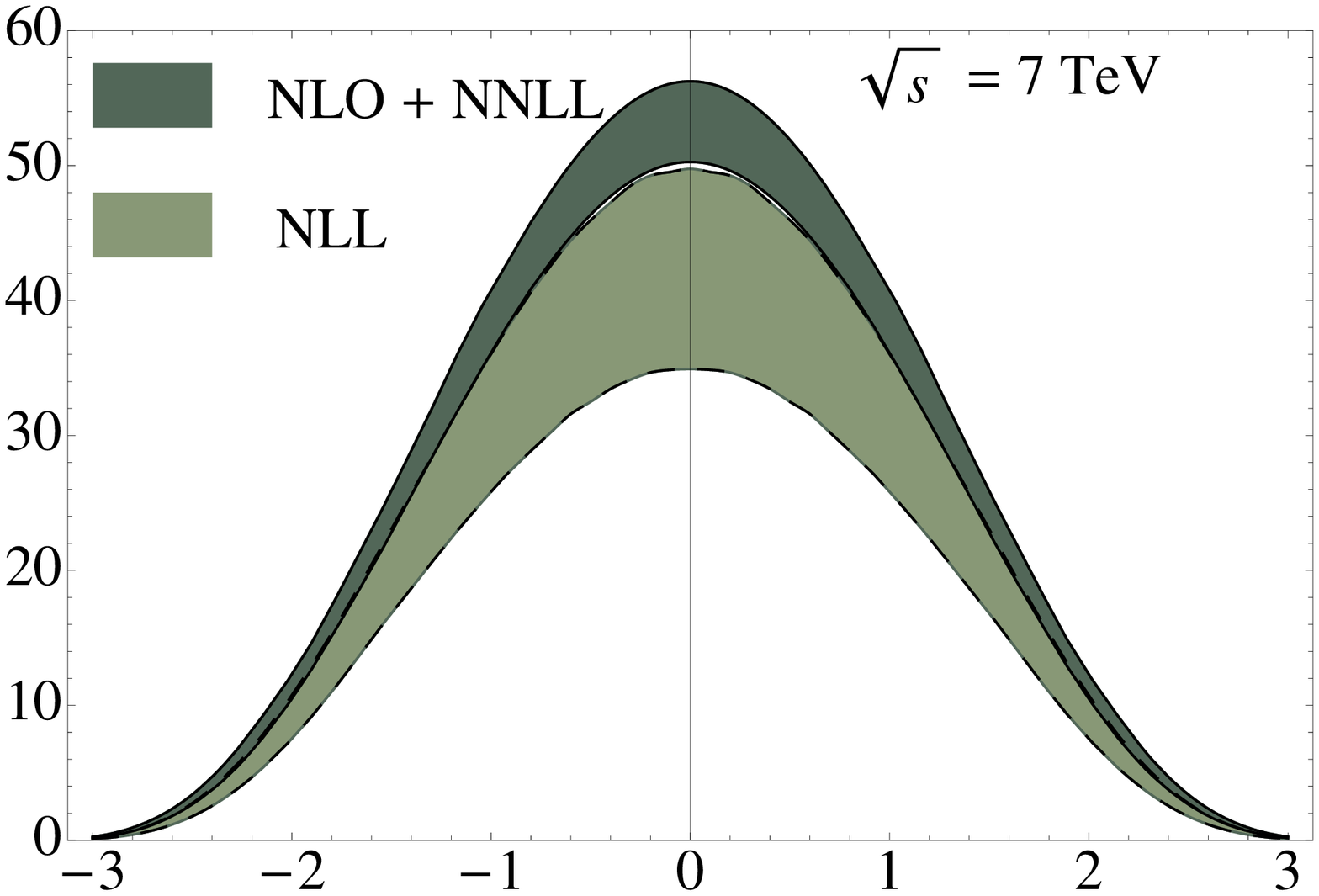}
\\[3mm]
\psfrag{y}[][][1][90]{$d\sigma/dy$ [pb]}
\psfrag{x}[]{$y$ }
\psfrag{z}[][][0.85]{$\sqrt{s}=7$\,TeV}
\includegraphics[width=0.43\textwidth]{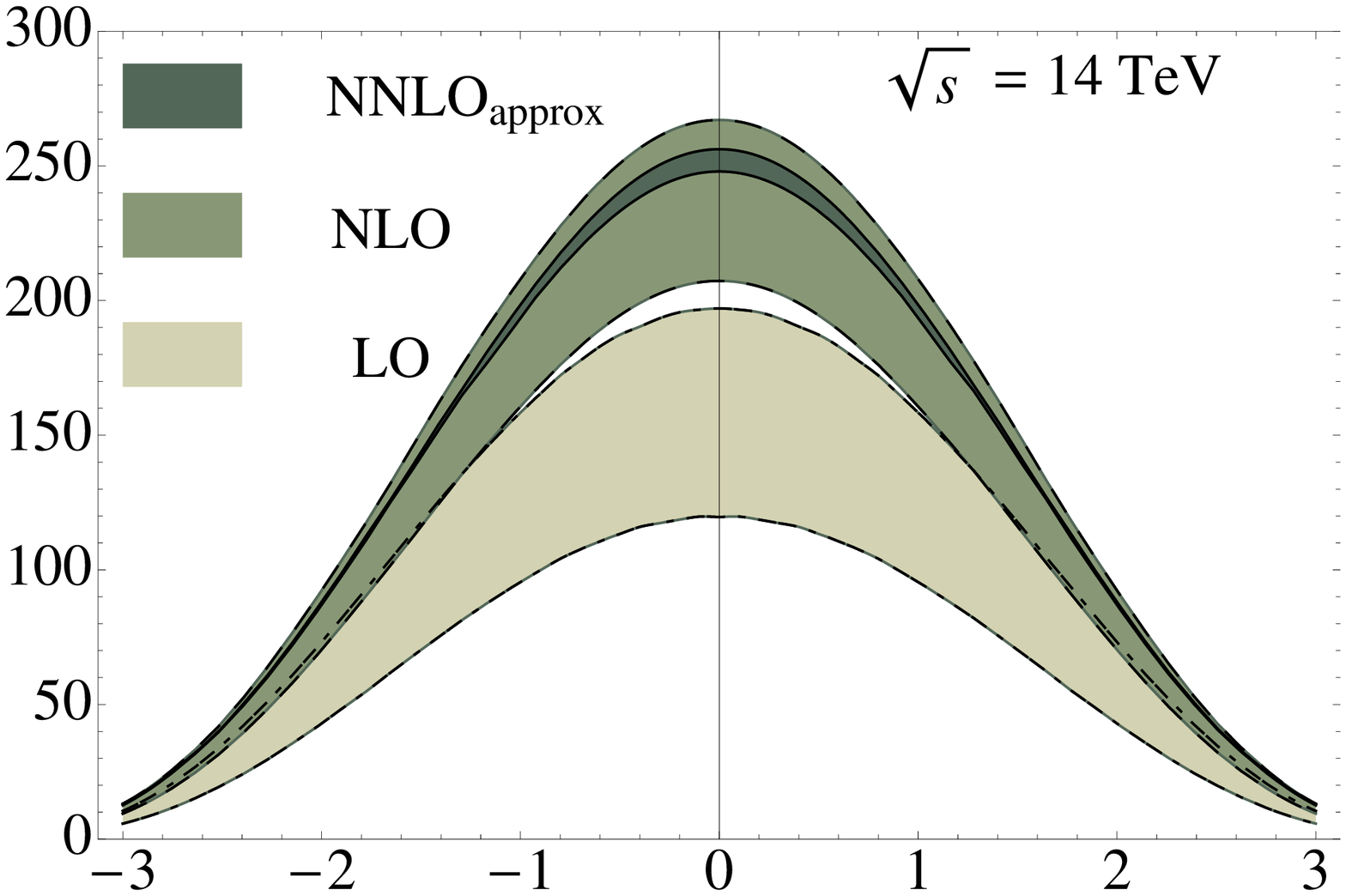}
&
\psfrag{y}[][][1][90]{$d\sigma/dy$ [pb]}
\psfrag{x}[]{$y$ }
\psfrag{z}[][][0.85]{$\sqrt{s}=7$\,TeV}
\includegraphics[width=0.43\textwidth]{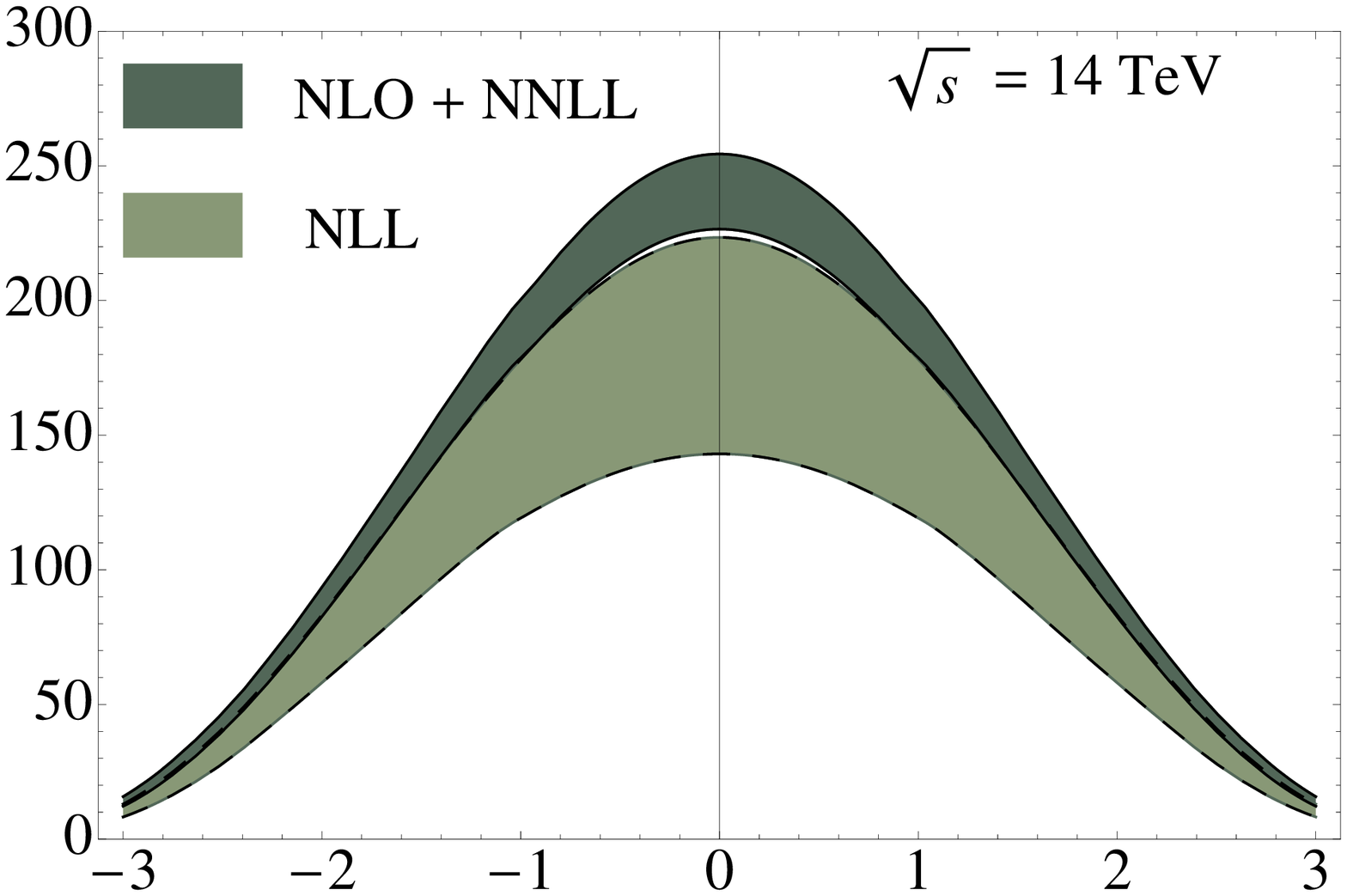}
\end{tabular}
\end{center}
\vspace{-2mm}
\caption{\label{fig:ydistributions} Left: Fixed-order predictions for the rapidity
  distribution at LO (light), NLO (darker), and approximate NNLO (dark bands) for the
  Tevatron (top) and LHC (bottom). Right: Corresponding predictions at NLL (light) and
  NLO+NNLL (darker bands) in resummed perturbation theory. The width of the bands reflects
  the uncertainty of the distributions under variations of the matching and factorization
  scales, as explained in the text.}
\end{figure}

\begin{figure}
\begin{center}
\begin{tabular}{lr}
\psfrag{y}[][][1][90]{$d\sigma/dp_T$ [fb/GeV]}
\psfrag{x}[]{$p_T$ [GeV]}
\psfrag{z}[][][0.85]{$\sqrt{s}=1.96$\,TeV}
\includegraphics[width=0.43\textwidth]{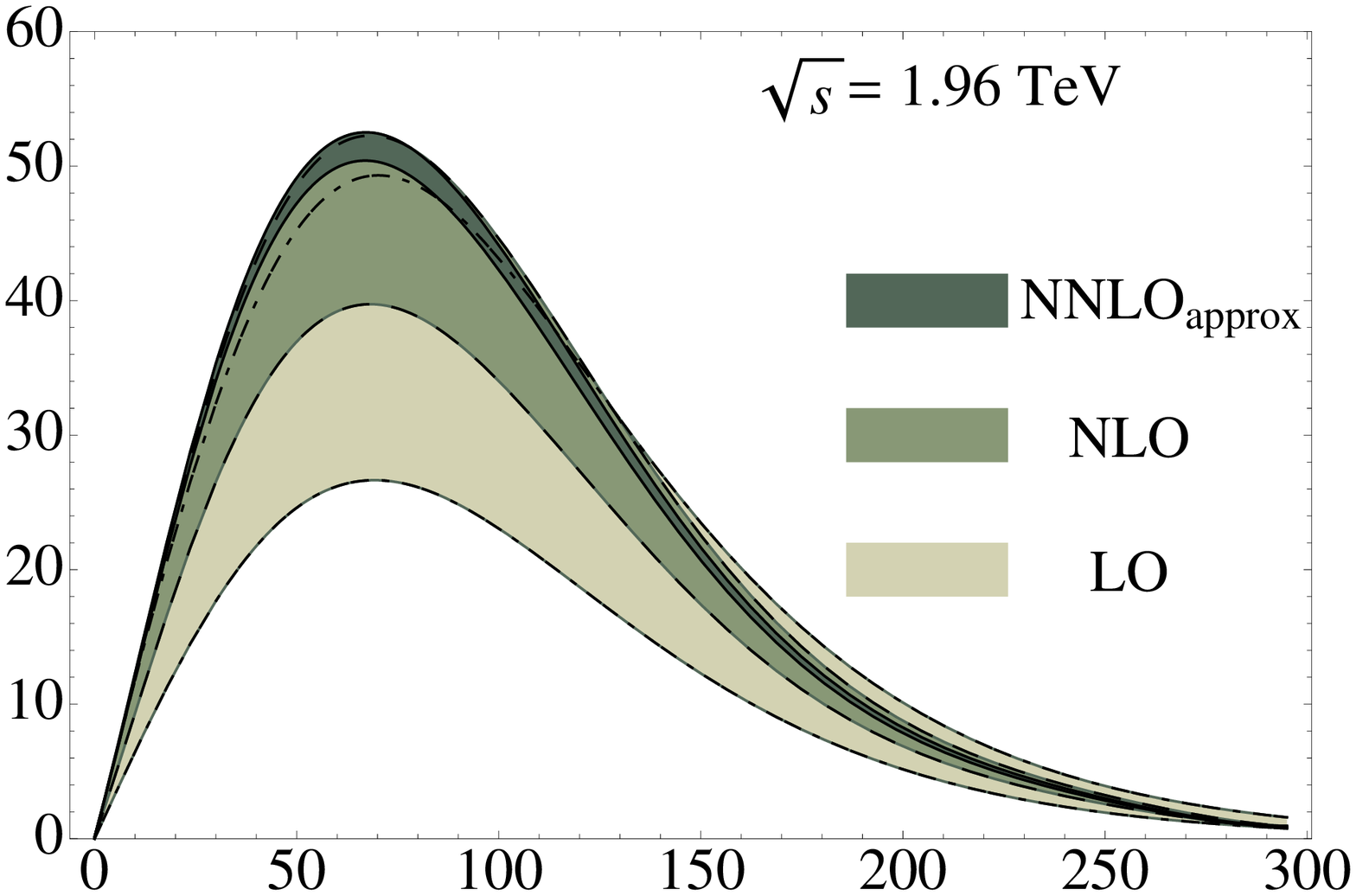}
&
\psfrag{y}[][][1][90]{$d\sigma/dp_T$ [fb/GeV]}
\psfrag{x}[]{$p_T$ [GeV]}
\psfrag{z}[][][0.85]{$\sqrt{s}=1.96$\,TeV}
\includegraphics[width=0.43\textwidth]{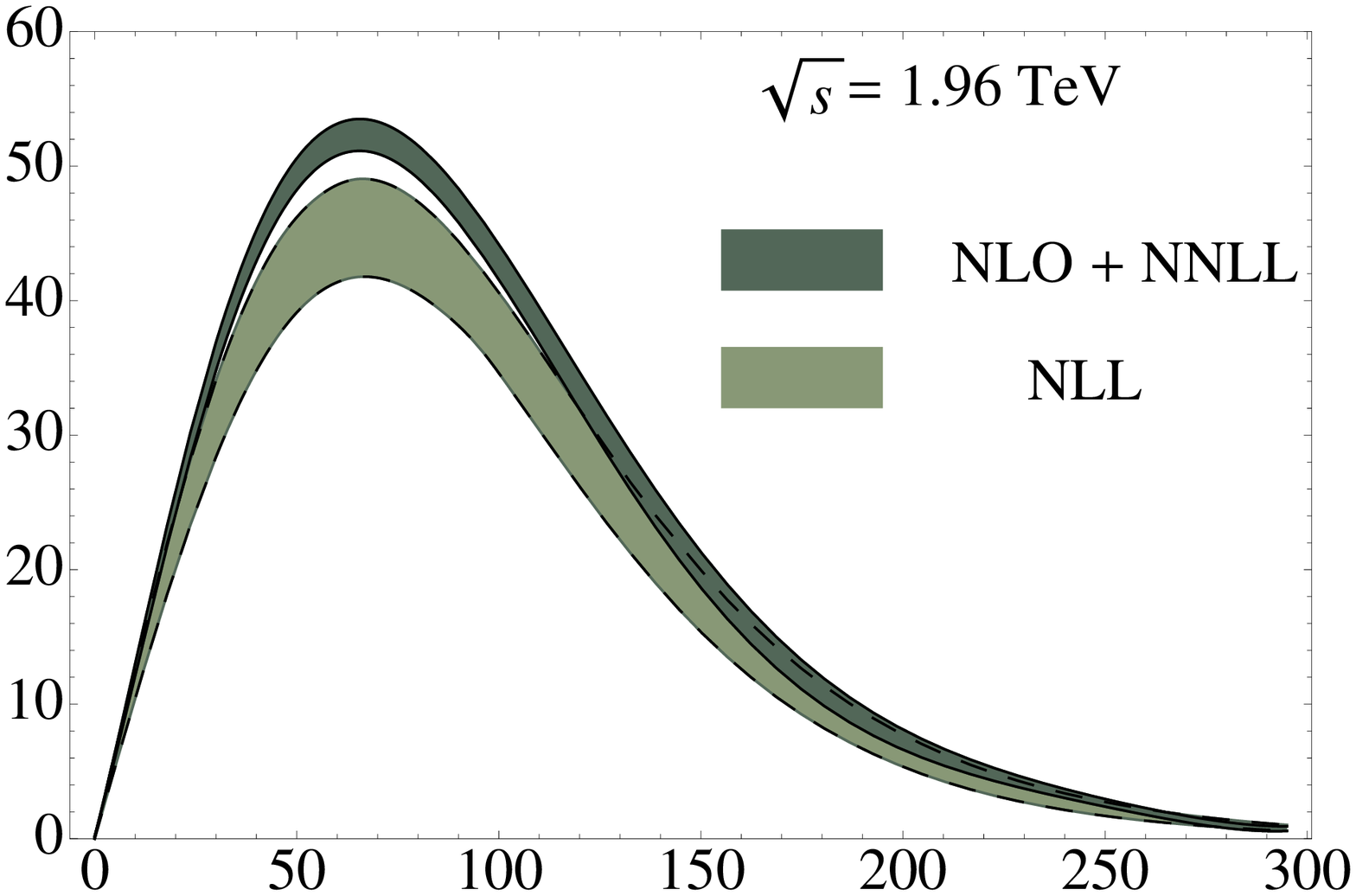}
\end{tabular}
\end{center}
\vspace{-2mm}
\caption{\label{fig:ptdistributions} Left: Fixed-order predictions for the $p_T$
  distribution at LO (light), NLO (darker), and approximate NNLO (dark bands) at the
  Tevatron. Right: Corresponding predictions at NLL (light) and NLO+NNLL (darker bands) in
  resummed perturbation theory. The width of the bands reflects the uncertainty of the
  distributions under variations of the matching and factorization scales, as explained in
  the text.}
\end{figure}

\begin{figure}
\begin{center}
\psfrag{y}[][][1][90]{$d\sigma/dp_T$ [fb/GeV]}
\psfrag{x}[]{$p_T$ [GeV]}
\psfrag{z}[][][0.85]{$\sqrt{s}=1.96$\,TeV}
\includegraphics[width=0.9\textwidth]{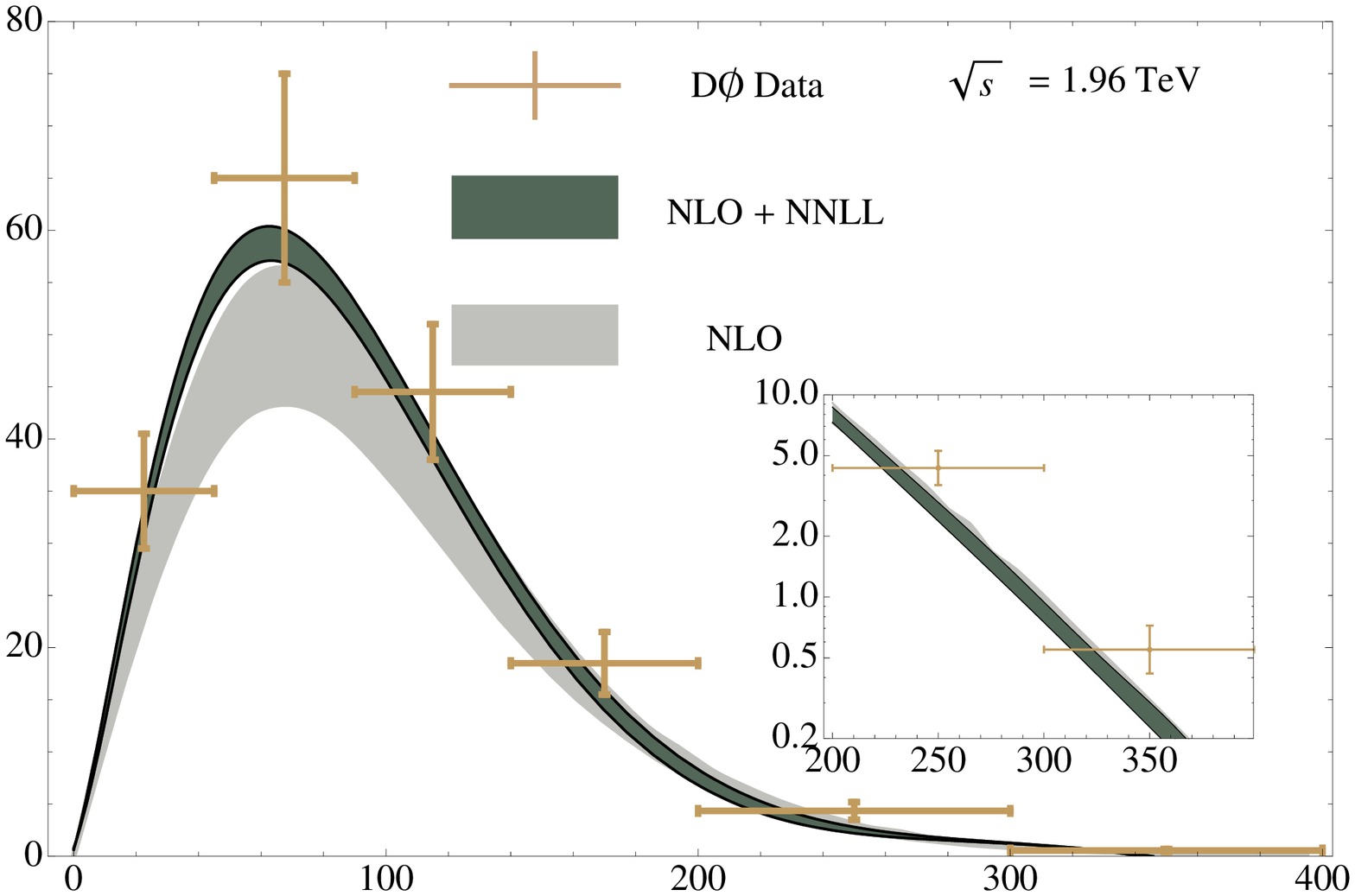}
\end{center}
\vspace{-2mm}
\caption{\label{fig:D0pt} Comparison between our NLO+NNLL predictions, NLO results and
  recent measurements from the D0 collaboration \cite{Abazov:2010js}. The error bands
  refers to perturbative uncertainties related to scale variations. Furthermore we have
  enlarged the region of bigger $p_T$ for better comparison.
}
\end{figure}

We now present results for the top-quark rapidity and transverse-momentum
distributions. We begin by studying rapidity distributions in
Figure~\ref{fig:ydistributions}, where we compare results from fixed-order and resummed
calculations in the $\scetopi$ scheme at the Tevatron and the LHC with $\sqrt{s}=7$ and
14~TeV. The results within a given perturbative approximation are represented as bands
indicating the theoretical uncertainties from scale variations. To make the bands in
fixed-order perturbation theory at a given point in $y$, we vary the value of the
factorization scale up and down from its default value at $\mu_f=2m_t$ by a factor of
two, and pick out the highest and lowest numbers at that point. To make the bands in
resummed perturbation theory, we vary individually $\mu_f$, $\mu_h$, or $\mu_s$ with the
others held fixed at their default values, pick out the highest and lowest numbers, and
then add the three highest and three lowest numbers in quadrature. The results in the
figure clearly show that the higher-order corrections contained in the approximate NNLO
and NLO+NNLL formulas tend to reduce the uncertainties due to scale dependence, and
slightly raise the central values at a given rapidity to the upper part of the fixed-order
NLO band. The error bands for the NLO+NNLL and approximate NNLO results are of similar
size at the Tevatron, but at the LHC the approximate NNLO bands are noticeably smaller, a
result which we will quantify in more detail when we study the total cross section.

Next, we consider the top-quark transverse-momentum distribution. In this case, we focus
our analysis on the Tevatron, where experimental measurements are available. In
Figure~\ref{fig:ptdistributions} we show our predictions for this distribution within the
different perturbative approximations, comparing the fixed-order and resummed results. As
with the rapidity distributions, the higher-order perturbative corrections serve to
decrease the scale dependence, and also to slightly raise the central values of the results at
a given $p_T$. We compare the NLO+NNLL results for the $p_T$ distribution with a recent
measurement at the Tevatron performed by the D0 collaboration using the lepton+jets channel
\cite{Abazov:2010js} in Figure~\ref{fig:D0pt}, showing also the NLO calculation for
illustration. Since the D0 analysis uses $m_t = 170$~GeV, for the purposes of this study
we deviate from our default choice and also adopt this value. We observe that the slight
increase in the the differential cross section due to resummation leads to a 
better agreement with the data compared to the NLO predictions. In general, the measured
spectrum and the NLO+NNLL theory prediction agree within the errors, both in the shape and
the normalization. This is true even at higher values of $p_T$, although one should keep
in mind that our scale-setting procedure was designed to work in areas where the
differential cross section is large, and at $p_T\sim 400$~GeV it would also be interesting
to perform a more involved analysis where $\mu_f,\mu_s$, and $\mu_h$ are chosen as
dynamical functions of $p_T$.

\subsection{The forward-backward asymmetry at the Tevatron}

The top-quark pair charge asymmetry is an important observable at the Tevatron. It
originates from the difference in the production rates for top and anti-top quarks at
fixed scattering angle or rapidity \cite{Kuhn:1998jr, Kuhn:1998kw}. Due to charge-conjugation invariance in QCD, it can also be interpreted as a forward-backward asymmetry
using (\ref{eq:cpi}). This asymmetry vanishes at LO in $\alpha_s$ and is only about 5\%--10\%
at NLO, so it is potentially sensitive to new physics contributions and has received much
interest for that reason.

The forward-backward asymmetry depends on the frame in which it is measured.
Experimentally the asymmetry has been measured both in the laboratory
frame and in the $t\bar{t}$ rest frame \cite{:2007qb, Aaltonen:2008hc}. In
\cite{Ahrens:2010zv}, we used results from PIM kinematics to calculate the asymmetry to
NLO+NNLL and approximate fixed order in the partonic center-of-mass frame, with the aim of
improving the NLO fixed-order calculations from \cite{Kuhn:1998kw} and the NLL results
from \cite{Almeida:2008ug}. In what follows we use our results for 1PI kinematics to
obtain NLO+NNLL and approximate NNLO predictions directly in the laboratory frame.

\begin{table}[t]
\begin{center}
\begin{tabular}{|l|c|c|c|c|}
  \hline
  & \multicolumn{2}{c|}{$0.2<\mu_f/{\rm TeV}<0.8$}
  & \multicolumn{2}{c|}{$m_t/2<\mu_f<2m_t$} 
  \\
  \cline{2-5} & $\Delta\sigma_{\rm FB}$ [pb] &  $A^t_{\text{FB}}$ [\%]
  & $\Delta\sigma_{\rm FB}$ [pb] & $A^t_{\text{FB}}$ [\%]
  \\
  \hline
  NLL, $\scetopi$ 
  & 0.234{\footnotesize $^{+0.099}_{-0.116}$} 
  & 4.12{\footnotesize $^{+2.11}_{-2.06}$} 
  & 0.238{\footnotesize $^{+0.105}_{-0.119}$} 
  & 4.24{\footnotesize $^{+2.17}_{-2.12}$}
  \\
  NLO leading, $\scetopi$
  & 0.164{\footnotesize $^{+0.077}_{-0.049}$} 
  & 4.40{\footnotesize $^{+0.38}_{-0.33}$}
  & 0.262{\footnotesize $^{+0.141}_{-0.085}$} 
  & 4.86{\footnotesize $^{+0.46}_{-0.39}$}
  \\
  NLO 
  & 0.163{\footnotesize $^{+0.076}_{-0.048}$} 
  & 4.36{\footnotesize $^{+0.36}_{-0.32}$} 
  & 0.260{\footnotesize $^{+0.140}_{-0.084}$} 
  & 4.81{\footnotesize $^{+0.45}_{-0.39}$}
  \\ \hline 
  NLO+NNLL, $\scetopi$
  & 0.295{\footnotesize $^{+0.026}_{-0.032}$}
  & 4.61{\footnotesize $^{+0.27}_{-0.26}$}
  & 0.312{\footnotesize $^{+0.027}_{-0.035}$}
  & 4.88{\footnotesize $^{+0.20}_{-0.23}$}
  \\
  NNLO approx, $\scetopi$
  & 0.241{\footnotesize $^{+0.023}_{-0.030}$} 
  & 4.27{\footnotesize $^{+0.31}_{-0.26}$}
  & 0.270{\footnotesize $^{+0.037}_{-0.023}$} 
  & 4.01{\footnotesize $^{+0.54}_{-0.00}$}
  \\
  \hline    
\end{tabular}
\end{center}
\vspace{-2mm}
\caption{\label{tab:FB} 
  The asymmetric cross section and forward-backward asymmetry at the Tevatron,
  evaluated at different orders in perturbation theory in the $p\bar p$ 
  center-of-mass frame. The errors refer to perturbative uncertainties 
  related to scale variations, as explained in the text.}
\end{table}

We define the forward-backward asymmetry as the ratio
\begin{align}
  A^t_{\text{FB}} = \frac{\Delta\sigma_{\text{FB}}}{\sigma} \,,
\end{align}
where the asymmetric cross section for the difference between the production of top quarks in
the forward and backward directions is
\begin{align}
  \label{eq:fbasym}
  \Delta\sigma_{\text{FB}} \equiv \int_{0}^{y_+} dy \int_0^{p_T^{\rm max}}\! d p_T\,
   \frac{d^2 \sigma^{p\bar{p} \to t X_{\bar t}}}{dp_Tdy}- \int_{-y_+}^{0} dy \int_0^{p_T^{\rm max}}\!
  d p_T \,\frac{d^2 \sigma^{p\bar{p} \to t X_{\bar t}}}{dp_Tdy} \,,
\end{align}
with
\begin{gather}
  y_+ = \frac{1}{2} \ln\frac{1+\sqrt{1-4m_t^2/s}}{1-\sqrt{1-4m_t^2/s}}\,,
  \qquad p_T^{\rm max} = \frac{\sqrt{s}}{2}\sqrt{ \frac{1}{\cosh^2 y}-\frac{4m_t^2}{s}} \,.
\end{gather}
In Table~\ref{tab:FB}, we show the results for the asymmetric cross section and
forward-backward asymmetry obtained in different perturbative approximations. In
calculating the asymmetry, we first evaluate the numerator and denominator of the ratio
$A^t_{\text{FB}} = \Delta\sigma_{\text{FB}}/\sigma$ to the order indicated in the table
(using our scheme for the PDFs from Table~\ref{tab:PDForder}), and then further expand the
ratio itself, as appropriate at that order. We have obtained the fixed-order NLO results
using the formulas from Appendix~A of \cite{Kuhn:1998kw} and cross-checked them using
MCFM. As seen from the table, the higher-order corrections contained in the NLO+NNLL and
approximate NNLO formulas stabilize the asymmetric cross section at values not far
from the NLO calculation with $\mu_f = m_t$. The forward-backward asymmetry shows a wider
spread, and its value is not correlated with the asymmetric cross section in a
straightforward way. For instance, at $\mu_f = m_t$, the central value of the asymmetric
cross section in fixed order is essentially unchanged but has noticeably smaller scale
uncertainties at approximate NNLO compared to NLO. Yet the asymmetry itself decreases and
the errors are not reduced in proportion to those for the asymmetric cross section.
However, simply taking the higher and lowest numbers from the NLO+NNLL and approximate
NNLO numbers leads to predictions in the range of roughly 4\%--5.6\%, so the message of this
analysis is that the NLO prediction is rather stable under radiative corrections. 
For comparison, the most recent measurement made by the CDF collaboration 
of the inclusive forward-backward asymmetry
in the laboratory frame is $A^{t}_{\text{FB}}  = (15.0 \pm 5.5) \%\,  \mbox{(stat.+sys.)}$
\cite{Aaltonen:2011kc}; roughly, the experimental result exceeds the theoretical predictions 
it Table~\ref{tab:FB} by two standard deviations. The paper \cite{Aaltonen:2011kc} also reports 
a measurement of the  asymmetry in the $t \bar{t}$ frame as a function of the pair invariant 
mass; in the high mass region the experimental measurement exceeds the NLO QCD prediction by more
than three standard deviations. It is possible to calculate the forward-backward asymmetry as a 
function of the invariant mass up to NLO+NNLL accuracy by employing the results of \cite{Ahrens:2010zv}. The results of such an analysis will be presented elsewhere \cite{inprep}.

\subsection{Total Cross Section \label{sec:totalcs}}


\begin{table}[t!]
\begin{center}
\begin{tabular}{|l|c|c|c|c|}
  \hline
  & Tevatron &  LHC (7\,TeV)  & LHC (8\,TeV) &  LHC (14\,TeV)
  \\
  \hline
  $\sigma_{\text{NLO leading, $\scetopi$}}$  & 
  5.92{\footnotesize $^{+0.74}_{-0.80}$}{\footnotesize $^{+0.33}_{-0.22}$} &
  149{\footnotesize $^{+13}_{-16}$}{\footnotesize $^{+8}_{-8}$}  &
  214{\footnotesize $^{+16}_{-22}$}{\footnotesize $^{+10}_{-10}$}  &
  853{\footnotesize $^{+35}_{-65}$}{\footnotesize $^{+29}_{-30}$}
  \\
  $\sigma_{\text{NLO leading, $\scetpim$}}$   & 
  5.50{\footnotesize $^{+0.78}_{-0.78}$}{\footnotesize $^{+0.31}_{-0.20}$} &
  134{\footnotesize $^{+16}_{-17}$}{\footnotesize $^{+7}_{-7}$} &
  192{\footnotesize $^{+21}_{-23}$}{\footnotesize $^{+9}_{-9}$} &
  761{\footnotesize $^{+64}_{-75}$}{\footnotesize $^{+25}_{-26}$}
  \\
  $\sigma_{\text{NLO, $q\bar q+gg$}}$ &
  5.89{\footnotesize $^{+0.77}_{-0.81}$} &
  142{\footnotesize $^{+14}_{-17}$}  &
  203{\footnotesize $^{+21}_{-23}$}  &
  801{\footnotesize $^{+67}_{-77}$}
  \\
  $\sigma_{\rm NLO}$ &
  5.79{\footnotesize $^{+0.79}_{-0.80}$}{\footnotesize $^{+0.33}_{-0.22}$} &
  133{\footnotesize $^{+21}_{-19}$}{\footnotesize $^{+7}_{-7}$}  &
  192{\footnotesize $^{+30}_{-27}$}{\footnotesize $^{+9}_{-9}$}  &
  761{\footnotesize $^{+105}_{-96}$}{\footnotesize $^{+26}_{-27}$}
  \\
  \hline
  $\sigma_{\text {NLO+NNLL, $\scetopi$}}$  & 
  6.53{\footnotesize $^{+0.14}_{-0.17}$}{\footnotesize $^{+0.32}_{-0.23}$} &
  157{\footnotesize $^{+7}_{-11}$}{\footnotesize $^{+8}_{-8}$}  &
  223{\footnotesize $^{+9}_{-15}$}{\footnotesize $^{+10}_{-11}$}  &
  845{\footnotesize $^{+27}_{-67}$}{\footnotesize $^{+27}_{-29}$}
  \\
  $\sigma_{\text {NLO+NNLL, $\scetpim$}}$  & 
  6.29{\footnotesize $^{+0.19}_{-0.20}$}{\footnotesize $^{+0.31}_{-0.23}$} &
  149{\footnotesize $^{+7}_{-6}$}{\footnotesize $^{+8}_{-8}$} &
  212{\footnotesize $^{+10}_{-9}$}{\footnotesize $^{+10}_{-10}$} &
  820{\footnotesize $^{+40}_{-44}$}{\footnotesize $^{+28}_{-29}$}
  \\
  $\sigma_{\text{NNLO approx, $\scetopi$}}$  & 
  6.30{\footnotesize $^{+0.30}_{-0.39}$}{\footnotesize $^{+0.32}_{-0.23}$} &
  153{\footnotesize $^{+2}_{-3}$}{\footnotesize $^{+8}_{-8}$}  &
  219{\footnotesize $^{+2}_{-3}$}{\footnotesize $^{+10}_{-11}$}  &
  847{\footnotesize $^{+6}_{-0}$}{\footnotesize $^{+28}_{-30}$}
  \\
  $\sigma_{\text{NNLO approx, $\scetpim$}}$  & 
  6.12{\footnotesize $^{+0.43}_{-0.47}$}{\footnotesize $^{+0.31}_{-0.23}$} &
  145{\footnotesize $^{+8}_{-7}$}{\footnotesize $^{+8}_{-8}$}  &
  207{\footnotesize $^{+11}_{-9}$}{\footnotesize $^{+10}_{-10}$}  &
  811{\footnotesize $^{+38}_{-25}$}{\footnotesize $^{+27}_{-29}$}
  \\
  \hline
 \end{tabular}
\end{center}
\vspace{-2mm}
\caption{\label{tab:CrossSections400}
  Results for the total cross section in pb, using the default choice $\mu_f=400$\,GeV.
  The first set of errors refers to perturbative uncertainties associated with scale
  variations, and the second to PDF uncertainties.}
\vspace{4mm}
\begin{center}
\begin{tabular}{|l|c|c|c|c|}
  \hline
  & Tevatron &  LHC (7\,TeV)  &     LHC (8\,TeV)  & LHC (14\,TeV)
  \\
  \hline
  $\sigma_{\text{NLO leading, $\scetopi$}}$  & 
  6.79{\footnotesize $^{+0.20}_{-0.70}$}{\footnotesize $^{+0.38}_{-0.24}$} &
  163{\footnotesize $^{+0}_{-11}$}{\footnotesize $^{+9}_{-9}$} &
  232{\footnotesize $^{+0}_{-14}$}{\footnotesize $^{+11}_{-12}$} &
  887{\footnotesize $^{+0}_{-66}$}{\footnotesize $^{+30}_{-32}$}
  \\
  $\sigma_{\text{NLO leading, $\scetpim$}}$ &
  6.42{\footnotesize $^{+0.42}_{-0.76}$}{\footnotesize $^{+0.35}_{-0.23}$} &
  152{\footnotesize $^{+7}_{-15}$}{\footnotesize $^{+8}_{-8}$} &
  217{\footnotesize $^{+8}_{-20}$}{\footnotesize $^{+10}_{-11}$} &
  836{\footnotesize $^{+18}_{-60}$}{\footnotesize $^{+29}_{-30}$}
  \\
  $\sigma_{ \text{NLO, $q\bar q+gg$}}$ &
  6.80{\footnotesize $^{+0.27}_{-0.73}$} &
  160{\footnotesize $^{+5}_{-15}$} &
  228{\footnotesize $^{+6}_{-20}$} &
  879{\footnotesize $^{+21}_{-62}$}
  \\
  $\sigma_{\rm NLO}$  & 
  6.72{\footnotesize $^{+0.36}_{-0.76}$}{\footnotesize $^{+0.37}_{-0.24}$} &
  159{\footnotesize $^{+20}_{-21}$}{\footnotesize $^{+8}_{-9}$} &
  227{\footnotesize $^{+28}_{-30}$}{\footnotesize $^{+11}_{-12}$} &
  889{\footnotesize $^{+107}_{-106}$}{\footnotesize $^{+31}_{-32}$} 
  \\
  \hline
  $\sigma_{\text{NLO+NNLL, $\scetopi$}}$  & 
  6.55{\footnotesize $^{+0.16}_{-0.14}$}{\footnotesize $^{+0.32}_{-0.24}$} &
  150{\footnotesize $^{+7}_{-7}$}{\footnotesize $^{+8}_{-8}$} &
  214{\footnotesize $^{+10}_{-10}$}{\footnotesize $^{+10}_{-11}$} &
  824{\footnotesize $^{+41}_{-44}$}{\footnotesize $^{+28}_{-30}$}
  \\
  $\sigma_{\text{NLO+NNLL, $\scetpim$}}$ & 
  6.46{\footnotesize $^{+0.18}_{-0.19}$}{\footnotesize $^{+0.32}_{-0.24}$} &
  147{\footnotesize $^{+7}_{-6}$}{\footnotesize $^{+8}_{-8}$} &
  210{\footnotesize $^{+10}_{-8}$}{\footnotesize $^{+10}_{-11}$} &
  811{\footnotesize $^{+45}_{-42}$}{\footnotesize $^{+29}_{-30}$}
  \\
  $\sigma_{\text{NNLO approx, $\scetopi$}}$ &
  6.63{\footnotesize $^{+0.00}_{-0.27}$}{\footnotesize $^{+0.33}_{-0.24}$} &
  155{\footnotesize $^{+3}_{-2}$}{\footnotesize $^{+8}_{-9}$} &
  222{\footnotesize $^{+5}_{-3}$}{\footnotesize $^{+11}_{-11}$} &
  851{\footnotesize $^{+25}_{-5}$}{\footnotesize $^{+29}_{-31}$}
  \\
  $\sigma_{\text{NNLO approx, $\scetpim$}}$ &
  6.62{\footnotesize $^{+0.05}_{-0.40}$}{\footnotesize $^{+0.33}_{-0.24}$} &
  155{\footnotesize $^{+8}_{-8}$}{\footnotesize $^{+8}_{-9}$} &
  221{\footnotesize $^{+12}_{-12}$}{\footnotesize $^{+11}_{-12}$} &
  860{\footnotesize $^{+46}_{-43}$}{\footnotesize $^{+30}_{-33}$}
  \\
  \hline
  \end{tabular}
\end{center}
\vspace{-2mm}
\caption{\label{tab:CrossSections} 
Same as Table~\ref{tab:CrossSections400}, but with the 
scale choice $\mu_f=m_t$.}
\end{table}

Our main new results concerning the total cross section are the NLO+NNLL and approximate
NNLO expressions in the $\scetopi$ scheme, obtained here for the first
time.  In addition to presenting these numbers, we also pay attention
to a comparison with the $\scetpim$ results, and a main outcome of
this study is that the total cross sections obtained within the two
types of kinematics are in good agreement, as long as the subleading
terms identified within the SCET formalism are included.  This is
arguably not the case in the traditional 1PI or PIM schemes used in previous
calculations, a point we discuss in more detail in
Section~\ref{subsec:1PIcompare}.

Our results for the total cross section within different
approximations are summarized in Tables~\ref{tab:CrossSections400} and
\ref{tab:CrossSections}. We have also shown the scale dependence of
the $\scetopi$ results at NLO+NNLL and approximate NNLO in more detail
in Figure~\ref{fig:cs_with_muvar}.  The main phenomenological results
are the NLO+NNLL and approximate NNLO numbers in the bottom half of
the tables, but for comparison we also give the NLO predictions
obtained within the different approximations. In that case, we also
present the sum of the $q\bar q$ and $gg$ channels alone, without the
extra piece from the $gq$ and $g\bar q$ channels; this allows for  a more
direct comparison with the leading-singular results in the $\scetopi$ and $\scetpim$ schemes at NLO, neither
of which includes those pieces.  In the fixed-order results, the scale
uncertainties are obtained by varying the factorization scale up and
down by a factor of two.  In the resummed results, we vary
independently $\mu_h,\mu_s$ and $\mu_f$ up and down by a factor of two
with the other scales held fixed, and then add the three uncertainties
together in quadrature to obtain the perturbative uncertainties shown
in the tables. The resummed results in the $\scetpim$ scheme are
obtained with the choices of $\mu_h$ and $\mu_s$ used in
\cite{Ahrens:2010zv}.  We have included PDF uncertainties obtained by
evaluating the cross section with the set of MSTW2008 PDFs at 90\%
confidence level.

In all cases, the results in the $\scetopi$ and $\scetpim$ schemes within a
given perturbative approximation are compatible with one another, once
the uncertainties estimated from scale variations are taken into
account.  The $\scetopi$ results tend to come out higher and the
$\scetpim$ results lower, but the differences are not dramatic.
At NLO, the central values of the exact results always fall between the 
predictions obtained by retaining only the leading singular pieces in the two types of kinematics.  The $\mu_f$ dependence
is also taken into account relatively well by the two types of kinematics,   
although at higher collider energies the $qg$ and $\bar q g$ channels give an important
contribution to the exact NLO result at lower values of $\mu_f$, which is not taken
into account by the leading-singular pieces of the threshold expansions. 
The NLO+NNLL and 
approximate NNLO formulas include these NLO corrections through the 
matching, and the fact that these higher-order approximations have 
a rather small scale dependence even without the NNLO corrections
from the $qg$ channel can be taken as an indication that such corrections
are small, but this is a point which should nonetheless be kept in mind
when applying approximate formulas to the LHC with $\sqrt{s}=14$~TeV.
   
Another noticeable pattern has to do not with the difference between
the $\scetopi$ and $\scetpim$ schemes, but rather between the NLO+NNLL and
approximate NNLO results: the NLO+NNLL results are higher for
$\mu_f=400$~GeV, but lower for $\mu_f=m_t$. We can learn more about
the scale variations in the NLO+NNLL and approximate NNLO results in
the $\scetopi$ scheme by examining Figure~\ref{fig:cs_with_muvar}.  Compared to
both the NLO+NNLL approximations and the approximate NNLO numbers in the 
 $\scetpim$ scheme, 
the approximate NNLO results in the $\scetopi$ scheme have very small scale
uncertainties around a given $\mu_f$, particularly at the LHC.  When
$\mu_f$ is considered in the entire range $m_t/2<\mu_f<800$~GeV, on
the other hand, the approximate NNLO results are in good agreement
with the resummed results in that same range.  This is not an
unreasonable means of comparison, since at a given $\mu_f$ the
resummed results probe scales ranging from $\mu_s^{\rm def}\sim
75$~GeV to $\mu_h \sim 400$~GeV, so in estimating errors in fixed-order
one should arguably focus on a similar range instead of  
just using $m_t/2 < \mu_f <2m_t$, as is often done in the literature.  

\begin{figure}
\begin{center}
\begin{tabular}{lr}
\psfrag{y}[][][1][90]{$\sigma$ [pb]}
\psfrag{x}[]{$m_t$ [GeV]}
\psfrag{z}[][][0.85]{$\sqrt{s}=1.96$\,TeV}
\includegraphics[width=0.43\textwidth]{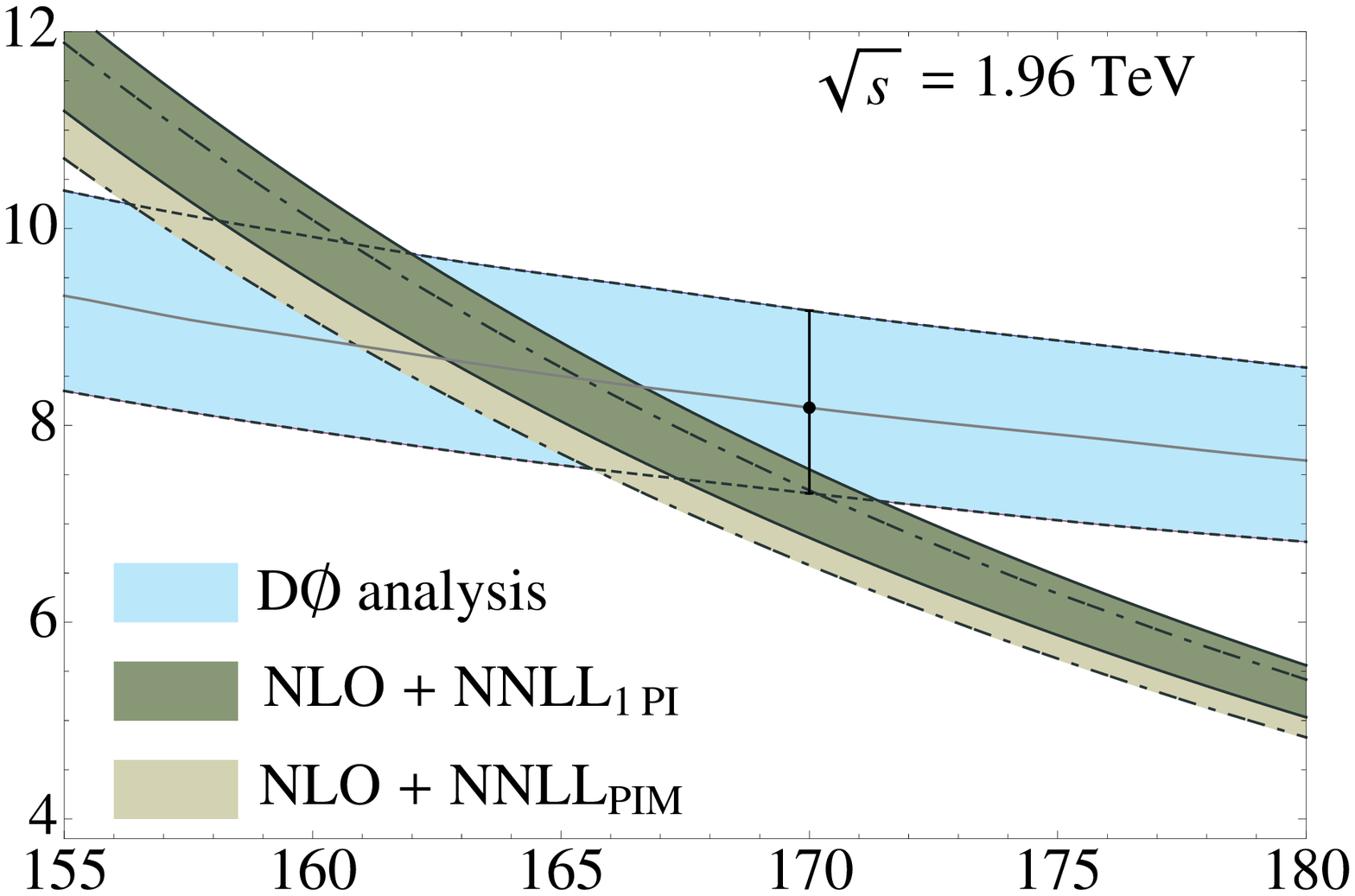}
&
\psfrag{y}[][][1][90]{$\sigma$ [pb]}
\psfrag{x}[]{$m_t$ [GeV]}
\psfrag{z}[][][0.85]{$\sqrt{s}=1.96$\,TeV}
\includegraphics[width=0.44\textwidth]{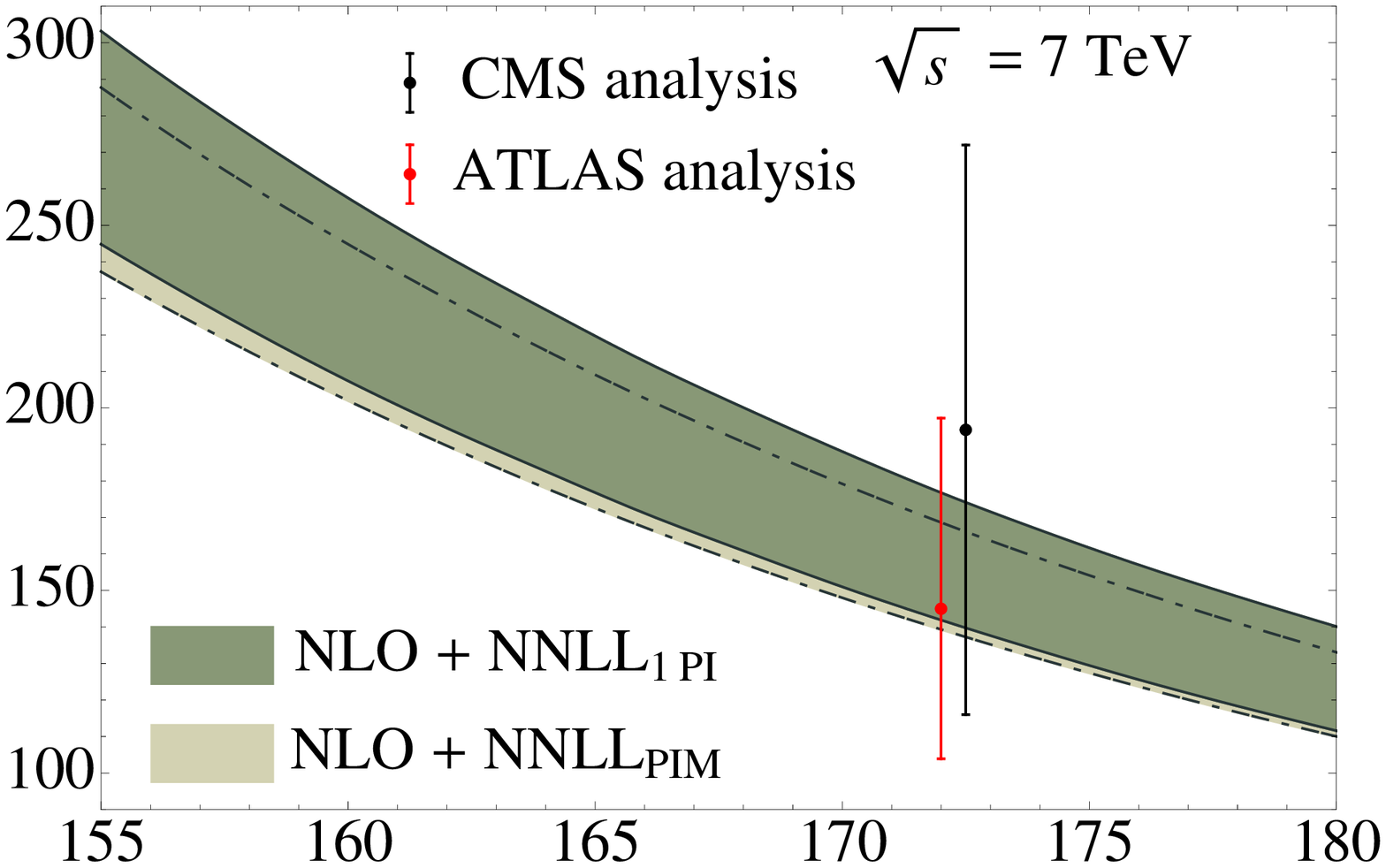}
\end{tabular}
\end{center}
\vspace{-4mm}
\caption{\label{fig:mtdepcs} Left: Dependence of the total cross section on top-quark mass defined
in the pole-mass scheme. The  $\scetopi$ and $\scetpim$ NLO+NNLL bands reflect the linearly combined
scale and PDF uncertainties. The blue band shows the dependence of a D0 measurement of the total 
cross section on $m_t$ \cite{Abazov:2009ae}. Right: The same for the LHC but with a comparison to recent CMS \cite{Khachatryan:2010ez} and ATLAS \cite{Aad:2010ey} measurements.}
\end{figure}

As emphasized in Section~\ref{sec:comparison}, due to our method for determining 
the soft scale $\mu_s$, the NLO+NNLL
predictions contain slightly different information than the
approximate NNLO formulas.
In particular, the NNLO expansion of the resummed formulas differ in
the structure of the $P_n$ distributions.  For instance, the
approximate NNLO results contain $P_3$ distributions, but the
equivalent terms in the direct NNLO expansion of the resummed formula
are of the form $P_2 \ln(\mu_s/\mu_f)$. These contribute at the same
order in the counting of RG-improved perturbation theory, but there
are obviously numerical differences between the two forms of the
expansion. This issue is discussed in more detail in the Appendix. 
Since the analysis in Figure~\ref{fig:SoftScales} was done
at NLO and at that order one encounters at most $P_1$
distributions, it is worthwhile to ask whether the choice of scale of
$\mu_s$ deduced there is really appropriate to account for the
mismatch between the two approximations.  If this were not the case,
our NLO+NNLL predictions would become unstable upon the inclusion of
the higher-order matching corrections and it would make more sense to
use the approximate fixed-order NNLO results.  To address this issue,
we have also calculated the cross section where we add the NNLO matching
coefficients on top of the NLO+NNLL resummation.  More precisely, we
include the pieces of the matching functions specified in
(\ref{eq:NNLOscheme}), but still including the NLL evolution matrices
in the trace, as in (\ref{eq:resC}).  In the case
where $\mu_f=\mu_h=\mu_s$, this approximation reduces to the
approximate NNLO result (compared to the NLO+NNLL result, which
reduces to the NLO result in this limit). We have checked that the numerical results in this ``NNLO+NNLL'' approximation are within the uncertainties
estimated by the NLO+NNLL calculation. This provides evidence
that our scale-setting procedure is indeed appropriate for 
effectively including the higher-order corrections.

In Figure~\ref{fig:mtdepcs} we show the total cross section calculated in the $\scetopi$ and 
$\scetpim$ schemes as a function of the top-quark mass defined in the pole-mass scheme. The bands 
represent linearly combined errors from scale and PDF uncertainties. On the left side we compare these 
with the dependence of a D0 measurement on $m_t$ at the Tevatron \cite{Abazov:2009ae}. We
have converted our error bands to $68$\% CL ($1\sigma$) to match the confidence level of the experimental errors. 
On the right side we present our predictions for the $m_t$-dependent cross section in combination
with new measurements done by the CMS \cite{Khachatryan:2010ez} and ATLAS \cite{Aad:2010ey} 
collaborations. Both analyses use a top-quark mass of 172.5~GeV, but we have 
set them aside in Figure~\ref{fig:mtdepcs} for clarity. Both for the Tevatron and LHC there
is good agreement between theoretical predictions and measurements. 
It would be interesting to use a more physical mass scheme like $\overline{\text{MS}}$ both
in theoretical and experimental analyses. This will be discussed in a forthcoming article \cite{inprep}.

\subsection{Comparison with previous results}
\label{subsec:1PIcompare}

\begin{figure}
\begin{center}
\begin{tabular}{ll}
\includegraphics{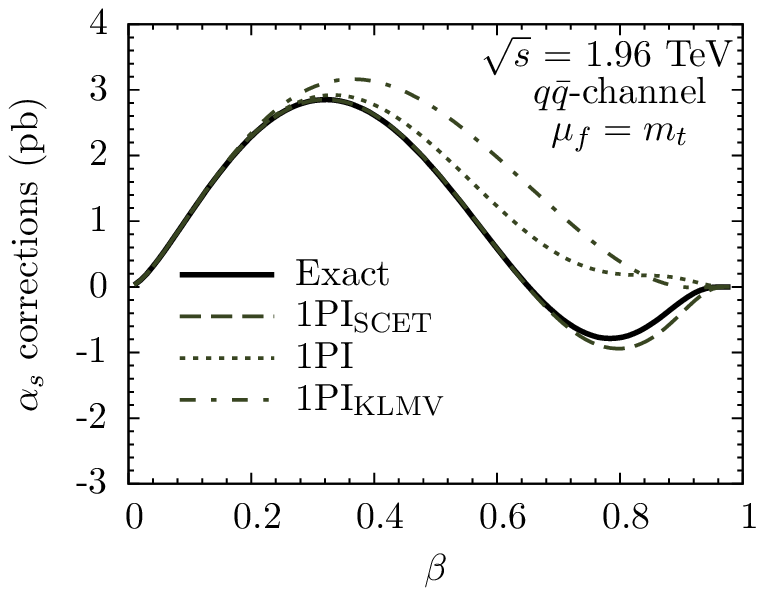} & 
\includegraphics{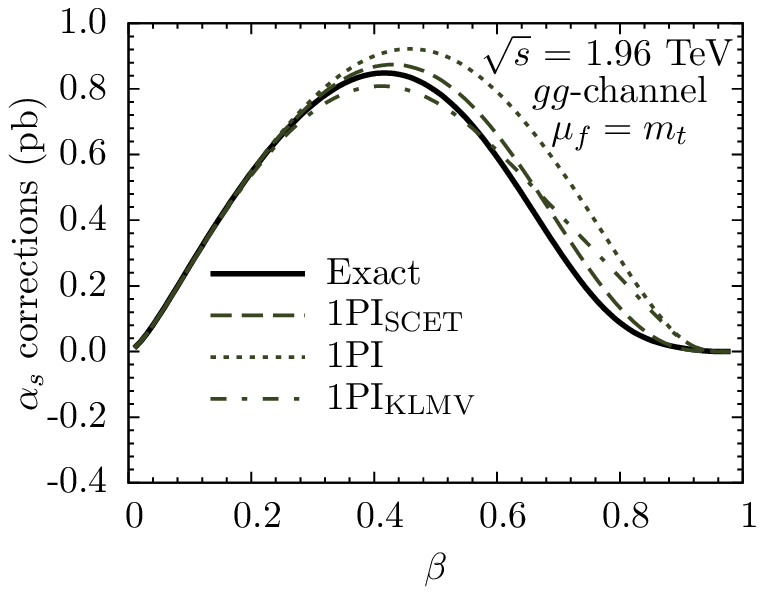}
\\
\includegraphics{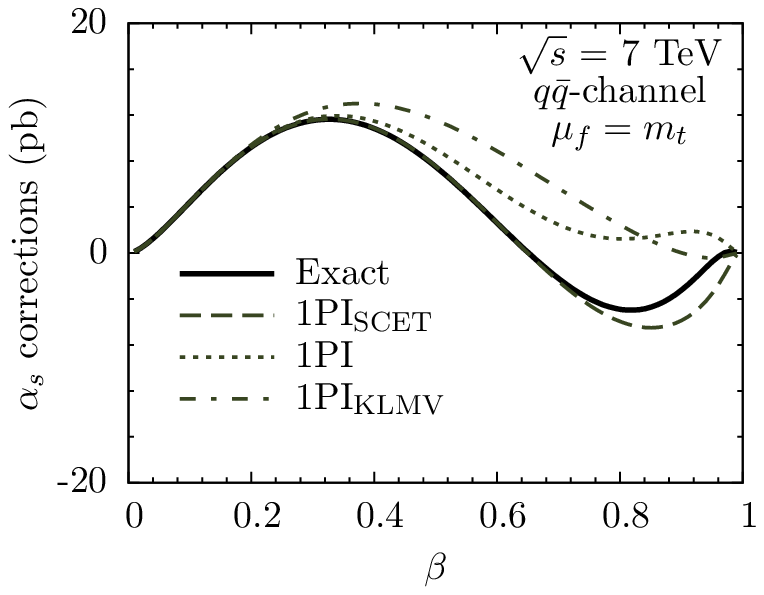} & 
\includegraphics{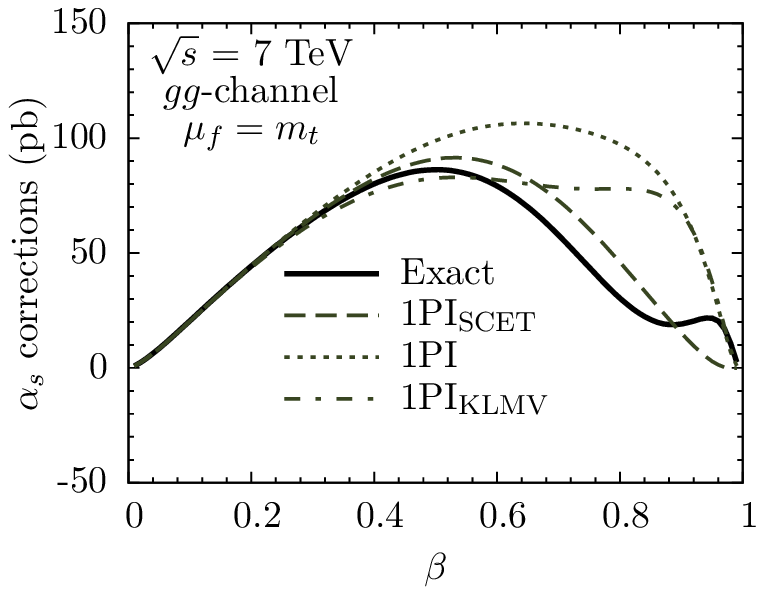} 
\\
\includegraphics{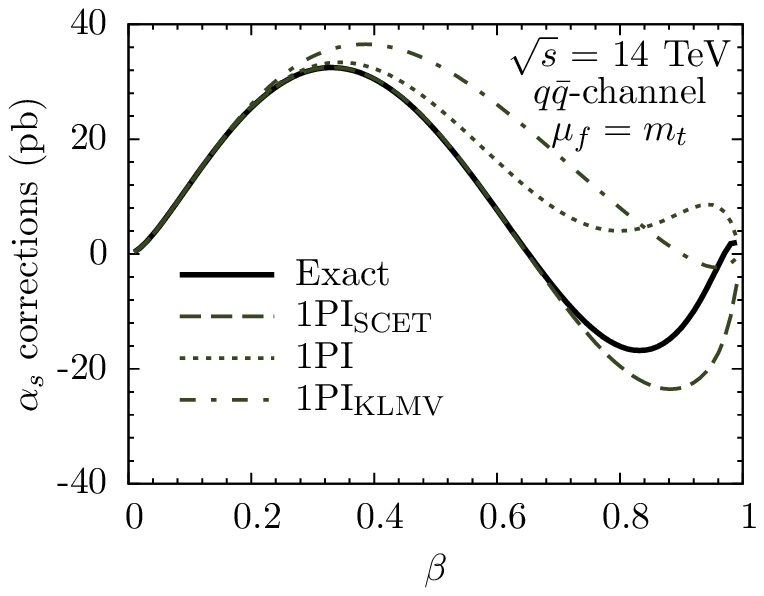} & 
\includegraphics{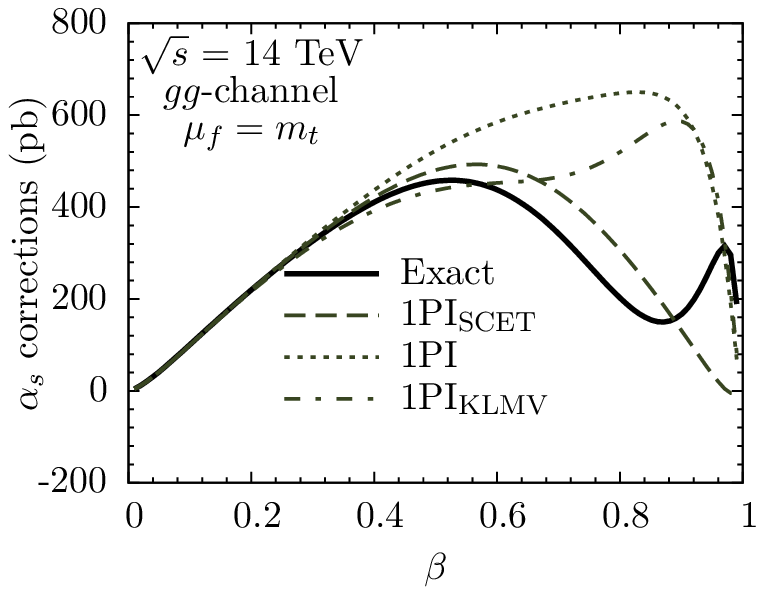} 
\end{tabular}
\end{center}
\vspace{-1ex}
\caption{\label{fig:beta-plot_kmlv} The $\alpha_s$ corrections 
to $d\sigma/d\beta$ for the choice $\mu_f=m_t$. The 1PI and 
$\scetopi$ results refer to the scheme used in the present work,
while the 1PI$_{\rm KLMV}$ results uses that from  
\cite{Kidonakis:2001nj}.}
\end{figure}

Recently, approximate NNLO cross sections and $p_T$ distributions in
1PI kinematics have been calculated in \cite{Kidonakis:2010dk}.
While
the resummation formalism leading to these results is different, we
have checked wherever possible that the structure of the NNLO
expansion as written in (\ref{eq:C2}) is the same. Therefore, our 
results in the 1PI scheme (but {\it not} the $\scetopi$ scheme, which
includes a set of corrections in the $R(s_4)$ term in that equation),
should in principle be the same.  
However, a direct comparison is complicated for two main reasons.
First, no explicit analytic results for the coefficients appearing in
(\ref{eq:C2}) were given in \cite{Kidonakis:2010dk},
which focused instead
on recalculating the two-loop anomalous dimension matrix obtained in
 \cite{Becher:2009kw,Ferroglia:2009ep,Ferroglia:2009ii}.
 As explained in the
text after that equation, there are ambiguities in what terms to
include in the coefficient multiplying the $\delta$-function term.
Second, even if the scheme were fixed and the analytic expressions
agreed, there would still be ambiguities in the numerical
implementation, as explained in Section~\ref{sec:numerical}.  In
particular, depending on the choice of the variables $\hat{s}'$, $\hat{t}_1'$
and $\hat{u}_1'$, the corrections stemming from the
plus-distributions can be different.  Since analytic results and the
choice of scheme were not specified in \cite{Kidonakis:2010dk}, we
cannot compare directly with the numerical results given there.

\begin{table}[t]
\begin{center}
\begin{tabular}{|l|c|c|c|}
  \hline
  & Tevatron &  LHC (7\,TeV)  & LHC (14\,TeV)
  \\
  \hline
  $\sigma_{\rm NLO~leading ,~1PI~(\scetopi)}$  & 
  7.23{\footnotesize $^{+0.45}_{-0.86}$} (6.79{\footnotesize $^{+0.20}_{-0.70}$}) &
  183{\footnotesize $^{+6}_{-18}$} (163{\footnotesize $^{+0}_{-11}$}) &
  1024{\footnotesize $^{+0}_{-67}$} (887{\footnotesize $^{+0}_{-66}$})
  \\
  $\sigma_{\rm NLO~leading,~PIM~(\scetpim)}$  & 
  6.20{\footnotesize $^{+0.28}_{-0.69}$} (6.42{\footnotesize $^{+0.42}_{-0.76}$}) &
  143{\footnotesize $^{+1}_{-12}$} (152{\footnotesize $^{+7}_{-15}$})  &
  771{\footnotesize $^{+0}_{-42}$} (836{\footnotesize $^{+18}_{-60}$})
  \\
  \hline
  $\sigma_{\rm NNLO~approx,~1PI~(\scetopi)}$  & 
  7.06{\footnotesize $^{+0.00}_{-0.29}$} (6.63{\footnotesize $^{+0.00}_{-0.27}$}) &
  180{\footnotesize $^{+3}_{-8}$}  (155{\footnotesize $^{+3}_{-2}$}) &
  1009{\footnotesize $^{+40}_{-54}$} (851{\footnotesize $^{+25}_{-5}$})
  \\
   $\sigma_{\rm NNLO~approx,~PIM~(\scetpim)}$  & 
  6.46{\footnotesize $^{+0.18}_{-0.45}$} (6.62{\footnotesize $^{+0.05}_{-0.40}$}) &
  148{\footnotesize $^{+14}_{-11}$} (155{\footnotesize $^{+8}_{-8}$})  &
  823{\footnotesize $^{+78}_{-67}$} (860{\footnotesize $^{+46}_{-43}$}) 
  \\
  \hline
  \end{tabular}
\end{center}
\vspace{-3mm}
\caption{\label{tab:CrossSectionsComparison} 
The total cross section in pb in the 1PI and PIM schemes, where the 
subleading terms produced through the SCET expansion are not included.  
The uncertainties are estimated from scale variation in the 
range $m_t/2<\mu_f<2m_t$. The results in the $\scetopi$ and $\scetpim$ schemes (also shown  in Table~\ref{tab:CrossSections}) are listed in parentheses for 
comparison. 
}
\end{table}

Instead, we shall present our own results in the 1PI and PIM schemes,
and use those to compare with the $\scetopi$ and $\scetpim$
results. For the default scale choice $\mu_f=m_t=173.1$~GeV, the
numbers in these schemes are shown in
Table~\ref{tab:CrossSectionsComparison}.  Comparing with
Table~\ref{tab:CrossSections}, we notice that the numbers in
the $\scetpim$ and PIM schemes are in much better agreement than those in
the $\scetopi$ and 1PI schemes.  Actually, the $\scetpim$,
$\scetopi$, and PIM results are compatible with each other within the
uncertainties estimated through scale variations, while the 1PI results are much higher, especially at the LHC. Since all of our numerical
studies have indicated that predictions in 1PI kinematics at LHC
energies are susceptible to large power corrections, we are rather
hesitant to give much weight to those particular results. We consider
the results in PIM kinematics more reliable, and if different
kinematics is to be used as a criterion for estimating uncertainties,
we believe the situation is more accurately reflected in
Tables~\ref{tab:CrossSections400} and~\ref{tab:CrossSections} than in
Table~\ref{tab:CrossSectionsComparison}.

The approximate NNLO numbers in
Table~\ref{tab:CrossSectionsComparison} are very close to those
of \cite{Kidonakis:2010dk} for the Tevatron, but at 
the LHC they are  roughly 10\% higher than those in \cite{Kidonakis:2010dk}.  
While we are unable to pinpoint the
source of this discrepancy for the reasons explained above, we can
compare the leading singular terms in our 1PI expansion at NLO with
those obtained in \cite{Kidonakis:2001nj}.  As explained in
Section~\ref{sec:numerical}, these were obtained using a different
scheme to evaluate the equivalent analytic results at threshold, so
the difference between two numbers gives a 
rough 
measure of the
power-suppressed ambiguities inherent to the formulas at threshold.
In Figure~\ref{fig:beta-plot_kmlv} we compare the results of
\cite{Kidonakis:2001nj} with 
 those obtained in our 
1PI scheme, in the $\scetopi$
scheme, and with the exact results. 
As seen from the figure, the leading terms in the $\scetopi$ scheme
provide the best approximation to the full results, and differences due
to the choice of scheme get larger as the collider energy
increases. The numbers in our 1PI scheme  are
higher than those from \cite{Kidonakis:2001nj}, which may account for
the discrepancy with \cite{Kidonakis:2010dk}. For reference,
the leading singular terms at NLO with $\mu_f=m_t$ in the scheme of
\cite{Kidonakis:2001nj} yield cross sections of $7.42$~pb at the
Tevatron, $174$~pb at the LHC with $\sqrt{s}=7$~TeV, and $968$~pb at
the LHC with $\sqrt{s}=14$~TeV.  In \cite{Kidonakis:2010dk}, it was
stated that in the $gg$ channel the leading singular terms in 1PI
kinematics account for over 98\% of the NLO correction at both the
Tevatron and the LHC.  For this NLO {\it correction}, the exact
results at $\mu_f=m_t$ are $0.42$~pb at the Tevatron and $280$~pb at
the LHC with $\sqrt{s}=14$~TeV, compared to $0.44$~pb and $360$~pb for
the leading singular pieces in the 1PI scheme of
\cite{Kidonakis:2001nj}.  The number at the LHC is quite a bit higher
than the exact result, and we therefore are not able to confirm the
statement made in \cite{Kidonakis:2010dk}, although we cannot rule out
that they were obtained in yet another (unspecified) scheme for
evaluating the formulas at threshold.

\section{Conclusions}
\label{sec:conclusions}

We have used techniques from SCET to study
higher-order perturbative corrections to 1PI observables in top-quark
pair production at hadron colliders.  In particular, we have calculated and collected the perturbative ingredients needed for an analysis at 
NLO+NNLL and approximate NNLO, and applied them to transverse-momentum and
rapidity distributions as well as the total production cross sections at the Tevatron and LHC, and to the forward-backward asymmetry at the Tevatron.
The details of factorization and resummation within our approach were
mainly taken over from the analogous study in PIM kinematics
presented in \cite{Ahrens:2010zv}.  However, differences arise from
the phase-space integrations for real emission in the soft limit. They 
affect the matrix-valued soft matching function, which must be
calculated from scratch.  We  presented results for this matching
function at NLO in fixed-order perturbation theory, which were previously unknown, and also addressed the differences in its RG equation needed for the
resummation of soft gluon corrections to all orders.  Along with the
anomalous dimensions and matrix-valued hard matching functions already
known from \cite{Ahrens:2010zv}, these are the essential ingredients
for an NLO+NNLL calculation.

Even apart from the new perturbative information included in our
results, an important aspect of the SCET approach to threshold
resummation is its ability to include a class of logarithmic
corrections, which are formally subleading in the exact threshold
limit but nonetheless give non-negligible numerical
contributions. These arise in the SCET formalism by noticing that in a
suitable reference frame the single dimensionful parameter
characterizing soft radiation is the total soft energy $E_s$ of the
emitted gluons, and expressing the logarithmic corrections in terms
of $\ln(E_s/\mu)$, without further expansion.  The subleading terms
generated in this way are of the form $\ln(1+s_4/m_t^2)/s_4$ in 1PI
kinematics, and $\ln z/(1-z)$ in PIM kinematics (with
$z=M^2/\hat{s}$).  We have implemented these new terms in our
numerical analysis summarized below, where results including them are
referred to as the $\scetopi$ and $\scetpim$ schemes.  Results
neglecting them are referred to as the 1PI and PIM schemes and should
correspond with previous results at a given order in perturbation
theory.  

Our improved calculations of the hard-scattering kernels near
threshold are {\it a priori} only important in regions of phase space
where kinematic restrictions imply the partonic threshold limit
$s_4\to 0$.  However, the differential cross section in such regions
is very small. Beyond that, the improved predictions are important if
a dynamical threshold enhancement occurs, due to a sharp fall-off of
the PDFs away from integration regions where $s_4\to 0$.  In
Section~\ref{sec:threshold} we performed a very detailed discussion of
the mechanism of threshold enhancement, and compared its effectiveness
in 1PI and PIM kinematics at the level of differential distributions
and the total cross section.  The main finding was that the leading
power corrections to the threshold limit in 1PI kinematics, which
scale as $E_s/m_t\sim s_4/m_t^2$, are in general larger than the
leading corrections in PIM kinematics, which scale as $E_s/M\sim
(1-z)$.  At the LHC with $\sqrt{s}=14$~TeV these corrections can
become quite large. This makes us somewhat reserved about the utility of
the threshold expansion in 1PI kinematics, especially at larger values of $p_T$.
Including the extra subleading terms unique to the $\scetopi$
approximation seemed to account for the power corrections to a good
degree of accuracy in the $q\bar q$ channel and also to some level in
the $gg$ channel, at least in regions of phase space where the
differential cross section is large. 
Since the top-quark pair production cross section at the Tevatron is dominated 
by the quark-antiquark channel, the predictions obtained in the $\scetopi$ scheme
at the Tevatron center-of-mass energy are expected to provide a reliable approximation to the exact results. In all cases, but especially at
the LHC with $\sqrt{s}=14$~TeV, the pure 1PI expansion gave larger
corrections compared to any other approximation, and we are rather
skeptical of any results obtained in that scheme.
For the reasons mentioned above, we limited the phenomenological
analysis to areas where the differential distributions are 
fairly large, or else to integrated quantities such as the
forward-backward asymmetry and the total cross section.  Our results
for the rapidity and transverse-momentum distributions can be found in
Figures~\ref{fig:ydistributions} and~\ref{fig:ptdistributions}.  As
expected, including the higher-order perturbative corrections in the
form of NLO+NNLL or approximate NNLO predictions leads to a
stabilization of the results under variations of the different
perturbative scales.  Our main results for the total cross section can
be found in Tables~\ref{tab:CrossSections400} and~\ref{tab:CrossSections}.
Although we consider the $\scetpim$ numbers more trustworthy, it is reassuring that the
$\scetopi$ numbers are actually in very good agreement with them. This is not the case for
the pure 1PI numbers, which come out significantly larger in all cases. If different types
of kinematics is to be used as a source of uncertainty in predictions based on soft-gluon resummation, we believe the true results are more accurately reflected in
Tables~\ref{tab:CrossSections400} and~\ref{tab:CrossSections}.

In Section~\ref{sec:totalcs} and the Appendix we analyzed in detail the
origin of the numerical differences between NLO+NNLL and approximate NNLO calculations. The structure of the plus-distributions considered in the two approaches is different, since in the resummation approach plus-distributions and logarithms of scale ratios are formally of the same order. We showed why our choice of the soft and hard scales for the NLO+NNLL calculations is sound and how the resummed results are stable with respect to reasonable variations of these scales.

In conclusion, we obtained NLO+NNLL predictions for the top-pair transverse-momentum and rapidity distributions at hadron colliders. In the case of the transverse-momentum distribution at the Tevatron, the predictions presented here can be already compared with experimental measurements. By integrating the results for the differential distributions we obtained predictions for fully inclusive observables such as the forward-backward asymmetry and the total cross section. For the total cross section in particular, we analyzed the relations among the $\scetopi$ results obtained here, the $\scetpim$ results of \cite{Ahrens:2010zv}, and the ones obtained with a different resummation approach \cite{Kidonakis:2010dk}. All the evidence gathered shows that the SCET-based approach provides consistent and reliable results in both types of kinematics.

\section*{Acknowledgments}

We would like to thank Rikkert Frederix, Sven Moch for useful discussions. This research was supported in part by the State of Rhineland-Palatinate via the Research Center {\em Elementary Forces and Mathematical Foundations},  
by the German Federal Ministry for Education and Research under grant 05H09UME: {\em
  Precision Calculations for Collider and Flavour Physics at the LHC}, by the German
Research Foundation under grant NE398/3-1, 577401: {\em Applications of Effective Field
  Theories to Collider Physics}, by the European Commission through the `LHCPhenoNet'
Initial Training Network PITN-GA-2010-264564, and by the Schweizer Nationalfonds under
grant 200020-124773.
 
\appendix
\section{Appendix\label{appx}}

In our numerical analysis we considered two different ways of
including higher-order perturbative corrections to the hard-scattering
kernels: constructing approximate expansions to NNLO in fixed order,
and all-orders resummmation to NNLL accuracy.  We emphasized that for
a fixed numerical value of $\mu_s$, the truncation of the NNLL series
to NNLO in $\alpha_s$ contains a different structure of corrections
than the approximate NNLO formula.  Here we explain this
statement in more detail, and show explicitly what types of
higher-order corrections the master formula (\ref{eq:resC}) resums. To
do so, it is sufficient to ignore the matrix structure of the RG
equations, and just consider hard and soft functions which are simple
functions of their arguments. Then the treatment of the hard function
is straightforward, and the complication for the soft function is its
non-locality.

To explain the issue, it is simplest to start with the hard function.
Ignoring its matrix structure and dependence on kinematic invariants, 
its RG equation is of the form
\begin{equation}
\mu \frac{d}{d\mu}\,H(L_h,\alpha_s(\mu))=\left( 2\Gamma_{\rm cusp} \ln
  \frac{M^2}{\mu^2}+2\gamma^h\right) H(L_h, \alpha_s(\mu)) \,,
\end{equation}
where $L_h\equiv \ln (M^2/\mu^2)$. The parameter $M$ is to be understood
as a generic hard scale; in the specific case of 1PI kinematics, it would
be $M=\sqrt{\hat{s}}$.  The RG equation can be used to generate 
higher-order terms in the perturbative expansion of $H$. For instance,
we can solve the equation as a fixed-order series in $\alpha_s$ using
the ansatz 
\begin{equation}
H(L_h,\alpha_s(\mu))=\alpha_s^2(\mu)\left[ h^{(0,0)} 
+ \frac{\alpha_s(\mu)}{4\pi}\,\sum_{n=0}^2 h^{(1,n)}L_h^n
   +\dots\right] .
\end{equation}
In terms of the lowest-order coefficient and the anomalous dimensions and
$\beta$-function, whose expansions we define as 
\begin{align}
  \label{eq:gammaexp}
  \Gamma_{\text{cusp}}(\alpha_s) &= \Gamma_0 \, \frac{\alpha_s}{4\pi} + \Gamma_1 \left(
    \frac{\alpha_s}{4\pi} \right)^2 + \Gamma_2 \left( \frac{\alpha_s}{4\pi} \right)^3 +
  \ldots \, , \nonumber
  \\
  \beta(\alpha_s) &= -2\alpha_s \left[ \beta_0 \, \frac{\alpha_s}{4\pi} + \beta_1 \left(
      \frac{\alpha_s}{4\pi} \right)^2 + \beta_2 \left( \frac{\alpha_s}{4\pi} \right)^3 +
    \ldots \right],
\end{align}
and similarly for $\gamma^h$, the expansion coefficients at NLO can be constructed as 
 \begin{equation}
\label{eq:NLOH}
H(L_h, \alpha_s(\mu))=\alpha_s^2(\mu)\left\{h^{(0,0)}
+\frac{\alpha_s(\mu)}{4\pi}\left[
-h^{(0,0)}\left(\frac{\Gamma_0}{2} L_h^2+ (\gamma^h_0 + 2\beta_0)L_h\right)+h^{(1,0)}\right]+\dots 
\right\} .
\end{equation}
One can obviously generalize this to any order in $\alpha_s$ and calculate
the coefficients of the logarithms at a given order in terms of the 
anomalous dimensions and lower-order matching coefficients. This is 
the method used in constructing approximate fixed-order expansions. 
  
In the effective-theory analysis, one assumes the presence of a second, 
widely separated scale $\mu_s \ll M$ and uses the counting of RG-improved
perturbation theory, i.e.\ $\ln(\mu_s/M)\sim 1/\alpha_s$.  Then the higher-order
corrections contain large logarithms, which can be resummed to all-orders
by using the exact solution to the RG equation.  This solution reads 
\cite{Bosch:2003fc}
\begin{equation}
\label{eq:resH}
H(L_h, \alpha_s(\mu))=e^{4S(\mu_h,\mu)-2a_{\gamma^h}(\mu_h,\mu)}\,
H\bigg(\ln\frac{M^2}{\mu_h^2},\alpha_s(\mu_h)\bigg)
\left( \frac{M}{\mu_h}\right)^{-2\eta_h} ,
\end{equation}
where $\eta_h=2a_\Gamma(\mu_h,\mu)$. The Sudakov exponent and 
normal anomalous exponent are
\begin{align}
  \label{eq:Sa}
  S(\mu_h,\mu) = -\int\limits_{\alpha_s(\mu_h)}^{\alpha_s(\mu)} \!d\alpha\,
  \frac{\Gamma_{\text{cusp}}(\alpha)}{\beta(\alpha)} \int\limits_{\alpha_s(\mu_h)}^\alpha
  \!\frac{d\alpha'}{\beta(\alpha')} \, , \qquad a_\Gamma(\mu_h,\mu) =
  -\int\limits^{\alpha_s(\mu)}_{\alpha_s(\mu_h)}\!
  d\alpha\,\frac{\Gamma_{\text{cusp}}(\alpha)}{\beta(\alpha)} \, ,
\end{align}
and  similarly for $a_{\gamma^h}$. 

The all-orders solution does not actually depend on $\mu_h$, as
indicated by the notation. The same is true if the matching
coefficient and the exponentials are consistently re-expanded as a
series in $\alpha_s(\mu)$ in fixed order, in which case one just gets
back the approximate formulas above.  In practice, however, one must
truncate the result at a given level of accuracy (e.g.\ NNLL), and
beyond that level a residual dependence on $\mu_h$ remains.  To avoid
large logarithms in the matching coefficients, one chooses $\mu_h\sim
M$ and runs to the scale $\mu_f$ using the all-orders solution.  Then
the exponential factors resum logarithms which count as $\ln (M^2/\mu_f^2)$
and are large for $\mu_f\sim \mu_s$.  To see this explicitly, we get
rid of $\alpha_s(\mu_h)$ everywhere by using (with $a(\mu_f)\equiv
\alpha_s(\mu_f)/4\pi$)
\begin{equation}
a(\mu_h)=\frac{a(\mu_f)}{X}-\frac{a(\mu_f)^2}{X^2}\,\frac{\beta_1}{\beta_0}\,\ln
X+\dots\,; \qquad X=1+\beta_0\,a(\mu_f)\,\ln\frac{\mu_h^2}{\mu_f^2} \,,
\end{equation}
and re-expand the solution in (\ref{eq:resH}) as a series in $a(\mu_f)$.
For reference, the expansion of the Sudakov factor
and anomalous exponent to NNLO read
\begin{align}
S(\mu_h,\mu_f)&=-a(\mu_f)\,\frac{\Gamma_0}{8}\,  L_{hf}^2+
a(\mu_f)^2\left(\frac{\beta_0\Gamma_0}{24}L_{hf}^3-\frac{\Gamma_1}{8}L_{hf}^2\right)+\dots \,,
\\ \label{eq:aGexp}
a_\Gamma(\mu_h,\mu_f)&=a(\mu_f)\,\frac{\Gamma_0}{2}\, L_{hf}+
a(\mu_f)^2\left(-\frac{\beta_0\Gamma_0}{4}L_{hf}^2+\frac{\Gamma_1}{2}L_{hf}\right)+\dots \,,
\end{align}
where $L_{hf}\equiv \ln(\mu_h^2/\mu_f^2)$. The expansion of $a_\gamma$ is identical to the one in (\ref{eq:aGexp}), with the replacements $\Gamma_i \to \gamma_i$.

Using the above equations to expand (\ref{eq:resH}) to NLO, one recovers the NLO solution (\ref{eq:NLOH}).  All the dependence on $\mu_h$ drops
out to that order, as long as we keep the one-loop matching correction.  
However, this is {\it not\/} the case if we expand our NLL approximation of 
the resummed hard function to NLO.  In that case, the NLO matching
coefficient is of higher-order in the counting and not included in the 
formula, so after expansion to NLO dependence on the scale $\mu_h$ remains.  
The direct expansion of our NLL formula
at NLO in fixed order reads 
\begin{equation}
\alpha_s^2(\mu_f)\,a(\mu_f) \left\{
-h^{(0,0)}\ln\frac{\mu_h^2}{\mu_f^2}
\left[\frac{\Gamma_0}{2} \ln\frac{\mu_h^2}{\mu_f^2}
+ \Gamma_0 \ln\frac{M^2}{\mu_h^2} + (\gamma^h_0 + 2\beta_0)\right]\right\} .
\end{equation}
This is necessarily different than the ``approximate NLO'' formula one
would deduce by dropping the coefficient $h^{(1,0)}$ from
(\ref{eq:NLOH}), because it depends on $\mu_h$, but if we set
$\mu_h=M$ it is the same, and it is for this reason that one can still
say the NLL solution ``resums logarithms of the form $\ln(M/\mu_f)$ to all
orders'', although a more accurate statement would be that it ``resums
logarithms of the form $\ln(\mu_h/\mu_f)$ to all orders'', which includes
the possibility of other choices such as $\mu_h=m_t$.  Given this 
fact, it makes little sense to construct an approximate formula 
for a quantity such as the hard function: if there are large logarithms, it 
is just as easy to sum them to all orders as it is to construct the 
fixed-order expansion, and if the logarithms are not large, there is no reason
to include that subset of the higher-order corrections without the full
answer. 

We can repeat the analysis above to compare the structure of approximate
fixed-order expansions and resummed formulas for the soft function.  In this
case the RG equation is non-local, and to solve for the momentum-space soft
function one uses the technique of Laplace transforms \cite{Becher:2006nr}.  The
solution for the resummed momentum-space soft function is
\begin{equation}
\label{eq:softsol}
S(\omega,\mu_f)=e^{-4S(\mu_s,\mu_f)+2a_{\gamma^s}(\mu_s,\mu_f)}\,\tilde{s}\left(\partial_\eta,\mu_s\right)\frac{1}{\omega}\left(\frac{\omega}{\mu_s}\right)^{2\eta}\frac{e^{-2
    \gamma_E \eta}}{\Gamma(2\eta)}\,,
\end{equation}
where $\eta=2a_\Gamma(\mu_s,\mu_f)$, and $\tilde{s}$ is the Laplace-transformed
function, which satisfies the local RG equation 
\begin{equation}
\label{eq:RGS}
\frac{d}{d\ln\mu}\,\tilde{s}\bigg(\ln\frac{M^2}{\mu^2},\alpha_s(\mu)\bigg)=-\left(2\Gamma_{\rm
  cusp}\ln\frac{M^2}{\mu^2}+2\gamma_s\right)\tilde{s}\bigg(\ln\frac{M^2}{\mu^2},\alpha_s(\mu)\bigg) \,.
\end{equation}
In this case, approximate formulas in fixed order are obtained by first
constructing the solution to $\tilde{s}$ using the local RG equation.
To NNLO, we use the ansatz
\begin{equation}
\label{eq:Sansatz}
\tilde{s}(L,\alpha_s(\mu))= 1
+\frac{\alpha_s(\mu)}{4\pi}\,\sum_{n=0}^2 s^{(1,n)}L^n +
\left(\frac{\alpha_s(\mu)}{4\pi}\right)^2 \sum_{n=0}^4 s^{(2,n)}L^n + \dots \,,
\end{equation}
where we set $s^{(0,0)}=1$ for simplicity. The explicit solution to NNLO reads 
\begin{align}
\label{eq:stexp}
\tilde{s}(L,\alpha_s(\mu )) &= 1+\frac{\alpha_s(\mu)}{4\pi}
\left[\frac{\Gamma_0}{2}L^2 + L \gamma^s_0+s^{(1,0)}\right] \nonumber \\
&+\left(\frac{\alpha_s(\mu)}{4\pi}\right)^2
\bigg[\frac{\Gamma_0^2}{8}L^4 +\left(-\frac{\beta_0\Gamma_0}{6}+\frac{\Gamma_0
    \gamma^s_0}{2}\right)L^3 +\frac{1}{2}\left(\Gamma_1-\beta_0\gamma^s_0+
(\gamma^s_0)^2 + \Gamma_0 s^{(1,0)}\right)L^2 \nonumber \\
&\hspace{2.6cm}+(\gamma^s_1-\beta_0 s^{(1,0)}+\gamma^s_0 s^{(1,0)})L + s^{(2,0)}\bigg].
\end{align}
To turn this into an approximate NNLO formula for the momentum-space
soft function $S(\omega,\mu_f)$, one must take the limit $\mu_s=\mu_f$
and derive replacement rules analogous to 
(\ref{eq:conversion}). This is readily done using the expansion
\begin{equation}
\frac{1}{\omega}\left(\frac{\omega}{\mu_s}\right)^{2\eta}
=\frac{1}{2\eta}\,\delta(\omega)+\sum_{n=0}^{\infty}\frac{2^n}{n!}\,
D_n(\omega)\,\eta^n\,,
\end{equation}
where the $D_n$ are defined as  
\begin{equation}
D_n(\omega)=\left[\frac{1}{\omega}\,\ln^n\frac{\omega}{\mu_f}\right]_+ .
\end{equation} 

As was the case with the hard function, the all-orders solution for the
resummed soft function does not actually depend on the scale $\mu_s$, but its
truncation to a given logarithmic order (e.g.\ NNLL) introduces residual scale
dependence.  As explained earlier in the paper, our method is to choose $\mu_s$
to be close to the numerical value where the corrections from the soft
function to the (differential) cross section are minimal.  We then adopt the
parametric counting $\mu_s\sim \omega$ and apply RG-improved perturbation
theory with $\ln \mu_s/\mu_f\sim 1/\alpha_s$, and the exponential factors
resum logarithms of the form $\ln \mu_s/\mu_f$ to all orders. Since the scale
$\mu_s$ is dynamically generated through the numerical analysis, it does not
appear in the fixed-order calculation, so the resummation formula deals with
different types of corrections than the approximate fixed-order
formulas.\footnote{In \cite{Becher:2007ty} it was shown that the soft scale
decreases as the PDFs fall off more quickly away from values of $x$ where
$\omega\sim 0$, so the formulas effectively resum logarithms of the 
slopes of the PDFs.} For this reason, the structure of $D_n$ 
distributions appearing at a
given order is not the same in the two approaches.

We now show this in more detail, working first to NLO.  In this case, 
the ``approximate NLO'' formula derived from the
solution (\ref{eq:stexp}) reads
\begin{equation}
S(\omega,\mu_f)\approx 1+a(\mu_f)\left[4 \Gamma_0 D_1(\omega)+2\gamma^s_0 D_0(\omega)
-\frac{\pi^2}{3}\Gamma_0 \,\delta(\omega)\right].
\end{equation}
This should be compared with the expansion of  the NLL formula to 
NLO in fixed order, for which the NLO correction reads
(with $L_s\equiv\ln\mu_s^2/\mu_f^2$)
\begin{equation}
a(\mu_f)\left[2 \Gamma_0 L_s D_0(\omega)
+\left(\frac{1}{2}\Gamma_0 L_s^2 + \gamma_0^s L_s \right)\delta(\omega)\right]
+\dots \,.
\end{equation}
The ``approximate NLO'' formula has $D_1$ distributions, while the NLO
expansion of our NLL formula has only $D_0$ distributions.  
In the counting of RG-improved perturbation theory, however, $D_0
\ln(\mu_s/\mu_f)\sim D_1$, so the tower of logarithms produced by the
expansion of the NLL formulas is of course correct.  
The analogous formula at NNLO is rather lengthy, but to illustrate
its structure, we focus on the leading correction in the logarithmic
power counting, which reads 
\begin{equation}
\label{eq:LLterms}
a(\mu_f)^2 \Gamma_0^2\left[12 L_s D_2(\omega) -6 L_s^2 D_1(\omega)+L_s^3
  D_0-\frac{1}{8}L_s^4\,\delta(\omega)\right],
\end{equation}
while the  leading term of the ``approximate NNLO'' formula 
is
\begin{equation}
a(\mu_f)^2 \,8 \Gamma_0^2 \,D_3(\omega)\,.
\end{equation}
Again, for $\omega\sim \mu_s$ the terms in the two equations are of 
the same parameteric order but contain different types of 
distributions: the resummed formulas generate at most $D_2$ distributions,
while the approximate NNLO formulas generate $D_3$ distributions.

From the discussion above it should be obvious that our formula
does not literally resum the highest tower of $D_n$ distributions to 
all orders, but rather terms which count that way in RG-improved
perturbation theory, when the dynamically generated soft scale
satisfies $\mu_s\sim \omega$. In the case of the hard function, we noted that 
the fixed-order corrections produced by expanding the resummed formula
were equal to those in the  approximate formula for the special choice 
$\mu_h=M$.  For the soft function, there is no
numerical value of $\mu_s$ for which this would be true, but the two are 
equal if we replace $L_s\to \partial_\eta$ and take the derivatives 
before re-expanding $\eta$ in $\alpha_s(\mu_f)$.  This procedure
can be generalized to all orders by evaluating the formula
\begin{equation}
\label{eq:allorders}
S(\omega,\mu_f)=\left\{\left[e^{-4S(\mu_s,\mu_f)+2a_{\gamma^s}(\mu_s,\mu_f)}
\tilde{s}\left(0,\alpha_s(\mu_s)\right)\right]\bigg|_{\ln(\mu_s^2/\mu_f^2) \to \partial
  \eta }\right\}\frac{1}{\omega}\left(\frac{\omega}{\mu_f}\right)^{2\eta}\frac{e^{-2
    \gamma_E \eta}}{\Gamma(2\eta)}\bigg|_{\eta\to 0} \,,
\end{equation}
where the factor in the curly brackets is understood to be expanded to
all orders as a series in $\alpha_s(\mu_f)$ and so is a function only
of $\ln(\mu_s^2/\mu_f^2)$.  In this way, we exponentiate the derivatives
with respect to $\eta$, which are what generate the highest-order 
distributions, and the expansion of the above formula to any given 
accuracy in fixed-order reproduces the approximate formulas.  
For instance, if we include the exact one-loop matching (which is just
$s^{(1,0)}$) and the two-loop anomalous dimensions, the expansion of
the above object gives back our approximate NNLO formula for the soft
function, plus higher-order terms that resum all the higher-order
$D_n$ distributions at NNLL order, after converting the derivatives
with respect to $\eta$ with replacement rules.  While this procedure
generalizes choosing $\mu_h= M$ in the hard function, it is by no
means the same conceptually.  The exact hard function  is independent 
of $\mu_h$, so varying it around values $\mu_h\sim M$ 
gives a way of estimating the  higher-order terms.  However, it would make 
no sense to replace, for instance, $\ln(\mu_s^2/\mu_f^2)\to c_0 \partial_\eta$ 
in (\ref{eq:allorders}), with $c_0\neq 1$,
since the derivatives generate both $\mu$-independent and $\mu$-dependent terms.

The conclusion of this discussion is that results based on approximate
NNLO formulas contain different information than those based on
NLO+NNLL resummation.  This is not just due to a truncation of the
NNLL series to NNLO, but also to the fact that the resummation
formula exponentiates logarithms depending on the ratio $\ln\mu_s/\mu_f$ in
combination with higher-order logarithmic plus-distributions, with
$\mu_s$ a dynamically generated numerical soft scale. Such logarithms do
not appear in the fixed-order calculation, which is independent of
$\mu_s$.  Therefore, the choice between using approximate NNLO and
NLO+NNLL amounts to whether one takes seriously the improved
convergence of the soft function at a numerically small soft scale.
If so, one should use the resummed formulas, if not, one should use
the approximate fixed-order calculations. In practice, this question
can only be answered after a numerical analysis, and for this reason,
we have evaluated both types of formulas in the studies in
Sections~\ref{sec:threshold} and \ref{sec:pheno}.

\end{document}